\documentclass[twocolumn]{aastex63}

\graphicspath{{./}{figures/}}

\shorttitle{Molecular Gas and $\alpha_\mathrm{CO}$ in NGC~3351 Galaxy Center}
\shortauthors{Teng et al.}

\usepackage{amsmath}
\usepackage{bm}
\usepackage{lineno}

\usepackage{xparse}

\newcommand*\species[1]{%
    \ensuremath{\IfBold{\bm{\mathrm{#1}}}{\mathrm{#1}}}%
}

\newcommand*{\trans}[1]{%
    \ensuremath{\IfBold{\bm{\mytrans{#1}}}{\mytrans{#1}}}%
}

\newcommand*{\delim}{\text{\scalebox{0.7}[1.0]{\ensuremath{-}}}}
\newcommand*\mytrans[1]{%
    \ifnum#1=10 (1\delim0)\else%
    \ifnum#1=21 (2\delim1)\else%
    \ifnum#1=32 (3\delim2)\else%
    \ifnum#1=43 (4\delim3)\else%
    \ifnum#1=54 (5\delim4)\else%
    \ifnum#1=65 (6\delim5)\else%
    \ifnum#1=76 (7\delim6)\else%
    \ifnum#1=87 (8\delim7)\else%
    \ifnum#1=98 (9\delim8)\else%
    \ifnum#1=109 (10\delim9)\else#1%
    \fi\fi\fi\fi\fi\fi\fi\fi\fi\fi%
}

\makeatletter
\newcommand*{\IfBold}{%
  \ifx\f@series\my@test@bx
    \expandafter\@firstoftwo
  \else
    \expandafter\@secondoftwo
  \fi
}
\newcommand*{\my@test@bx}{bx}
\makeatother

\DeclareDocumentCommand\chem{ m g }{%
    {\species{#1}%
        \IfNoValueF {#2} {\,\trans{#2}}%
    }%
}

\begin{document}

\title{Molecular Gas Properties and CO-to-$\bm{\mathrm{H_2}}$ Conversion Factors in the Central Kiloparsec of NGC~3351}

\correspondingauthor{Yu-Hsuan Teng}
\email{yuteng@ucsd.edu}

\author[0000-0003-4209-1599]{Yu-Hsuan Teng}
\affiliation{Center for Astrophysics and Space Sciences, Department of Physics, University of California San Diego, \\
9500 Gilman Drive, La Jolla, CA 92093, USA}

\author[0000-0002-4378-8534]{Karin M. Sandstrom}
\affiliation{Center for Astrophysics and Space Sciences, Department of Physics, University of California San Diego, \\
9500 Gilman Drive, La Jolla, CA 92093, USA}

\author[0000-0003-0378-4667]{Jiayi Sun}
\affiliation{Department of Astronomy, The Ohio State University, 140 West 18th Avenue, Columbus, Ohio 43210, USA}

\author[0000-0002-2545-1700]{Adam K. Leroy}
\affiliation{Department of Astronomy, The Ohio State University, 140 West 18th Avenue, Columbus, Ohio 43210, USA}

\author[0000-0001-6421-0953]{L. Clifton Johnson}
\affiliation{Center for Interdisciplinary Exploration and Research in Astrophysics (CIERA) and Department of Physics and Astronomy, Northwestern University, 1800 Sherman Avenue, Evanston, IL 60201, USA}

\author[0000-0002-5480-5686]{Alberto D. Bolatto}
\affiliation{Department of Astronomy and Joint Space-Science Institute, University of Maryland, College Park, MD 20742, USA}
\affiliation{Visiting Scholar at the Flatiron Institute, Center for Computational Astrophysics, NY 10010, USA}
\affiliation{Visiting Astronomer at the National Radio Astronomy Observatory, VA 22903, USA}

\author[0000-0002-8804-0212]{J.~M.~Diederik~Kruijssen}
\affil{Astronomisches Rechen-Institut, Zentrum f\"ur Astronomie der Universit\"at Heidelberg, \\ M\"onchhofstra\ss e 12-14, D-69120 Heidelberg, Germany}

\author{Andreas Schruba}
\affiliation{Max-Planck-Institut f{\"u}r extraterrestrische Physik, Giessenbachstra{\ss}e~1, D-85748 Garching, Germany}

\author[0000-0003-1242-505X]{Antonio Usero}
\affil{Observatorio Astron{\'o}mico Nacional (IGN), C/Alfonso XII, 3, E-28014 Madrid, Spain}

\author[0000-0003-0410-4504]{Ashley T. Barnes}
\affiliation{Argelander-Institut f{\"u}r Astronomie, Universit{\"a}t Bonn, Auf dem H{\"u}gel 71, 53121 Bonn, Germany}

\author[0000-0003-0166-9745]{Frank Bigiel}
\affiliation{Argelander-Institut f{\"u}r Astronomie, Universit{\"a}t Bonn, Auf dem H{\"u}gel 71, 53121 Bonn, Germany}

\author[0000-0003-4218-3944]{Guillermo A. Blanc}
\affiliation{Observatories of the Carnegie Institution for Science, 813 Santa Barbara Street, Pasadena, CA 91101, USA}
\affiliation{Departamento de Astronom\'ia, Universidad de Chile, Camino del Observatorio 1515, Las Condes, Chile}

\author[0000-0002-9768-0246]{Brent Groves}
\affiliation{International Centre for Radio Astronomy Research, University of Western Australia, \\ 35 Stirling Highway, Crawley, WA 6009, Australia}

\author[0000-0002-6760-9449]{Frank P. Israel}
\affiliation{Sterrewacht Leiden, Leiden University, P.O. Box 9513, 2300 RA, Leiden, The Netherlands}

\author[0000-0001-9773-7479]{Daizhong Liu}
\affiliation{Max-Planck-Institut f{\"u}r extraterrestrische Physik, Giessenbachstra{\ss}e~1, D-85748 Garching, Germany}

\author[0000-0002-5204-2259]{Erik Rosolowsky}
\affiliation{Department of Physics, University of Alberta, Edmonton, AB T6G 2E1, Canada}

\author[0000-0002-3933-7677]{Eva Schinnerer}
\affiliation{Max-Planck-Institut f{\"u}r Astronomie, K{\"o}nigstuhl 17, D-69117, Heidelberg, Germany}

\author[0000-0003-1545-5078]{J.~D. Smith}
\affiliation{Department of Physics and Astronomy, University of Toledo, Ritter Obs., MS \#113, Toledo, OH 43606, USA}

\author[0000-0003-4793-7880]{Fabian Walter}
\affiliation{Max-Planck-Institut f{\"u}r Astronomie, K{\"o}nigstuhl 17, D-69117, Heidelberg, Germany}

\begin{abstract}

The CO-to-$\mathrm{H_2}$ conversion factor ($\alpha_\mathrm{CO}$) is critical to studying molecular gas and star formation in galaxies. The value of $\alpha_\mathrm{CO}$ has been found to vary within and between galaxies, but the specific environmental conditions that cause these variations are not fully understood. Previous observations on $\sim$kpc scales revealed low values of $\alpha_\mathrm{CO}$ in the centers of some barred spiral galaxies, including NGC~3351. We present new ALMA Band~3, 6, and~7 observations of $^{12}$CO, $^{13}$CO, and C$^{18}$O lines on $100$~pc scales in the inner ${\sim}2$~kpc of NGC~3351. Using multi-line radiative transfer modeling and a Bayesian likelihood analysis, we infer the $\mathrm{H_2}$ density, kinetic temperature, CO column density per line width, and CO isotopologue abundances on a pixel-by-pixel basis. Our modeling implies the existence of a dominant gas component with a density of $2{-}3\times10^3$~$\mathrm{cm^{-3}}$ in the central ${\sim}1$~kpc and a high temperature of $30{-}60$~K near the nucleus and near the contact points that connect to the bar-driven inflows. Assuming a $\mathrm{CO}/\mathrm{H_2}$ abundance of $3\times10^{-4}$, our analysis yields $\alpha_\mathrm{CO}{\sim}0.5{-}2.0$~$\mathrm{M_\odot\,(K~km~s^{-1}~pc^2)^{-1}}$ with a decreasing trend with galactocentric radius in the central ${\sim}1$~kpc. The inflows show a substantially lower $\alpha_\mathrm{CO}{\lesssim}0.1$~$\mathrm{M_\odot\,(K~km~s^{-1}~pc^2)^{-1}}$, likely due to lower optical depths caused by turbulence or shear in the inflows. Over the whole region, this gives an intensity-weighted $\alpha_\mathrm{CO}$ of ${\sim}1.5$~$\mathrm{M_\odot\,(K~km~s^{-1}~pc^2)^{-1}}$, which is similar to previous dust modeling based results at kpc scales. This suggests that low $\alpha_\mathrm{CO}$ on kpc scales in the centers of some barred galaxies may be due to the contribution of low optical depth CO emission in bar-driven inflows.   

\end{abstract}

\keywords{Barred spiral galaxies (136); CO line emission (262); Molecular gas (1073); Star forming regions (1565)}

\section{Introduction} \label{sec:intro}

Stars are born in cold and dense molecular clouds that are mainly composed of molecular hydrogen, $\mathrm{H_2}$. It is therefore known that molecular clouds play an important role in star formation and galaxy evolution \citep[see review by][]{2012ARA&A..50..531K}. However, emission from the most abundant molecule, $\mathrm{H_2}$, is not directly observable in cold molecular gas, since its lowest energy transition has an upper energy level $E/k \approx 510$~K and can only be seen in gas with temperatures above ${\sim}80$~K \citep{2016ApJ...830...18T}. Therefore, to trace cold molecular gas, the most common approach is to observe the low-$J$ rotational lines of the second most abundant molecule, carbon monoxide ($^{12}$C$^{16}$O; hereafter CO), and then apply a CO-to-$\mathrm{H_2}$ conversion factor to infer the total amount of molecular gas \citep[see review by][]{co-to-h2}. The CO-to-$\mathrm{H_2}$ conversion factor is formally defined as the ratio between $\mathrm{H_2}$ column density and the integrated intensity of CO 1--0:
\begin{equation}
X_\mathrm{CO} = \frac{N_\mathrm{H_2}}{I_\mathrm{CO(1-0)}}\ \rm \left(\frac{cm^{-2}}{K\ km~s^{-1}} \right)~.
\label{def_Xco}
\end{equation}
In mass units, Equation~\ref{def_Xco} can be rewritten as
\begin{equation}
\alpha_\mathrm{CO} = \frac{M_\mathrm{mol}}{L_\mathrm{CO(1-0)}}\ \rm \left(\frac{M_\odot}{K\ km~s^{-1}\ pc^2} \right)~,
\label{def_alpha_intro}
\end{equation}
where $M_\mathrm{mol}$ is the total molecular mass including the contribution from Helium ($M_\mathrm{mol} \sim 1.36\,M_\mathrm{H_2}$), and $L_\mathrm{CO(1-0)}$ is the CO 1--0 luminosity. $X_\mathrm{CO}$ can be converted to $\alpha_\mathrm{CO}$ by multiplying by a factor of $4.5\times10^{19}$ (see Section~\ref{sec:alpha_co} for more details). 

Various methods have been used to measure the value of $\alpha_\mathrm{CO}$ by estimating the total gas mass, including: using measured line-widths and sizes with the assumption of virial balance in giant molecular clouds (GMCs) \citep{1987ApJS...63..821S,1987ApJ...319..730S,2008ApJ...686..948B,2013ApJ...772..107D}; dust emission or extinction converted to gas mass with various assumptions on the dust-to-gas ratio \citep{2008ApJ...679..481P,2011ApJ...737...12L,2011A&A...536A..19P,2015ApJ...803...38I}; $\gamma$-ray emission converted to gas mass with knowledge of the cosmic ray flux \citep{2010ApJ...710..133A,2012A&A...538A..71A,2017A&A...601A..78R}; and the use of $^{13}$CO \citep{2018MNRAS.475.3909C}, [C\,{\small II}] \citep{2017MNRAS.470.4750A,2020ApJ...903...30B,2020A&A...643A.141M}, or the Kennicutt-Schmidt law based on star formation rate measurements \citep{2012AJ....143..138S,2013ApJ...764..117B}. These studies have found a roughly constant $\alpha_\mathrm{CO}$ value of ${\sim}4.4$ $\mathrm{M_\odot\ (K~km~s^{-1}~pc^2)^{-1}}$ (or $X_\mathrm{CO} \sim 2\times10^{20}$ $\mathrm{cm^{-2}\ (K~km~s^{-1})^{-1}}$ in molecular clouds in the disks of the Milky Way or other nearby spiral galaxies. 
As a result, many studies assume a constant $\alpha_\mathrm{CO}$ value similar to the Milky Way average.
However, $\alpha_\mathrm{CO}$ may depend on gas conditions such as density, temperature, metallicity, and velocity dispersion \citep[e.g.,][]{2012MNRAS.421.3127N,co-to-h2}, and the $\alpha_\mathrm{CO}$ values could vary by orders of magnitude in different environments as predicted from hydrodynamical simulations \citep{2012ApJ...747..124F,2020ApJ...903..142G}.
For instance, a low $\alpha_\mathrm{CO}$ value could be caused by conditions such as: (1) enhanced temperature and/or decreased density in GMCs that are virialized \citep[e.g.,][]{co-to-h2}, (2) GMCs with increased velocity dispersion relative to mass, which alters their virial balance \citep[e.g.,][]{2011MNRAS.411.1409W}, or (3) CO emission from molecular gas that is not associated with GMCs \citep[e.g.,][]{2015ApJ...801...25L}. On the other hand, $\alpha_\mathrm{CO}$ can also be extremely high in low-metallicity galaxies \citep[e.g.,][]{2018MNRAS.478.1716P,2020A&A...643A.141M} due to envelopes of ``CO-dark H$_2$'' \citep{2010ApJ...716.1191W}.

Studies have shown that galaxy centers tend to have lower $\alpha_\mathrm{CO}$, which may be due to higher temperatures and/or dynamical effects in galaxy centers \citep[e.g.,][]{2013ApJ...777....5S,2020A&A...635A.131I}. The value of $\alpha_\mathrm{CO}$ in the Galactic Center was found to be 3--10 times lower than in the Galactic disk \citep{1985A&A...143..267B,1995ApJ...452..262S,2001ApJ...562..348O,2012ApJ...750....3A}. In addition, previous kpc-scale observations also revealed substantially lower $\alpha_\mathrm{CO}$ values than the Milky Way average in many nearby galaxy centers, especially those with nuclear gas concentrations \citep{2004AJ....127.2069M,2009A&A...506..689I,2013ApJ...764..117B,2013ApJ...777....5S,2020A&A...635A.131I}. 
As stellar bars or spiral arms funnel gas to the centers and create concentrations of gas and star formation, galaxy centers frequently host the most active star formation in disk galaxies \citep[e.g.][]{2007ApJ...671.1388D,2021MNRAS.505.4310C}. Feedback from starburst and active galactic nuclei (AGNs) could also lead to significantly different physical conditions in the central kpc of galaxies. 
In (ultra-)luminous infrared galaxies (U/LIRGs) and/or galaxy mergers, $\alpha_\mathrm{CO}$ values were also found to be lower than those of the Milky Way clouds, with the gas usually being warmer, denser, and having higher CO isotopologue abundances  \citep{1998ApJ...507..615D,2012MNRAS.426.2601P,2014ApJ...796L..15S,2017ApJ...840....8S,2016A&A...594A..70K}.  
To understand molecular gas and star formation in galaxies, it is critical to study the variation of $\alpha_\mathrm{CO}$ and its relation to environmental conditions. Galaxy centers, having conditions of high gas surface density, high excitation, and/or altered molecular gas dynamics compared to the simple picture of isolated virialized GMCs, are thus ideal test beds for studying the influence of physical properties on $\alpha_\mathrm{CO}$.

In order to best diagnose the physical state of the molecular gas and reasons for $\alpha_\mathrm{CO}$ variation, it is necessary to observe the optically thin isotopologues of CO. Since the low-$J$ rotational lines of CO are optically thick in molecular clouds, this leads to a degeneracy in CO excitation and subsequently to uncertain column densities. Therefore, optically thin tracers are crucial for self-consistent measurements of excitation conditions or molecular gas column densities. One of the CO isotopologues, $^{13}$CO, has CO/$^{13}$CO abundance ratios ranging from ${\sim}20$ at the Galactic Center to ${\sim}70$ near the solar neighborhood \citep{1990ApJ...357..477L,2005ApJ...634.1126M}. With such a low abundance, $^{13}$CO is generally optically thin in molecular clouds except for the densest regions \citep{2011RAA....11..787T,2014A&A...564A..68S,2020MNRAS.497.1972B}. Another CO isotopologue, C$^{18}$O, is approximately an order of magnitude less abundant than $^{13}$CO, and is therefore optically thin with only few exceptions \citep{2008A&A...487..237W,2018A&A...612A.117A}. Several studies have used the lowest-$J$ transitions of CO isotopologues to constrain physical conditions in nearby galaxy centers at $\gtrsim$kpc scales \citep{1990ApJ...348..434E,2009A&A...506..689I,2017ApJ...836L..29J,2020A&A...635A.131I}. However, it is only recently that observations of CO isotopologues on cloud-scales (${\sim}100$~pc) have become routinely possible toward nearby galaxy centers, thanks to the high sensitivity and resolution of the Atacama Large Millimeter/\linebreak[0]{}submillimeter Array (ALMA). Therefore, it is now possible to obtain multiple rotational levels of optically thin CO isotopologue lines resolved in a galaxy center, which not only allows for more reliable physical results but also reveals the detailed variation of conditions in the galaxy center. 

In this study, we investigate cloud-scale variation of molecular gas properties and $\alpha_\mathrm{CO}$ using ALMA observations of CO isotopologues toward the center of NGC~3351. NGC~3351 is a barred spiral galaxy located at a distance of $9.96\pm0.33$~Mpc \citep{2021MNRAS.501.3621A}, with a (photometric) disk inclination of $45.1^\circ$ and position angle of $188.4^\circ$ \citep{2020ApJ...897..122L}. It has a total stellar mass of $2.3 \times 10^{10}~\mathrm{M_\odot}$ and a star formation rate of $1.3~\mathrm{M_\odot~\mathrm yr^{-1}}$ \citep{2021arXiv210407739L}. Early studies have found a circumnuclear ring\footnote{With higher resolution, it might be revealed as a tightly wound double spiral emanating from the inner region of the bar instead of a continuous ring-like structure. See \citet{1995ApJ...450...90H} for an example in NGC~1068.} with a diameter of ${\sim}20\arcsec$ (${\sim}1$~kpc) in the center of NGC~3351, harboring intense massive star formation activity \citep[e.g.,][]{1982A&AS...50..491A,2009AJ....137.4670L}. Previous observations in UV, H$\alpha$, Pa$\alpha$, and radio continuum also identified multiple star-forming complexes at $100$~pc scales along the ring \citep{1997ApJ...484L..41C,1997A&A...325...81P,2007MNRAS.378..163H,2020ApJS..248...25L,2021ApJ...913...37C}. The star formation rate and the stellar mass are estimated to be ${\sim}0.4~\mathrm{M_\odot~\mathrm yr^{-1}}$ and $3{-}10 \times 10^{8}~\mathrm{M_\odot}$ in the central few kpc region \citep{1997AJ....114.1850E,1997A&A...325...81P,2021ApJ...913...37C}. In addition, recent studies that analyzed the kinematics in the center of NGC~3351 have revealed bar-driven inflows that funnel gas/dust into the ring \citep[e.g.][]{2019MNRAS.488.3904L,2020ApJ...897..122L}. Notably, there are no signs of AGN activity in the nucleus of NGC~3351 \citep{2009MNRAS.398.1165G,2011ApJ...731...60G,2019MNRAS.482..506G}, while \citet{2018MNRAS.473.4582L} showed that the nucleus may be a recent star formation site. This indicates that emission from the center of NGC~3351 is dominated by star formation. \citet{2019MNRAS.488.3904L} also showed that stellar feedback processes may drive a nuclear outflow in NGC~3351 even without an AGN host.
At an angular resolution of ${\sim}40\arcsec$, \citet{2013ApJ...777....5S} found a $\alpha_\mathrm{CO}$ value ${\sim}6\times$ lower than the standard Milky Way value in the central $1.7$~kpc of NGC~3351 with an uncertainty of $<0.2$~dex. This region has an oxygen abundance similar to the solar value, so the contrast cannot be explained by a metallicity difference \citep{2010ApJS..190..233M}.

In this paper, we present new ALMA observations toward the inner ${\sim}2$~kpc of NGC~3351 with the CO isotopologues $^{12}$CO, $^{13}$CO, and C$^{18}$O in their lowest rotational transitions $J$=1--0, 2--1, and 3--2 at a matched angular resolution of $2.1\arcsec$ (${\sim}100$~pc). To understand what physical processes control $\alpha_\mathrm{CO}$ in galaxy centers, we study the distribution of environmental conditions and $\alpha_\mathrm{CO}$ values in this region and investigate the possible causes of such variations. In Section~\ref{sec:obs}, we describe the details of observations and data reduction. The resulting spectra, images and line ratios are shown in Section~\ref{sec:result}. In Section~\ref{sec:model}, we present non-local thermodynamic equilibrium (non-LTE) radiative transfer modeling and results. Section~\ref{sec:alpha_co} shows our analysis of $\alpha_\mathrm{CO}$ variation and its possible implications. Finally, we compare our results with other literature or observations in Section~\ref{sec:discussion}, and summarize our findings in Section~\ref{sec:conclusion}.

\section{Observations and Data Reduction} \label{sec:obs}

\begin{table*}
\centering
\caption{ALMA Observations \label{tab:obs}}
\begin{tabular}{lccccccc}
  \hline\hline
  Line &Project &Rest Frequency &$n_\mathrm{crit}$ at $20$~K &Array &Beam Size &LAS &rms per channel\\
  &Code &(GHz) &($\rm cm^{-3}$) & &($''$) &($''$) &($\mathrm{K~(2.5~km~s^{-1})^{-1}}$)\\ 
  \hline
  CO J=1--0 &(1) &115.27 &$2.2\times10^3$ &12m &2.1 &11.9 &0.320 \\ 
  CO J=2--1 &(2) &230.54 &$1.1\times10^4$ &12m+7m+TP &1.5 &-- &0.108 \\ 
  $\rm ^{13}CO$ J=2--1 &(1) &220.40 &$9.4\times10^3$ &12m &1.8 &10.4 &0.036 \\ 
  $\rm ^{13}CO$ J=3--2 &(3) &330.59 &$3.1\times10^4$ &12m &1.2 &7.6 &0.032\\ 
  $\rm C^{18}O$ J=2--1 &(1) &219.56 &$9.3\times10^3$ &12m &1.8 &10.4 &0.036\\ 
  $\rm C^{18}O$ J=3--2 &(3) &329.33 &$3.1\times10^4$ &12m &1.2 &7.6 &0.032\\ 
  \hline
\end{tabular}
\tablecomments{ALMA project codes: (1)~2013.1.00885.S, (2)~2015.1.00956.S, (3)~2016.1.00972.S. Critical densities ($n_\mathrm{crit}$) are calculated from the Einstein-A and collisional rate coefficients provided by the Leiden Atomic Molecular Database \citep{2005A&A...432..369S}. In optically thick cases with $\tau_0 = 5$, $n_\mathrm{crit}$ of CO 1--0 and 2--1 at $20$~K decrease to $6.1\times10^2$ and $3.1\times10^3$ $\mathrm{cm}^{-3}$, respectively \citep[see][]{2015PASP..127..299S}. The rms of CO 1--0 is higher due to it being observed in a more extended array configuration, which has lower surface brightness sensitivity.}
\end{table*}

\subsection{ALMA Observations} \label{subsec:alma_obs}

We obtained ALMA Band~3, 6, and~7 observations to capture CO, $^{13}$CO, and C$^{18}$O lines as well as millimeter continuum emission from the central ${\sim}2$~kpc of NGC~3351 (see Table~\ref{tab:obs}).
These observations were designed to resolve the gas structures around the starburst ring at $1{-}2\arcsec$ (${\sim}50{-}100$~pc) resolution. 
In this paper, we focus on the molecular line observations and their implied gas conditions. Results of the dust continuum will be included in a future work. 

Our Band~3 observation targets the CO\,1--0 line and $100{-}115$~GHz continuum.
It uses the C43-3 configuration of the ALMA 12m array to reach an angular resolution of $2.1\arcsec \times 1.3\arcsec$ for the CO line, and a single 12m pointing covered the entire central region (${\sim}30\arcsec$ in diameter).
The achieved root-mean-square (rms) noise level is $0.32$~K per $2.5$~km~s$^{-1}$ channel (averaged across the field-of-view, out to a primary beam response threshold of 0.25). We note that Band~3 has the lowest frequency, and thus it requires a more extended array configuration to achieve an angular resolution similar to the Band~6 or~7 observations. With a more extended array configuration, the surface brightness sensitivity decreases, which leads to a higher rms level for our Band~3 data compared to the other bands for a given integration time. Although a longer integration time in Band~3 observations can reduce the noise level, the rms of $0.32$~K per channel is already sufficient for secure detection of CO 1--0 over the central kpc region of NGC~3351.

Our Band~6 observation targets the $J$=2--1 transition of the $^{13}$CO and C$^{18}$O lines as well as the $215{-}235$~GHz continuum.
With the C43-1 configuration of the ALMA 12m array, it achieves an angular resolution of $1.8\arcsec\times1.2\arcsec$ and uses a 7-pointing mosaic to cover the central $30\arcsec \times 30\arcsec$ area.
The achieved rms noise level is $0.036$~K per $2.5$~km~s$^{-1}$ channel.

Our Band~7 observation targets the $J$=3--2 transition of the $^{13}$CO and C$^{18}$O lines and the $325{-}345$~GHz continuum.
The ALMA C40-1 configuration provides an angular resolution of $1.2\arcsec \times 1.0\arcsec$ and a 14-pointing mosaic covers the central $30\arcsec \times 30\arcsec$ area.
The achieved rms noise level is $0.032$~K per $2.5$~km~s$^{-1}$ channel.

To complement these observations obtained specifically for this project, we also include the PHANGS--ALMA CO\,2--1 data \citep{2021arXiv210407739L} in our analysis to aid the imaging and provide additional constraints on the gas temperature and isotopologue ratios.
This data set was observed in Cycle 3 (project code: 2015.1.00956.S) and it reaches a similar angular resolution as the other observations used in this project.
It combines ALMA 12m, 7m, and total-power (TP) observations to ensure a complete $u{-}v$ coverage. 
The characteristics of the PHANGS--ALMA data are described in detail in \citet{2021arXiv210407739L}.

Our Band~3, 6, and 7 observations were taken with the 12m array alone and thus lack short-spacing information (see Table~\ref{tab:obs} for the largest angular structure (LAS) recoverable for each line). To estimate how much emission these 12m-only observations can recover, we perform a test with the PHANGS CO 2--1 data, which includes coverage of all $u{-}v$ scales with 12m and 7m arrays as well as TP observations. We imaged the 12m-only PHANGS data and the 12m+7m+TP and compared the resulting maps to check flux recovery. By dividing the 12m-only image by the combined image, we find a flux recovered ratio of ${\sim}97\%$ over the entire observed region. The analyses presented in this paper include only the pixels within a flux recovered ratio of $1 \pm 0.3$.

\subsection{Calibration and Imaging}

We calibrate the raw data with the calibration scripts provided by the observatory. We use the appropriate versions of the CASA pipeline to calibrate the Cycle~2 and Cycle~5 observations as recommended by the observatory (version 4.2.2 and 5.1.1, respectively).

We then image the CO lines and continuum separately with the CASA task \texttt{tclean} in two steps, as inspired by the PHANGS-ALMA imaging scheme \citep{2021ApJS..255...19L}.
We weight the $u$--$v$ data according to the ``Briggs'' scheme with a robustness parameter
$r=0.5$, which offers a good compromise between noise
and resolution.
We first perform a shallow, multi-scale cleaning to pick up both compact and extended emission down to a signal-to-noise ($\mathrm{S/N}$) ratio of~$3$ throughout the entire field of view.
After that, we perform a deep, single-scale cleaning to capture all remaining emission down to $\mathrm{S/N} \sim 1$.
This second step is restricted to within a cleaning mask, which is constructed based on the presence of significant CO\,2--1 detection in the PHANGS--ALMA data.
This way, the deep cleaning process can efficiently recover most remaining signals at their expected position-position-velocity ($ppv$) locations.

These calibration and imaging procedures produce high-quality data cubes for the CO lines as well as 2D images for the continuum emission.

\subsection{Product Creation and Error Estimation}

From the CO line data cubes, we derive 2D maps of integrated line intensity (moment~0), intensity-weighted velocity (moment~1), and velocity dispersion estimated using the effective line width estimator, as well as their associated uncertainties. We describe our methodology here, and note that it is very similar to that adopted in the PHANGS--ALMA pipeline \citep[also see \citealt{2018ApJ...860..172S}]{2021ApJS..255...19L}.

First, we convolve all data cubes to the finest possible round beam ($\mathrm{FWHM} = 2.1\arcsec$; set by the CO 1--0 resolution) to ensure that all CO line data probe the same spatial scale.
Second, we estimate the local rms noise in the data cubes by iteratively rejecting CO line detection and calculating the median absolute deviation (MAD) for the signal-free $ppv$ pixels.
Third, we create a signal mask for each data cube by finding all $ppv$ positions with $\mathrm{S/N} > 5$ detection in more than two consecutive channels, and then expanding this signal mask to include all morphologically connected $ppv$ positions with $\mathrm{S/N} > 2$ detection in more than two consecutive channels.
Fourth, we combine the signal masks for all CO lines in ``union'' (i.e., logical OR) to generate a master signal mask.
Fifth, we collapse all CO line data cubes within this master signal mask to create moment maps, and calculate the associated uncertainties based on Gaussian error propagation.
Finally, we regrid all the data products such that the new grid Nyquist-samples the beam (i.e., grid spacing equals half of the beam FWHM).

This product creation scheme yields a coherent set of moment maps and uncertainty maps for all the CO lines, which have matched resolution and consistently cover the same $ppv$ footprint.

\section{Results} \label{sec:result}

\begin{figure*} \centering
\includegraphics[width=\linewidth]{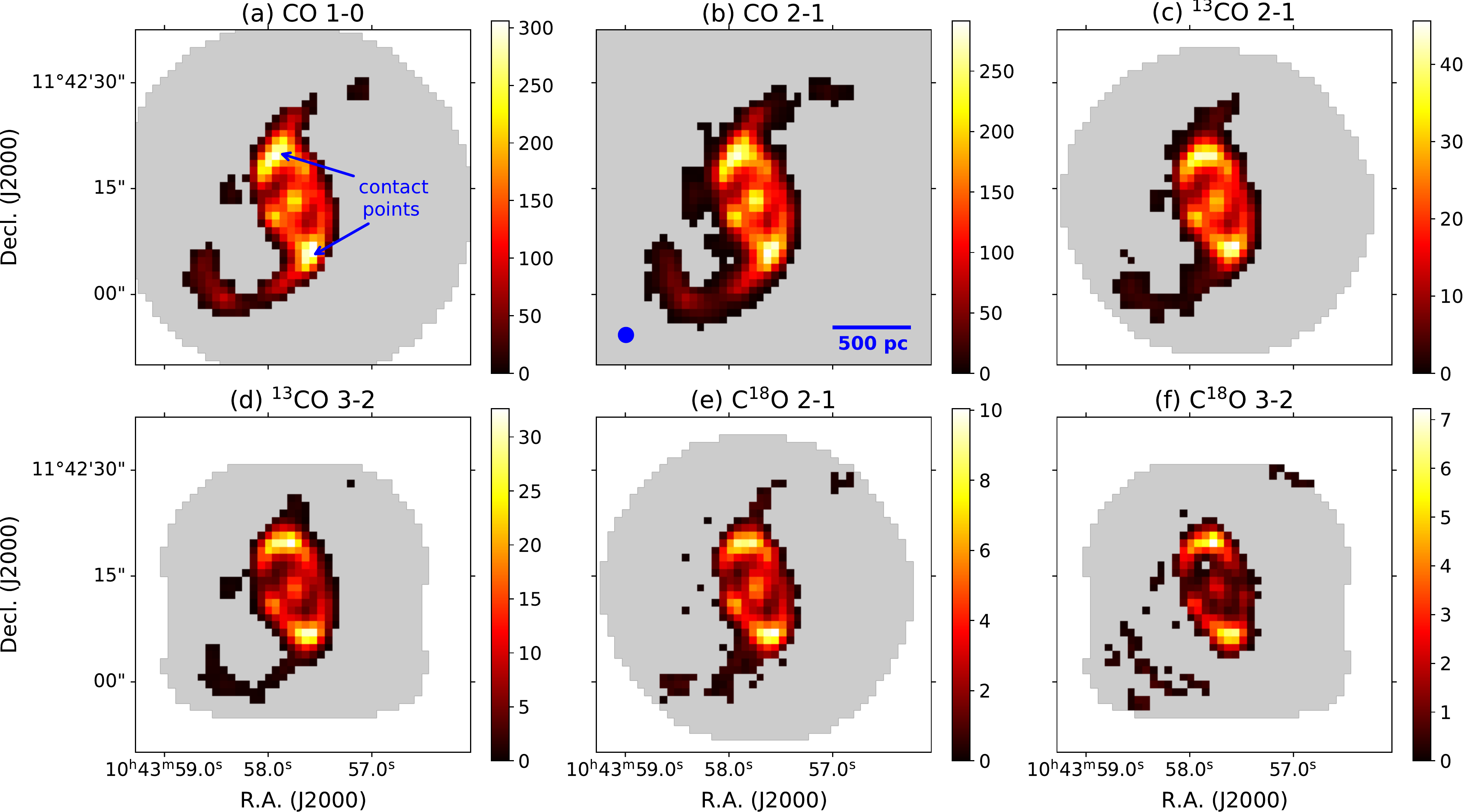} 
\caption{Integrated intensity maps of several CO isotopologues and rotational transitions (in units of $\mathrm{K~km~s^{-1}}$). The matched beam size of $2.1\arcsec$ and a scale bar of 500 pc is shown in panel (b). The white areas lie outside the field of view of ALMA observations, while the gray areas show the pixels without confident detection (i.e. $< 3\sigma$ for $^{12}$CO and $^{13}$CO and $< 1\sigma$ for C$^{18}$O). These images resolve a circumnuclear star-forming ring with ${\sim}20\arcsec$ or 1~kpc in diameter, a gap between the ring and the nucleus, and two bar-driven inflows connected to the ``contact points'' at the northern and southern parts of the ring.}
\label{fig:mom0}
\end{figure*}

\begin{figure*}
\begin{minipage}{.39\linewidth}
\centering
\includegraphics[width=\linewidth]{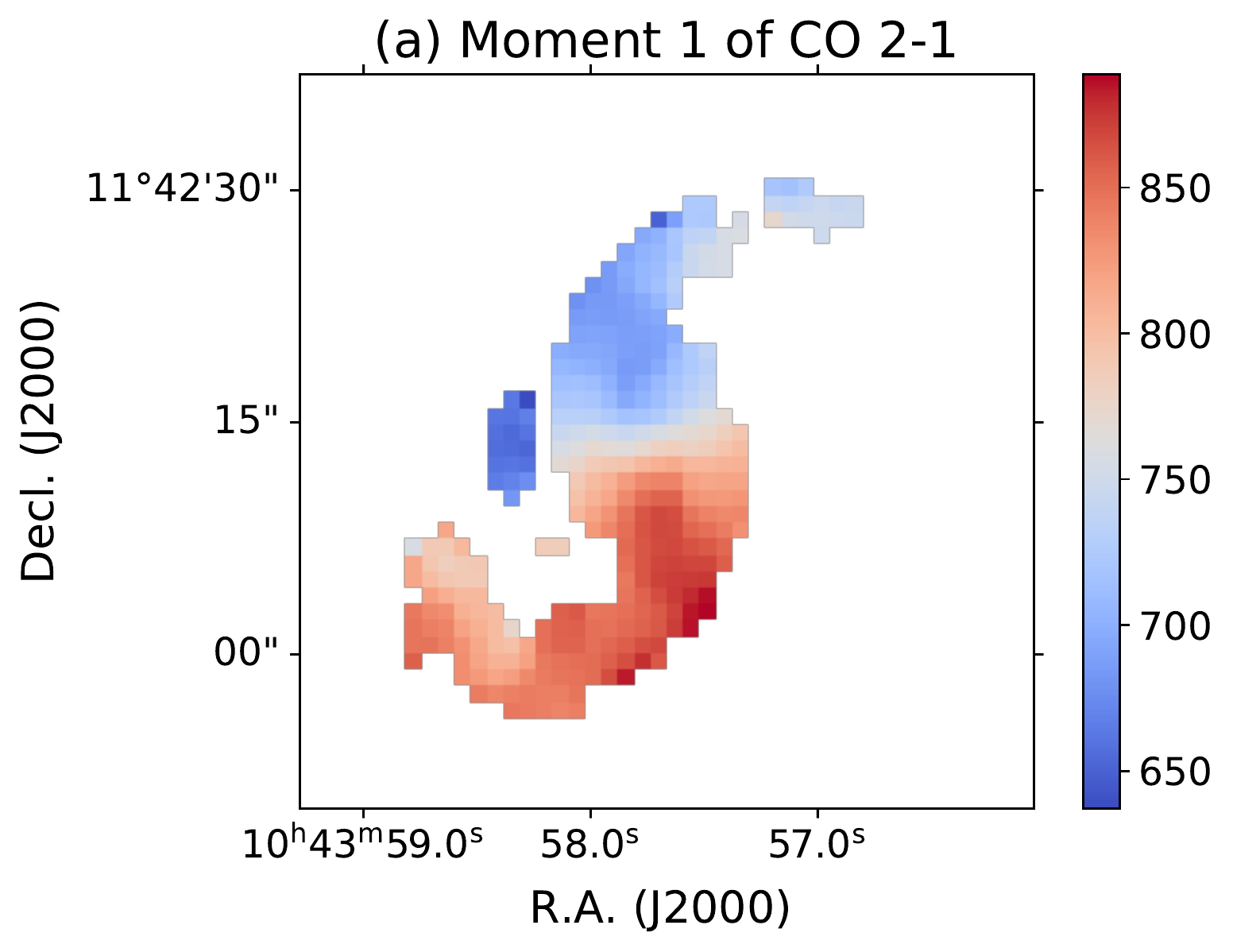}
\end{minipage}
\hfill
\begin{minipage}{.31\linewidth}
\centering
\includegraphics[width=\linewidth]{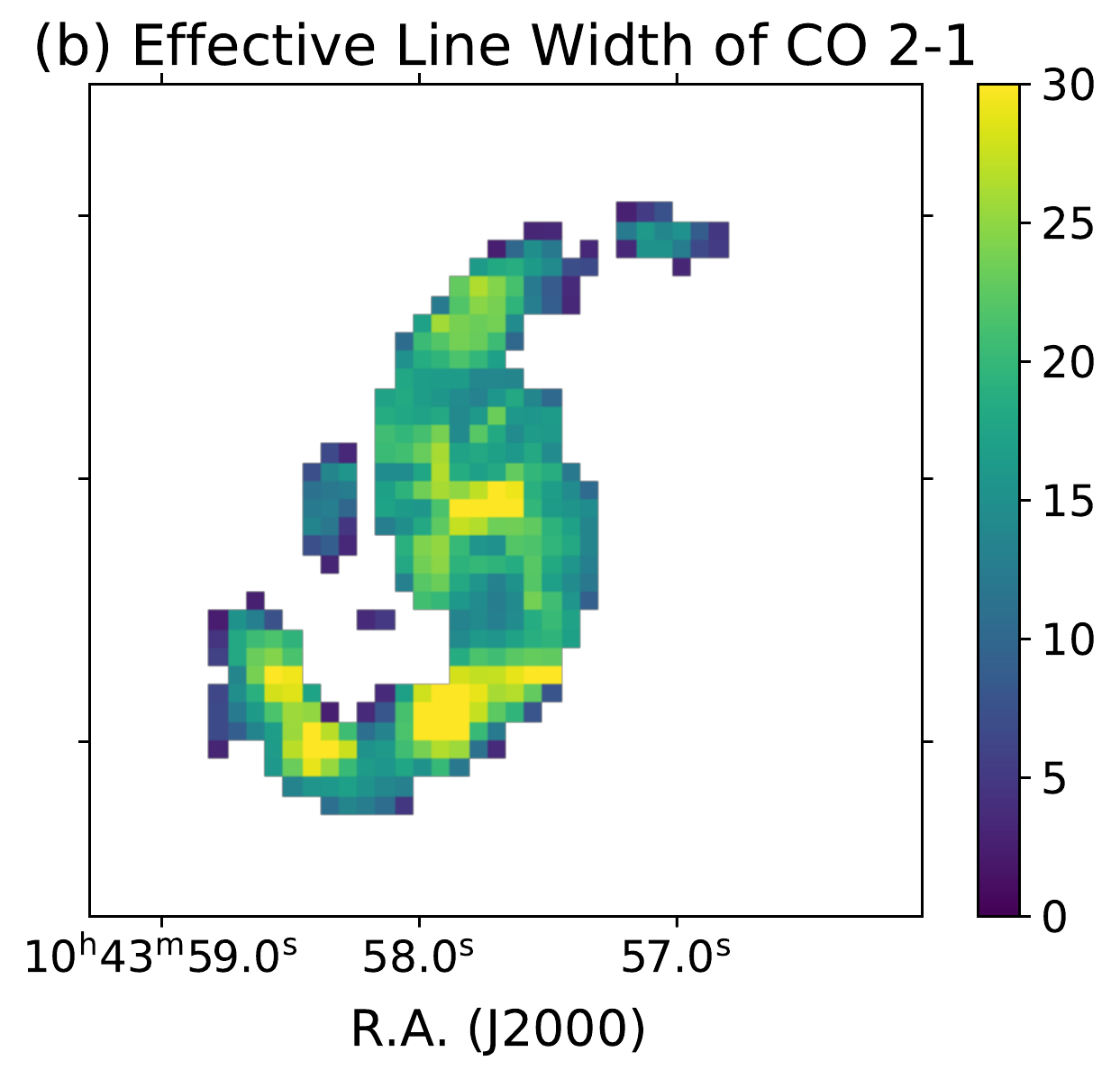}
\end{minipage}
\hfill
\begin{minipage}{.27\linewidth}
\centering
\includegraphics[width=\linewidth]{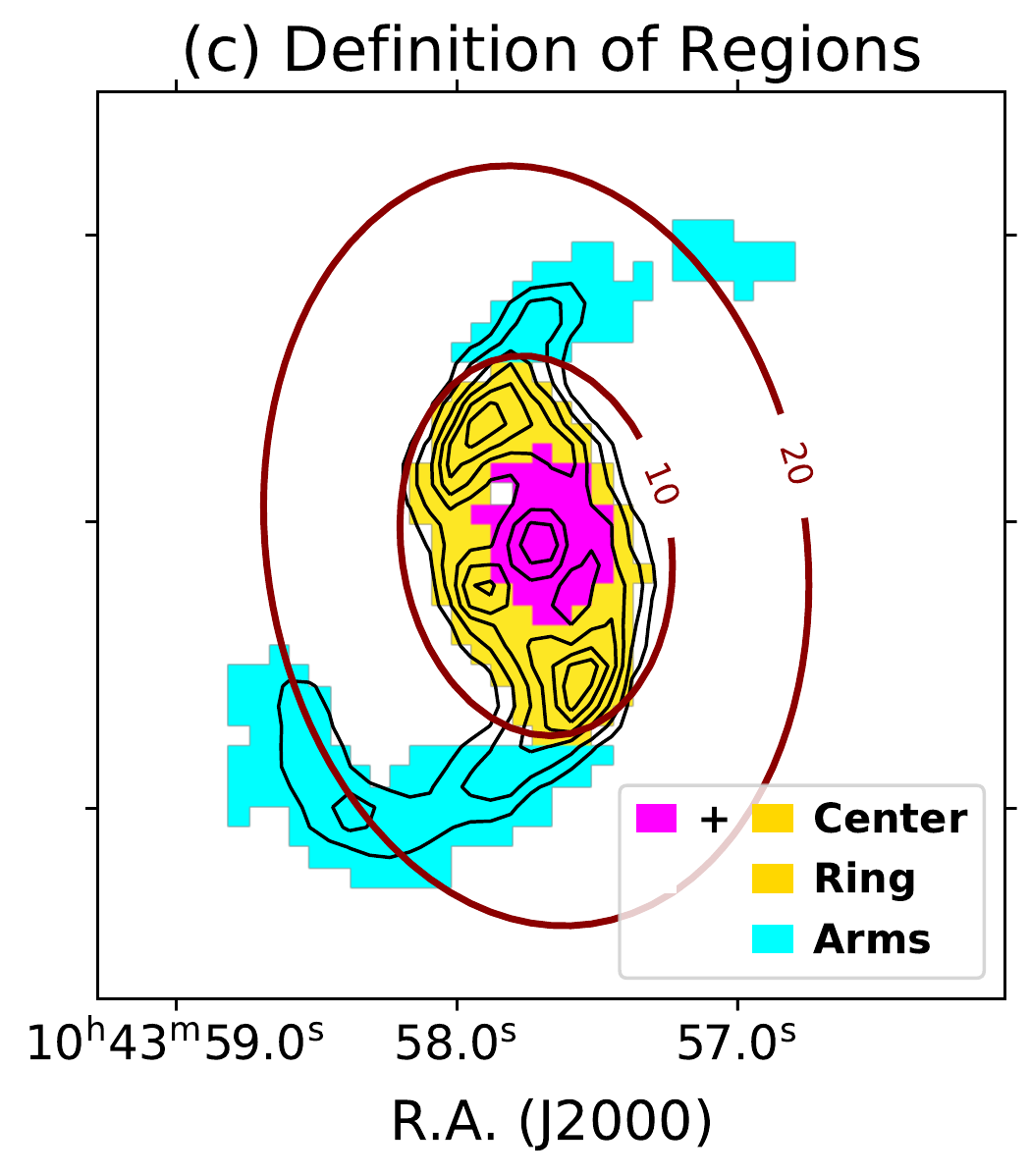}
\end{minipage}
\caption{(a) Intensity-weighted velocity (moment~1) and (b) effective line width of CO 2--1, both in units of km~s$^{-1}$. The moment 1 shows a clear sign of counterclockwise gas rotation, and the line widths are broader at the nucleus and along the arms. (c) Definition of the center, ring, and arm regions, overlaid with contour levels of the CO 2--1 integrated intensity at $I_{\rm CO(2-1)} = 20, 50, 100, 150, 200, 250$ K~km~s$^{-1}$ and the galactocentric radius at $10\arcsec$ and $20\arcsec$, respectively. We will present various regional statistics based on these defined regions.}
\label{fig:velocity}
\end{figure*}

\begin{figure*} 
\centering
\includegraphics[width=0.32\linewidth]{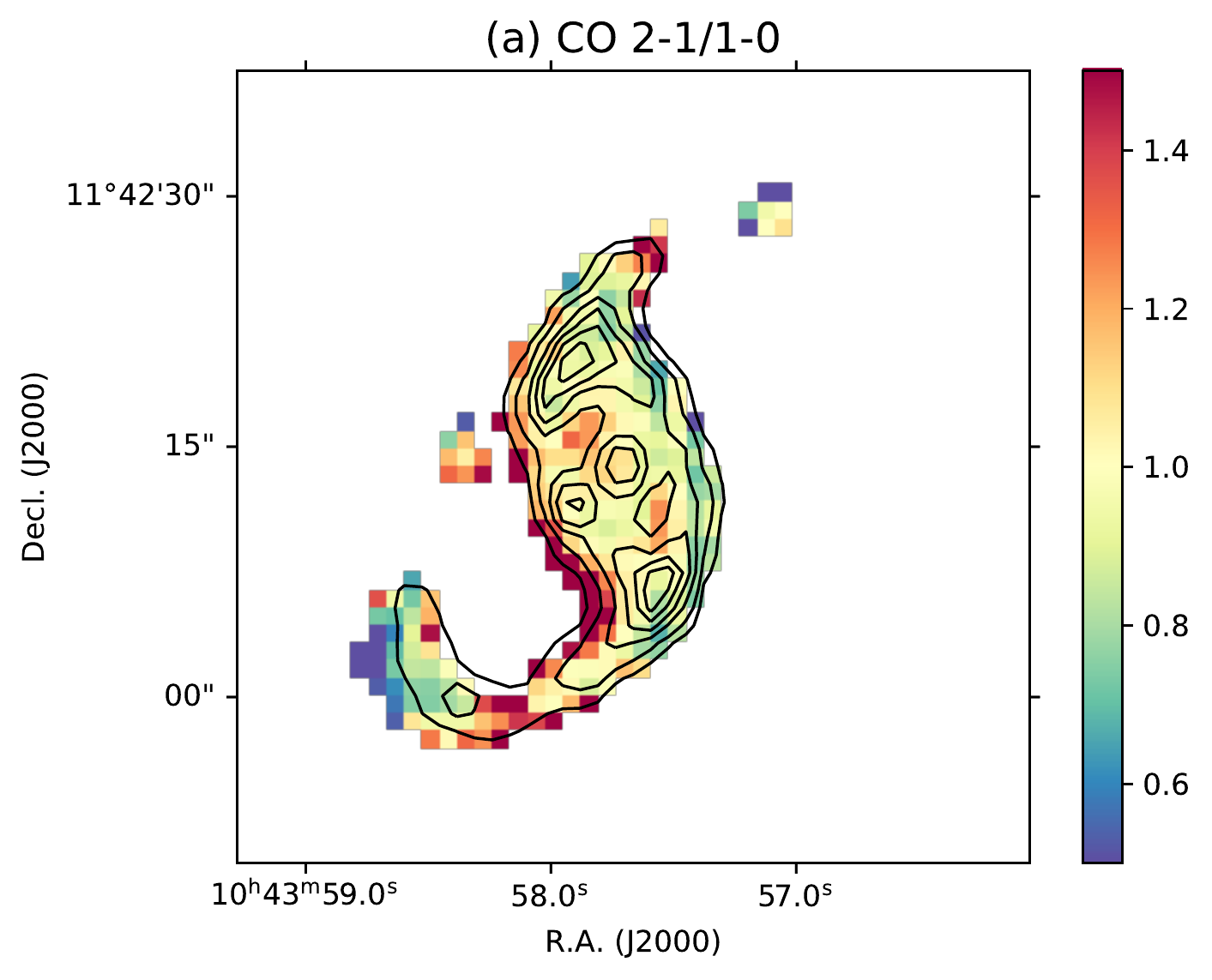} \includegraphics[width=0.32\linewidth]{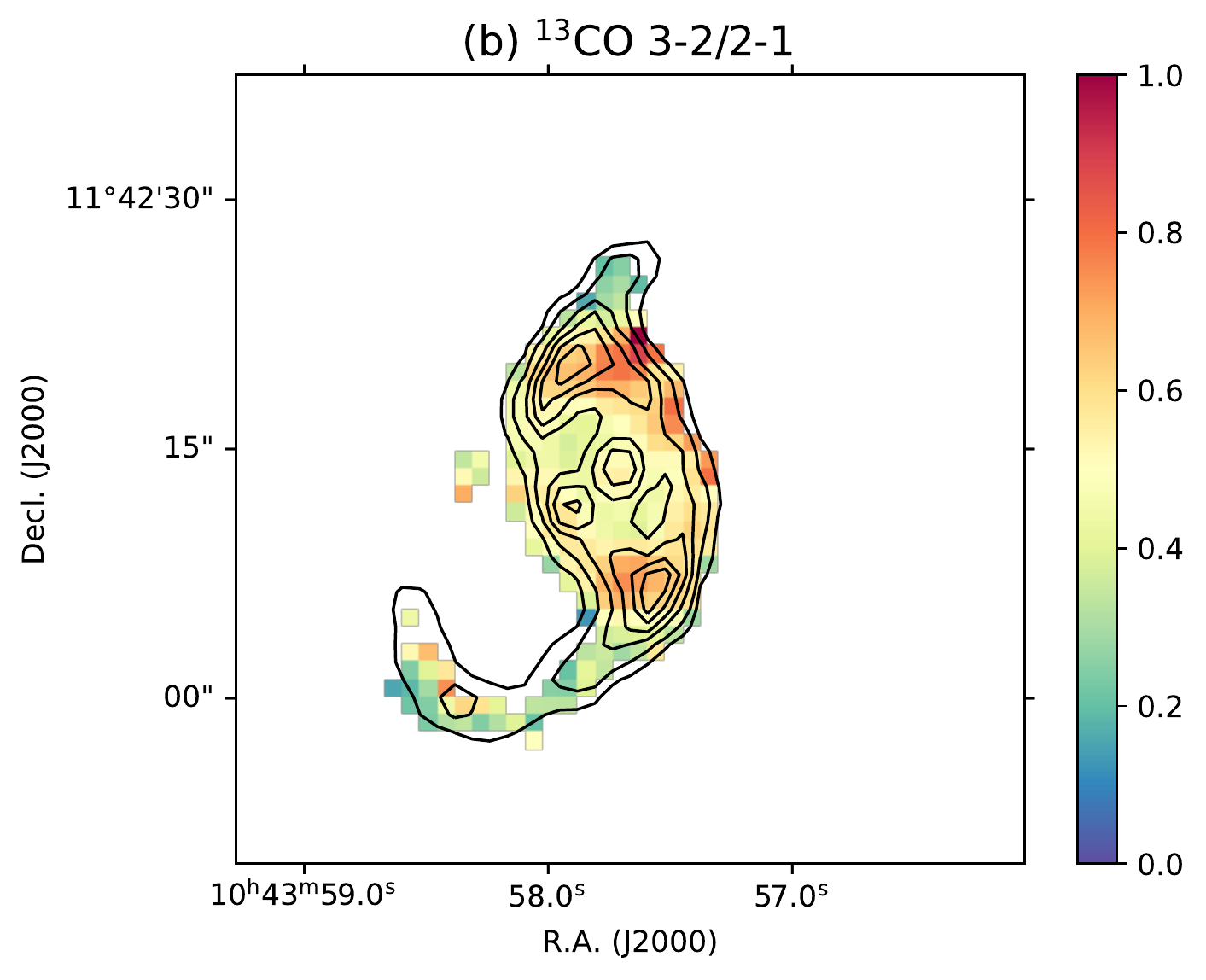} \includegraphics[width=0.32\linewidth]{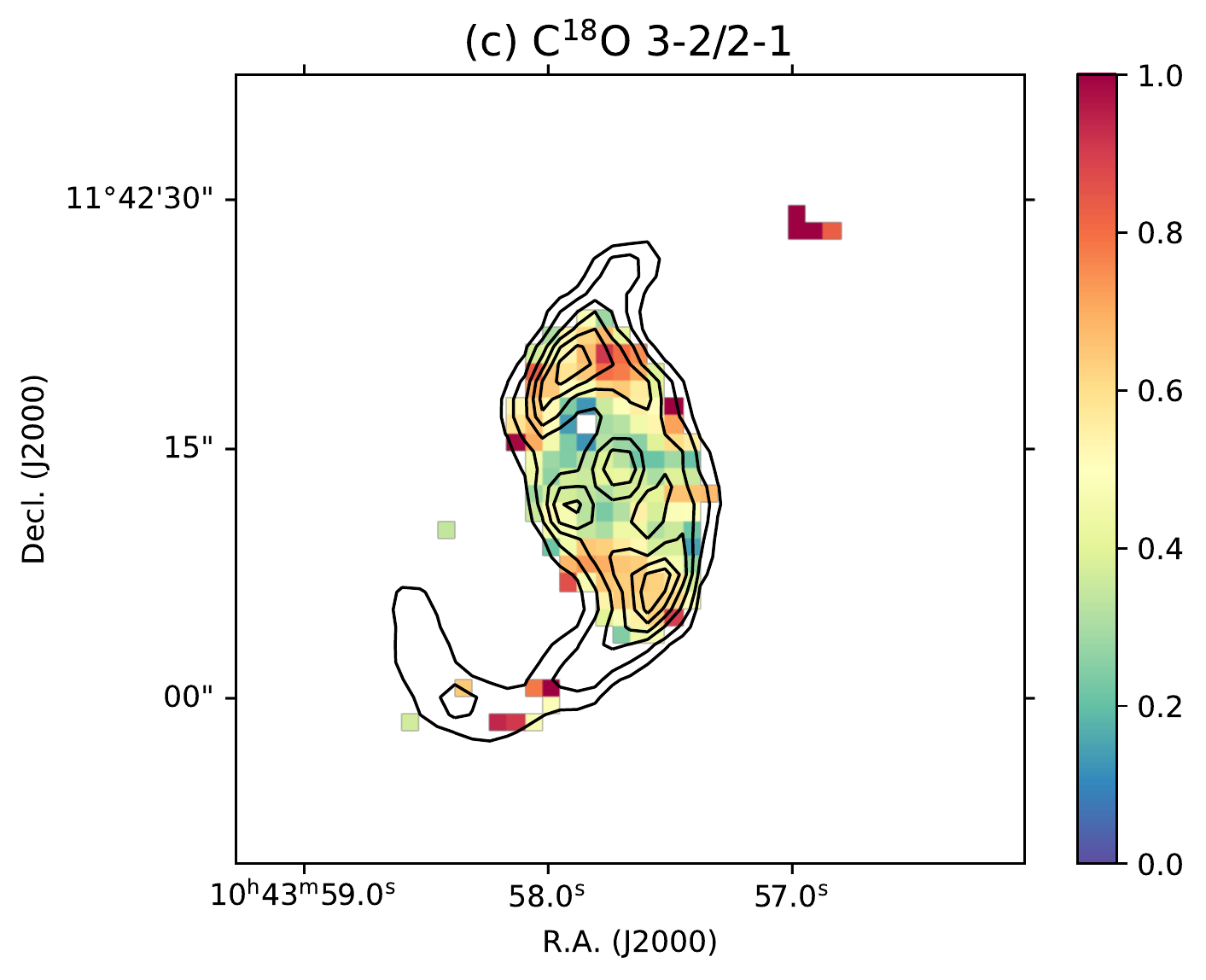} \\
\includegraphics[width=0.33\linewidth]{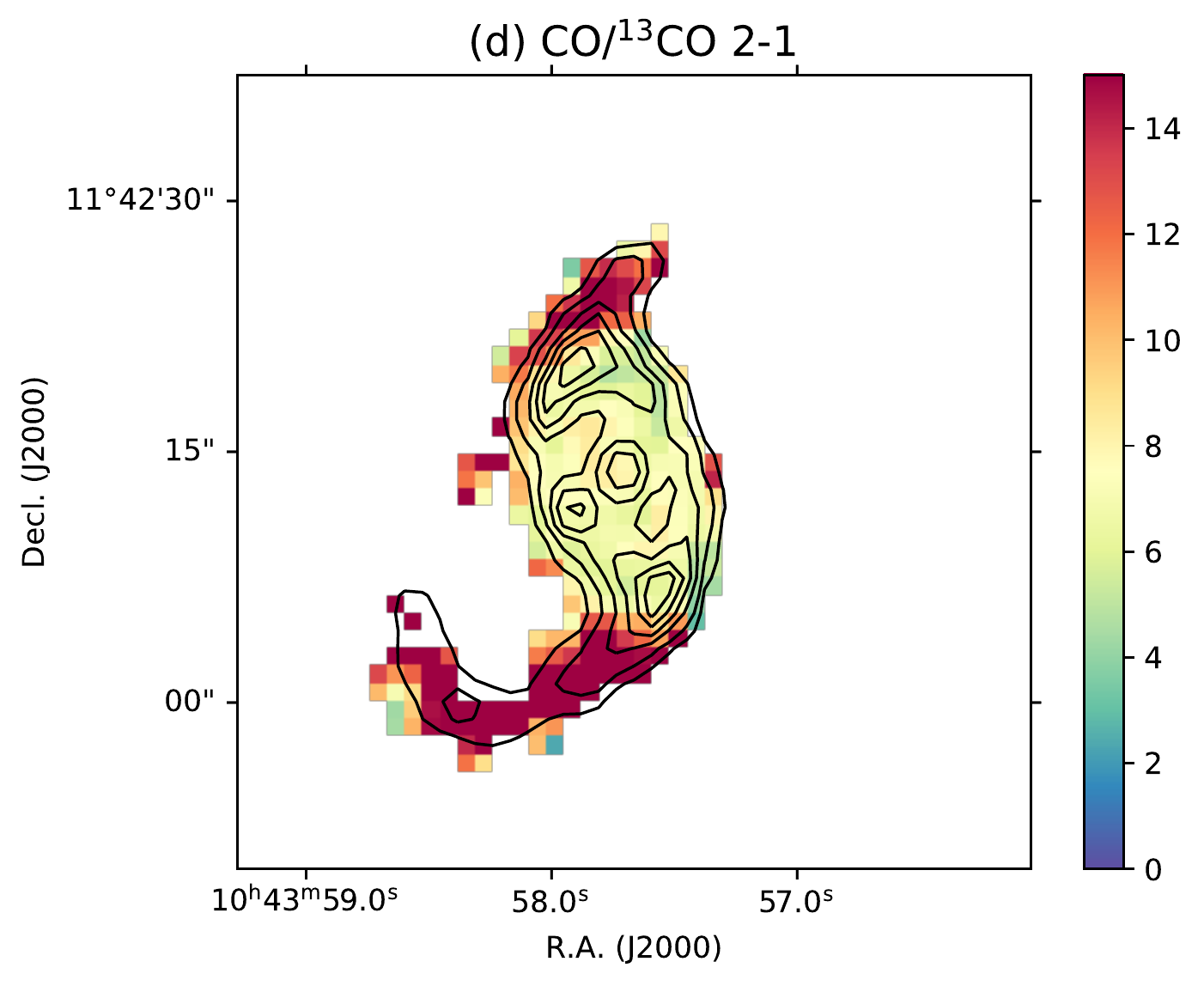} 
\includegraphics[width=0.33\linewidth]{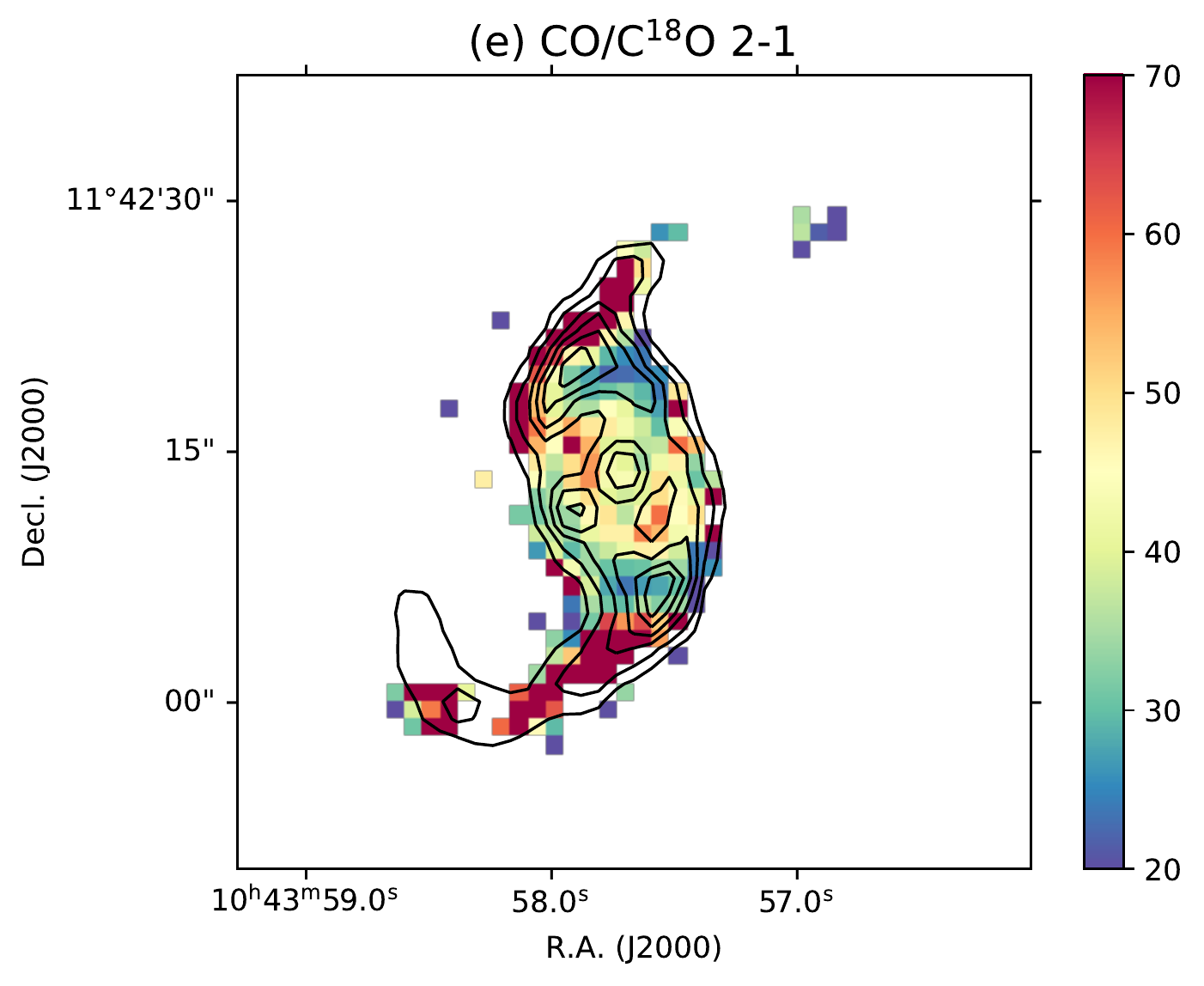} \\ \includegraphics[width=0.33\linewidth]{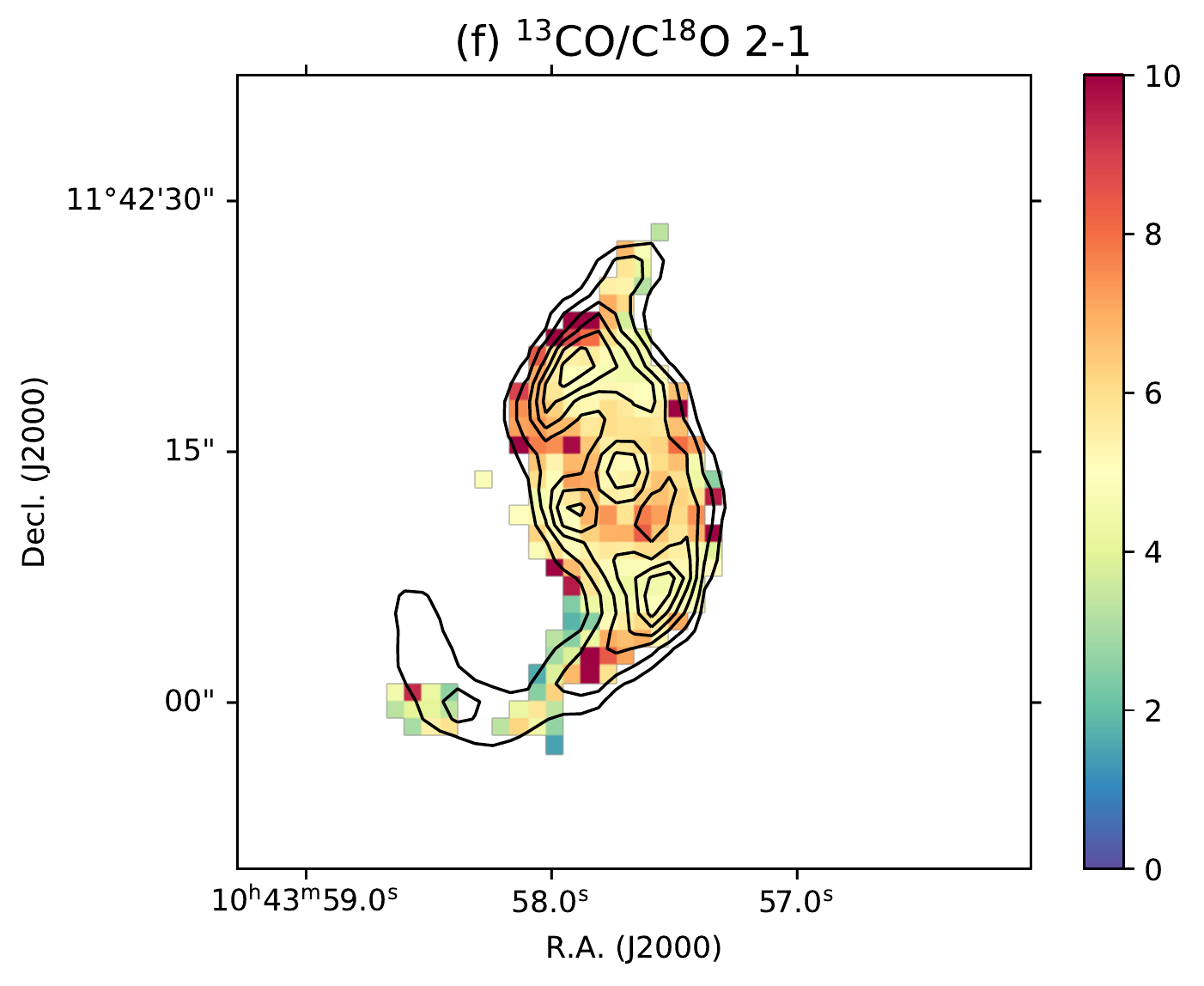} \includegraphics[width=0.33\linewidth]{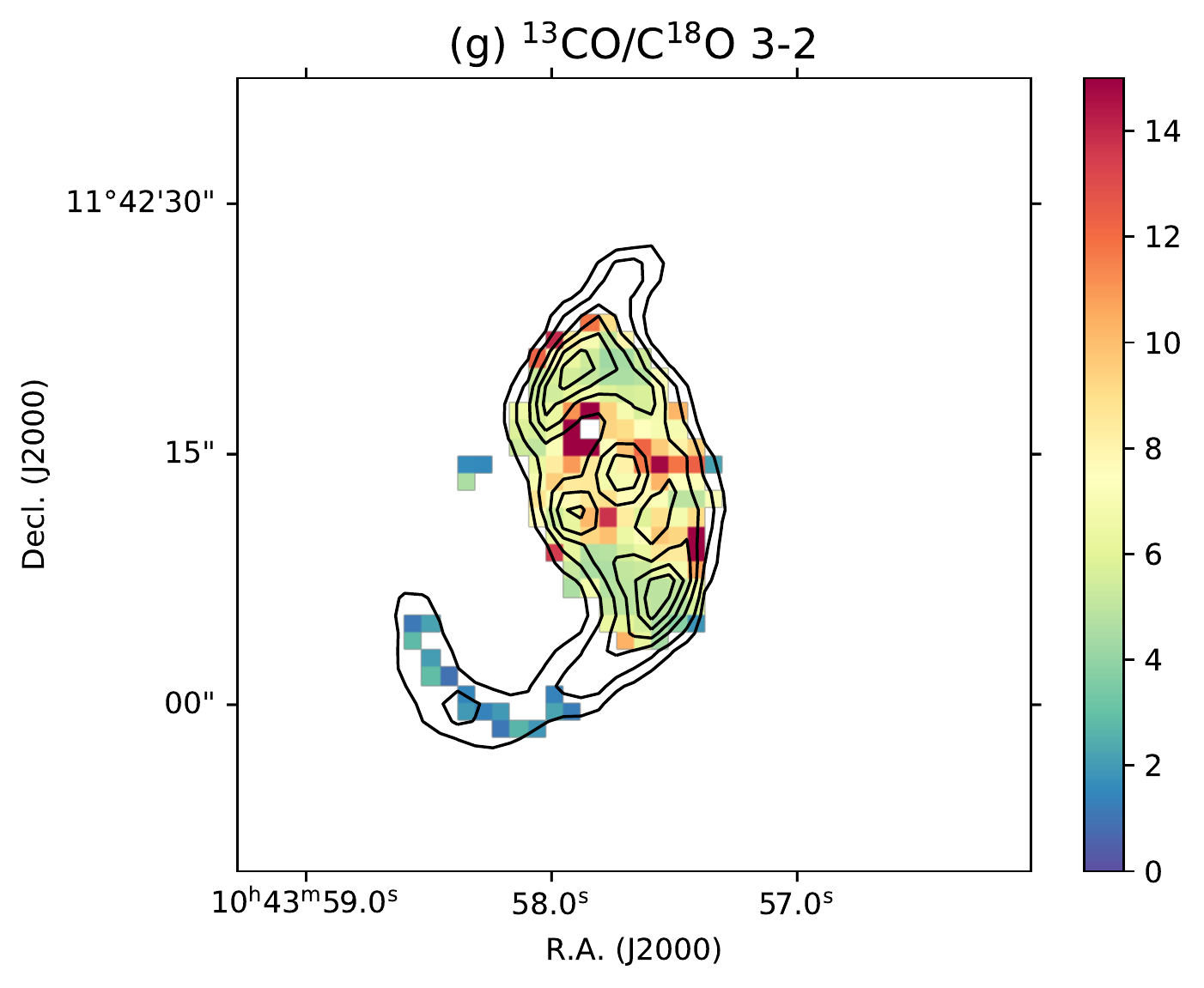}
\caption{Line ratio maps. A $3\sigma$ mask in both relevant lines is applied to each panel, except for the C$^{18}$O lines where a $1\sigma$ cutoff is applied. Contour levels of the CO 2--1 emission are the same as in Figure~\ref{fig:velocity}c. (a)--(c) show the primarily temperature-sensitive line ratios, and (d)--(g) show the line ratios primarily sensitive to isotopologue abundances or optical depths. The CO 2--1/1--0 ratio of ${\sim}1$ suggests optically thick and thermalized emission in this region. Notably, the arm regions show significantly higher CO/$^{13}$CO and CO/C$^{18}$O line ratios than the center.}
\label{fig:ratio}
\end{figure*}

\begin{figure*} \centering
\includegraphics[width=\linewidth]{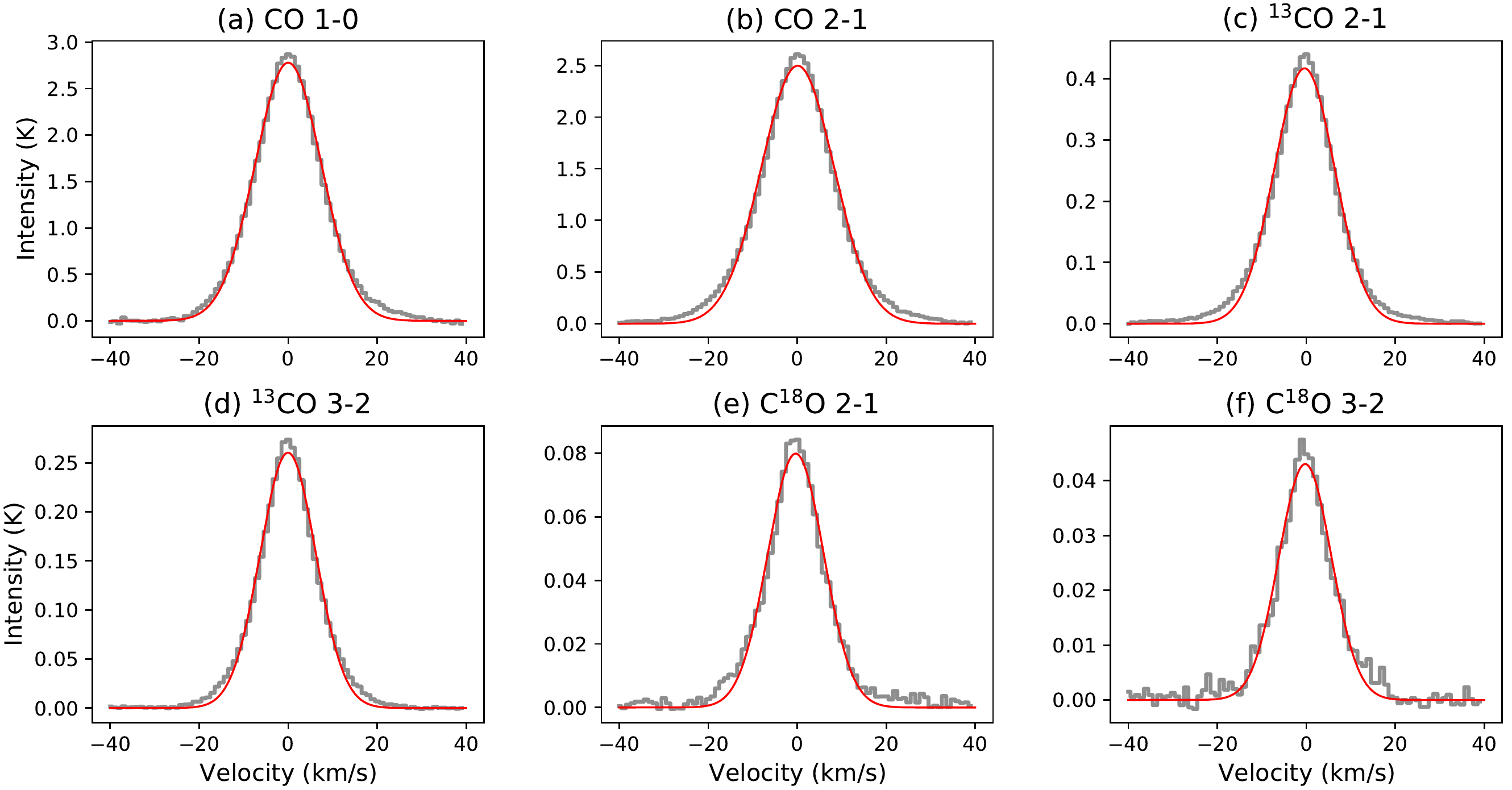} 
\caption{Shifted and stacked spectra over the center region, overlaid with the best-fit Gaussian profiles. All six lines show a single-peaked spectrum that can be well-described by a Gaussian function. The ring region also shows similar spectra, except having a slightly higher peak temperature and a ${\sim}1$~km~$\mathrm{s^{-1}}$ narrower line width. The averaged spectra for the arms region is presented in Figure~\ref{fig:specfit_arms}.}
\label{fig:specfit_center}
\end{figure*}

\begin{table*}
\begin{center}
\caption{Regional Line Ratios in NGC~3351. \label{tab:ratio}}
\begin{tabular}{lcccccccc}
  \hline\hline
  Region & &CO $\frac{2-1}{1-0}$ &$\rm{}^{13}CO$ $\frac{3-2}{2-1}$ &$\rm{C}^{18}O$ $\frac{3-2}{2-1}$ &$\rm \frac{CO}{{}^{13}CO}$ 2--1 &$\rm \frac{CO}{C^{18}O}$ 2--1 &$\rm \frac{{}^{13}CO}{C^{18}O}$ 2--1 &$\rm \frac{{}^{13}CO}{C^{18}O}$ 3--2\\
  \hline
  Whole &Median &0.99  &0.51 &0.47 &7.99 &42.86 &5.69 &6.48\\
  &Mean &1.03 &0.50 &0.51 &10.67 &53.63 &5.92 &7.08\\
  &Std.~Dev. &0.31 &0.14 &0.23 &5.67 &38.56 &2.10 &3.63\\
  &Integrated Mean &1.01 &0.55 &0.50 &8.06 &45.68 &5.67 &6.22 \\
  \hline
  Center &Median &0.99 &0.54 &0.46 &7.02 &41.10 &5.75 &6.89\\ 
  &Mean &1.01 &0.55 &0.48 &7.38 &46.21 &6.06 &7.75\\
  &Std.~Dev. &0.21 &0.10 &0.18 &1.86 &23.45 &1.51 &3.30\\
  &Integrated Mean &0.98 &0.57 &0.51 &7.11 &39.41 &5.54 &6.25 \\
  \hline
  Ring &Median &0.99 &0.56 &0.52 &7.01 &40.33 &5.57 &6.37\\
  &Mean &1.01 &0.57 &0.52 &7.49 &47.24 &6.04 &7.27\\ 
  &Std.~Dev. &0.23 &0.10 &0.18 &2.10 &26.91 &1.69 &3.11\\ 
  &Integrated Mean &0.98 &0.60 &0.55 &7.13 &38.65 &5.42 &5.86 \\
  \hline
  Arms &Median &0.98 &0.33 &0.87 &15.67 &60.25 &4.53 &1.92\\
  &Mean &1.00 &0.35 &0.88 &16.71 &69.36 &5.11 &1.84\\ 
  &Std.~Dev. &0.33 &0.14 &0.38 &6.05 &49.30 &2.17 &0.60\\
  &Integrated Mean &1.06 &0.25 &0.52 &20.21 &141.97 &7.02 &3.33 \\
  \hline
\end{tabular} 
\tablecomments{Line ratios are calculated using the moment~0 maps in units of K~km~$\mathrm{s^{-1}}$. The integrated means are calculated by first averaging the integrated intensities in each region and then dividing to obtain the ratios, while the mean, median and standard deviation are for the individual pixels of the map. Due to poor detection of C$^{18}$O 3--2 in the arms, the statistics for C$^{18}$O 3--2/2--1 and $^{13}$CO/C$^{18}$O 3--2 in the arms are only based on a few pixels.}
\end{center}
\end{table*}

Figure~\ref{fig:mom0} shows the moment~0 maps of the observed CO lines. 
In these images, we blank pixels with $\mathrm{S/N} < 3$ in moment~0, except for the two C$^{18}$O images where pixels with $\mathrm{S/N} < 1$ are blanked.
All six images resolve the circumnuclear star-forming ring structure and a gap between the ring and the nucleus. The images also reveal two spiral arm-like structures, which are gas inflows driven by the bar \citep[also see][]{2019MNRAS.488.3904L,2020ApJ...897..122L}. 
Note that the pixels in the central ${\sim}20\arcsec$ region have $\mathrm{S/N} > 3$ even in the C$^{18}$O images, but a $\mathrm{S/N} > 1$ cutoff in C$^{18}$O further reveals parts of the inflow arms. We will show that constraints from the C$^{18}$O lines are not critical to our results for the arms (Section~\ref{subsec:arms}), so we do not exclude pixels from the analysis based on $< 3\sigma$ detection in C$^{18}$O lines.
The moment~1 map in Figure~\ref{fig:velocity}a shows blue-shifted spectra in the northern half of the galaxy and red-shifted spectra in the southern half, indicating a counterclockwise gas rotation along the ring.
Figure~\ref{fig:velocity}b shows the effective line widths\footnote{The effective line width is defined as $\Sigma I/(\sqrt{2 \pi}\,T_\mathrm{peak})$, where $\Sigma I$ is the integrated line intensity and $T_\mathrm{peak}$ is the line peak temperature in K. It is referred to as ``equivalent width'' in \citet{2001ApJ...551..852H} and subsequent works. \citet{2018ApJ...860..172S,2020ApJ...901L...8S} also introduced this quantity in detail.} of CO 2--1. The rotation also causes a large line width around the nucleus, likely due to strong, unresolved gas motions within the central beam. As marked up in Figure~\ref{fig:mom0}a, the two ``contact points'' connecting the ring and the inflows have the brightest emission in all the lines. 

Using the six moment~0 maps in units of K~km~s$^{-1}$, we generate seven line ratio maps as shown in Figure~\ref{fig:ratio}: the primarily temperature sensitive line ratios are presented in the top row (a--c) as CO 2--1/1--0, $^{13}$CO 3--2/2--1, and C$^{18}$O 3--2/2--1; the following rows (d--g) show CO/$^{13}$CO 2--1, CO/C$^{18}$O 2--1, $^{13}$CO/C$^{18}$O 2--1, and $^{13}$CO/C$^{18}$O 3--2, which are primarily sensitive to isotopologue abundance ratios and/or optical depths. 
There are notable variations in the line ratios throughout the map. The main trends appear to be increased $^{13}$CO 3--2/2--1 and C$^{18}$O 3--2/2--1 in regions of the star-forming ring, potentially revealing higher temperatures, though it is interesting to note that these trends are not mirrored in CO 2--1/1--0 likely due to the emission being optically thick and thermalized. 
In addition, the west side of the northern contact point near $(\mathrm{R.A.}, \mathrm{Decl.}) = (10^\mathrm{h}43^\mathrm{m}57\fs7, 11^\circ42\arcmin20\arcsec)$ shows enhanced ratios in 3--2/2--1 of $^{13}$CO and C$^{18}$O, which may imply a change in excitation conditions. The region also has lower ratios in the CO/$^{13}$CO and CO/C$^{18}$O 2--1 maps due to fainter emission in CO lines. 
There is a clear trend for both CO/$^{13}$CO and CO/C$^{18}$O 2--1 to increase in the arms, possibly indicating a significant change of abundance ratios or optical depths. As star formation occurs only in the ring/nucleus but not the inflows, enrichment of $^{13}$C or $^{18}$O could lower the CO/$^{13}$CO or CO/C$^{18}$O abundances and thus lower the CO/$^{13}$CO and CO/C$^{18}$O line ratios in the ring/nucleus. On the other hand, the enhanced velocity dispersion in the arms as shown in Figure~\ref{fig:velocity}b may also lower the optical depths, leading to more escape CO emission and higher CO/$^{13}$CO and CO/C$^{18}$O line ratios in the arms. Note that higher line ratios are not observed in the nucleus where the effective line widths are also broader, which is likely due to beam smearing effects. More details will be discussed in Section~\ref{subsec:arms}. 

We define three regions for our analysis: the center, ring, and arms. Figure~\ref{fig:velocity}c illustrates the definition of these regions. The center is defined as the central $20\arcsec$ or ${\sim}1$~kpc (in galactocentric diameter) where $\mathrm{S/N} > 3$ in the moment~0 maps of all six lines. The ring region shares the same outer edge as the center, but excludes the inner $9\arcsec$ or ${\sim}450$~pc region of the center. The arms region includes only the pixels outside a $20\arcsec$ diameter but excludes the ``blob'' that lies to the east of the center. Figure~\ref{fig:specfit_center} shows the averaged spectra over the center region, where we applied the stacking approach by \citet{stacking} and used the moment 1 of CO 2--1 as velocity centroids. The shapes of the averaged spectra over the ring are almost identical to the center, while they show slightly higher peak intensities and ${\sim}1$~km~$\mathrm{s^{-1}}$ narrower line widths. This is because the ring region excludes the emission-faint ``gap'' region and the nucleus which has high velocity dispersion. The spectra of the arms are presented and analyzed in Section~\ref{subsec:arms}.

Table~\ref{tab:ratio} lists the line ratios averaged over the whole map and for the center, ring, and arm regions separately. The means, medians and standard deviations are calculated with the ensemble of line ratio in each pixel, and we also present the integrated means where the line fluxes are first summed up in each region to calculate the ratios. 
We note that due to poor detection of C$^{18}$O in the arms, the averaged C$^{18}$O 3--2/2--1 ratio of the arms is based on only few pixels. As shown in Table~\ref{tab:ratio}, the CO 2--1/1--0 ratio is ${\sim}1.0$ in all regions, which would indicate optically thick, thermalized emission. The CO/$^{13}$CO and CO/C$^{18}$O ratios vary the most from the center/ring to the arms. Also, the $^{13}$CO 3--2/2--1 ratio is lower in the arms compared to the rest, which may indicate changes in temperature or density. 
As these line ratio variations may imply variations in environmental conditions, we investigate the spatial distribution of multiple physical parameters through joint analysis of all the observed lines using non-LTE radiative transfer modeling and Bayesian likelihood analyses in Section~\ref{sec:model}.

\section{Multi-line Modeling} \label{sec:model}

\subsection{Modeling Setup} \label{subsec:model_setup}

\begin{table}
\centering
\caption{Model Grid Parameters \label{tab:radex}}
\begin{tabular}{lcc}
  \hline\hline
  Parameter & Range & Step Size \\
  \hline
  & One-Component Models\\
  $\log(n_\mathrm{H_2}\,[\mathrm{cm}^{-3}])$ & $2.0{-}5.0$ & $0.2$~dex \\
  $\log(T_\mathrm{k}\,[\mathrm{K}])$ & $1.0{-}2.3$ & $0.1$~dex \\
  $\log(N_\mathrm{CO}\,[\mathrm{cm}^{-2}])$ & $16.0{-}21.0$ & $0.2$~dex \\
  $X_{12/13}$ & $10{-}200$ & $10$ \\
  $X_{13/18}$ & $2{-}20$ & $1.5$ \\
  $\Phi_\mathrm{bf}$ & $0.05{-}1.0$ & $0.05$ \\
  $\Delta v\,[\mathrm{km~s}^{-1}]$ & $15.0$ & -- \\
  \hline
  & Two-Component Models\\
  $\log(n_\mathrm{H_2}\,[\mathrm{cm}^{-3}])$ & $2.0{-}5.0$ & $0.25$~dex \\
  $\log(T_\mathrm{k}\,[\mathrm{K}])$ & $1.0{-}2.0$ & $0.1$~dex \\
  $\log(N_\mathrm{CO}\,[\mathrm{cm}^{-2}])$ & $16.0{-}19.0$ & $0.25$~dex \\
  $\log(\Phi_\mathrm{bf})$ & $-1.3 \ldots -0.1$ & $0.1$ \\
  $X_{12/13}$ & $25$ & -- \\
  $X_{13/18}$ & $8$ & -- \\
  $\Delta v\,[\mathrm{km~s}^{-1}]$ & $15.0$ & -- \\
  \hline
\end{tabular} 
\end{table}

We use the non-LTE radiative transfer code RADEX \citep{radex} to model the observed line intensities under various combinations of $\mathrm{H_2}$ density ($n_\mathrm{H_2}$), kinetic temperature ($T_\mathrm{k}$), CO column density per line width ($N_\mathrm{CO}/\Delta v$), $\rm CO/{}^{13}CO$ ($X_{12/13}$) and $\rm {}^{13}CO/{C}^{18}O$ ($X_{13/18}$) abundance ratios, and the beam-filling factor ($\Phi_\mathrm{bf}$). The beam-filling factor is a fractional area of the beam covered by the emitting gas, and thus it should be $\leq 1$. Tests of the recoverability of model parameters with RADEX fitting have shown that modeling the isotopologue abundance ratio as a free parameter leads to more accurate results than assuming a fixed ratio \citep{2016ApJ...819..161T}. With six measured lines, we construct two different models with a one-component or two-component assumption on gas phases. The setup for these models are described separately in Section~\ref{subsec:model_6d} and \ref{subsec:model_4d_2comp}. 

The input to RADEX includes a molecular data file specifying the energy levels, statistical weights, Einstein A-coefficients, and collisional rate coefficients of each specific molecule. The data files we use for CO, $\rm {}^{13}CO$, and $\rm C^{18}O$ are from the Leiden Atomic and Molecular Database \citep{2005A&A...432..369S} with collisional rate coefficients taken from \citet{2010ApJ...718.1062Y}. To run RADEX, we directly input the $T_\mathrm{k}$, $n_\mathrm{H_2}$, molecular species column density ($N$), and line width ($\Delta v$). The input $T_\mathrm{k}$ and $n_\mathrm{H_2}$ are used to set collisional excitation, while $N$ and $\Delta v$ are the radiative transfer parameters that determine the optical depth and escape probability for the modeled molecular emission line. 

RADEX assumes a homogeneous medium and uses radiative transfer equations based on the escape probability formalism \citep{1960mes..book.....S} to find a converged solution for the excitation and radiation field. Initially, RADEX guesses the population of each energy level under LTE assumption and then computes for each transition the resultant optical depth at the line center ($\tau_0$) by
\begin{equation}
\tau_0 = \frac{A_{ij}}{8 \pi \tilde{\nu}^3}\ \frac{N}{\Delta v} \left( x_j \frac{g_i}{g_j} - x_i \right)~,
\label{eqn_tau}
\end{equation}
where $A_{ij}$ is the Einstein A-coefficient, $\tilde{\nu}$ is the wave number, and $x_i$ and $g_i$ are the fractional population and statistical weight, respectively, in level~$i$. The optical depth then determines the escape probability ($\beta$) for a uniform sphere:
\begin{equation}
\beta = \frac{1.5}{\tau}\,\left[1 - \frac{2}{\tau^2} + \left(\frac{2}{\tau} + \frac{2}{\tau^2}\right) e^{-\tau}\right]~,
\label{eqn_beta}
\end{equation}
which estimates the fraction of photons that can escape the cloud \citep{2006agna.book.....O}. This directly constrains the radiation field, and thus a new estimation of level population and excitation temperature, leading to new optical depth values. This procedure is done iteratively until convergence is reached.

We first run RADEX with CO to build up a 3D grid with varying $N_\mathrm{CO}/\Delta v$, $n_\mathrm{H_2}$, and $T_\mathrm{k}$. To construct a model grid with $X_{12/13}$ and $X_{13/18}$ axes, we calculate the column densities of $\rm {}^{13}CO$ and $\rm C^{18}O$ that correspond to each varying $N_\mathrm{CO}/\Delta v$ and abundance ratios, and then re-run RADEX with the calculated $N_\mathrm{^{13}CO}/\Delta v$ and $N_\mathrm{C^{18}O}/\Delta v$ values. The RADEX output of the three molecules predicts a Rayleigh--Jeans equivalent temperature ($T_\mathrm{RJ}$) for each line under the varied parameters. Assuming a Gaussian profile, the predicted line fluxes can then be estimated using a Gaussian integral:  
\begin{equation}
\int T_\mathrm{RJ}\,\mathrm{d}v = \frac{\sqrt{\pi}}{2 \sqrt{\ln{2}}}\,T_\mathrm{RJ}\ \Delta v~,
\label{eqn_mom0}
\end{equation}
where $\Delta v$ is the FWHM line width.
Finally, we expand the grid with an additional axis of beam-filling factor ($\Phi_\mathrm{bf}$) by multiplying the predicted line integrated intensities with the varying $\Phi_\mathrm{bf}$. This assumes the same beam-filling factor among all six emission lines.
Note that since we are essentially fitting $N_\mathrm{CO}/\Delta v$ with RADEX (see Equation~\ref{eqn_tau}), the estimation of $N_\mathrm{CO}$ would vary with line widths in different regions. Thus, when $N_\mathrm{CO}$ values are needed in further analyses, we multiply $N_\mathrm{CO}/\Delta v$ with the observed CO 1--0 line width for each pixel. This assumes all six lines to have the same line width, which is consistent with our assumption of single-component gas in Section~\ref{subsec:model_6d}. We have checked that adopting the CO 2--1 line width also results in similar $N_\mathrm{CO}$, as the line widths of 1--0 and 2--1 agree to within 50\%. The high-S/N emission in $^{13}$CO also yields similar line widths to those from the low-J CO emission - e.g., the mean CO 1--0/$^{13}$CO 2--1 line width ratio is $1.07 \pm 0.21$. We do not consider the low-S/N pixels because the line widths are very uncertain at low S/N and therefore do not provide a strong constraint.

To evaluate the goodness of fit for each parameter set $\theta = (n_\mathrm{H_2}, T_\mathrm{k}, N_\mathrm{CO}/\Delta v, X_{12/13}, X_{13/18}, \Phi_\mathrm{bf})$, we compute the $\chi^2$ values at each point in the model grid:
\begin{equation}
\chi^2(\theta) = \sum_{i=1}^n \frac{{\left[ S^\mathrm{mod}_i(\theta) - S^\mathrm{obs}_i \right]}^2}{\sigma_i^2}~,
\label{eqn_chi2_r}
\end{equation}
where $S^\mathrm{mod}$ and $S^\mathrm{obs}$ are the modeled and observed line integrated intensities with uncertainty $\sigma_i$, and $n = 6$ since we consider six lines. 
To include both the measurement and calibration uncertainties, $\sigma_i$ can be written as $\sigma_i^2 = \sigma_{\mathrm{noise},\,i}^2 + \sigma_\mathrm{cal}^2$, where $\sigma_{\mathrm{noise},\,i}$ is simply the measurement uncertainty of the $i$th observed line intensity and $\sigma_\mathrm{cal}$ represents the calibration uncertainty. For ALMA Band~3, 6, and~7 observations, a flux calibration uncertainty of ${\gtrsim}5$\% is suggested for point sources \citep{2018MNRAS.478.1512B,2019MNRAS.485.1188B}, and an additional few percent should be added for extended sources \citep[e.g.,][]{2016ApJ...819..161T,2021ApJS..255...19L}. Thus, we adopt a $10$\% calibration uncertainty (i.e., $\sigma_\mathrm{cal} = 0.1\times\mathrm{flux}$) for each line and assume independent calibrations. Although there should exist correlations between the calibration uncertainties for lines observed in the same spectral setup with the same calibrator, assuming independent calibrations is a less stringent constraint and therefore more conservative for the modeling.

The best-fit parameter set can be determined by selecting the combination with the lowest $\chi^2$ value.
By assuming a multivariate Gaussian probability distribution, the $\chi^2$ value for each parameter set can be converted to the probability as
\begin{equation}
P(S^\mathrm{obs}|\theta) = \frac{1}{Q}\,\exp\left[-\frac{1}{2} \chi^2(\theta)\right]~,
\label{eqn_prob}
\end{equation}
where $Q^2 = \Pi_i (2 \pi \sigma_i)$.
We calculate this probability for each grid point in the model parameter space, and we make our grid space wide enough to cover all grid points with reasonably high likelihood. Next, we generate the marginalized 1D or 2D likelihood distributions for any given parameter(s) by summing the joint probability distribution over the full range of all other parameters except the one(s) of interest.
With such marginalized 1D likelihood distributions, we find the ``1DMax'' solutions that give the highest 1D likelihood in each parameter. We also determine the $50$th percentile values (i.e., medians) of the cumulative 1D likelihoods and compare them with the 1DMax values to better understand the probability distributions. This procedure of computing the $\chi^2$ values and the likelihood distributions is applied to every pixel of the galaxy image in order to generate maps of the galaxy showing the derived physical parameters.  
In Sections~\ref{subsec:model_6d} and~\ref{subsec:model_4d_2comp}, we present these results based on a one-component model with six free parameters and a two-component model with eight free parameters, respectively. 
For reproducibility, we release all the source code and parameters in a GitHub repository\footnote{\url{https://github.com/ElthaTeng/multiline-ngc3351}}.

\subsection{One-Component Models} \label{subsec:model_6d}

\begin{table*}
\centering
\caption{1DMax Solutions from One-Component Modeling \label{tab:1comp}}
\begin{tabular}{lccccccc}
  \hline\hline
  Region & &$\log \left(N_\mathrm{CO}\,\frac{\mathrm{15~km~s^{-1}}}{\Delta v}\right)$ &$\log T_\mathrm{k}$ &$\log n_\mathrm{H_2}$ &$X_{12/13}$ &$X_{13/18}$ &$\Phi_\mathrm{bf}$ \\
  & &($\rm cm^{-2}$) &(K) &($\rm cm^{-3}$) & & & \\
  \hline
  Whole &Median  &17.8 &1.2 &3.2 &30 &8.0 &0.20 \\ 
  &Mean &17.46 &1.35 &3.46 &51.46 &10.43 &0.23 \\ 
  &Std.~Dev. &1.13 &0.41 &0.76 &58.31 &6.48 &0.19 \\ 
  \hline
  Center &Median  &18.4 &1.3 &3.2 &20 &6.5 &0.25 \\ 
  &Mean &18.30 &1.34 &3.33 &22.42 &7.38 &0.26 \\ 
  &Std.~Dev. &0.59 &0.22 &0.40 &14.64 &1.66 &0.16 \\ 
  \hline
  Ring &Median  &18.4 &1.3 &3.2 &20 &6.5 &0.20 \\ 
  &Mean &18.21 &1.33 &3.38 &22.94 &7.12 &0.26 \\ 
  &Std.~Dev. &0.65 &0.22 &0.43 &16.66 &1.74 &0.18 \\
  \hline
  Arms &Median  &16.2 &1.1 &3.2 &40 &15.5 &0.15 \\ 
  &Mean &16.83 &1.35 &3.48 &78.56 &12.38 &0.21 \\ 
  &Std.~Dev. &0.99 &0.51 &0.93 &67.51 &7.86 &0.20 \\
  \hline
\end{tabular} 
\tablecomments{The statistics are determined for the ensemble of 1DMax solutions across the pixels. The standard deviation does not reflect the uncertainties in the 1D likelihoods of each pixel.}
\end{table*} 

\begin{figure*}
\centering
\includegraphics[width=\linewidth]{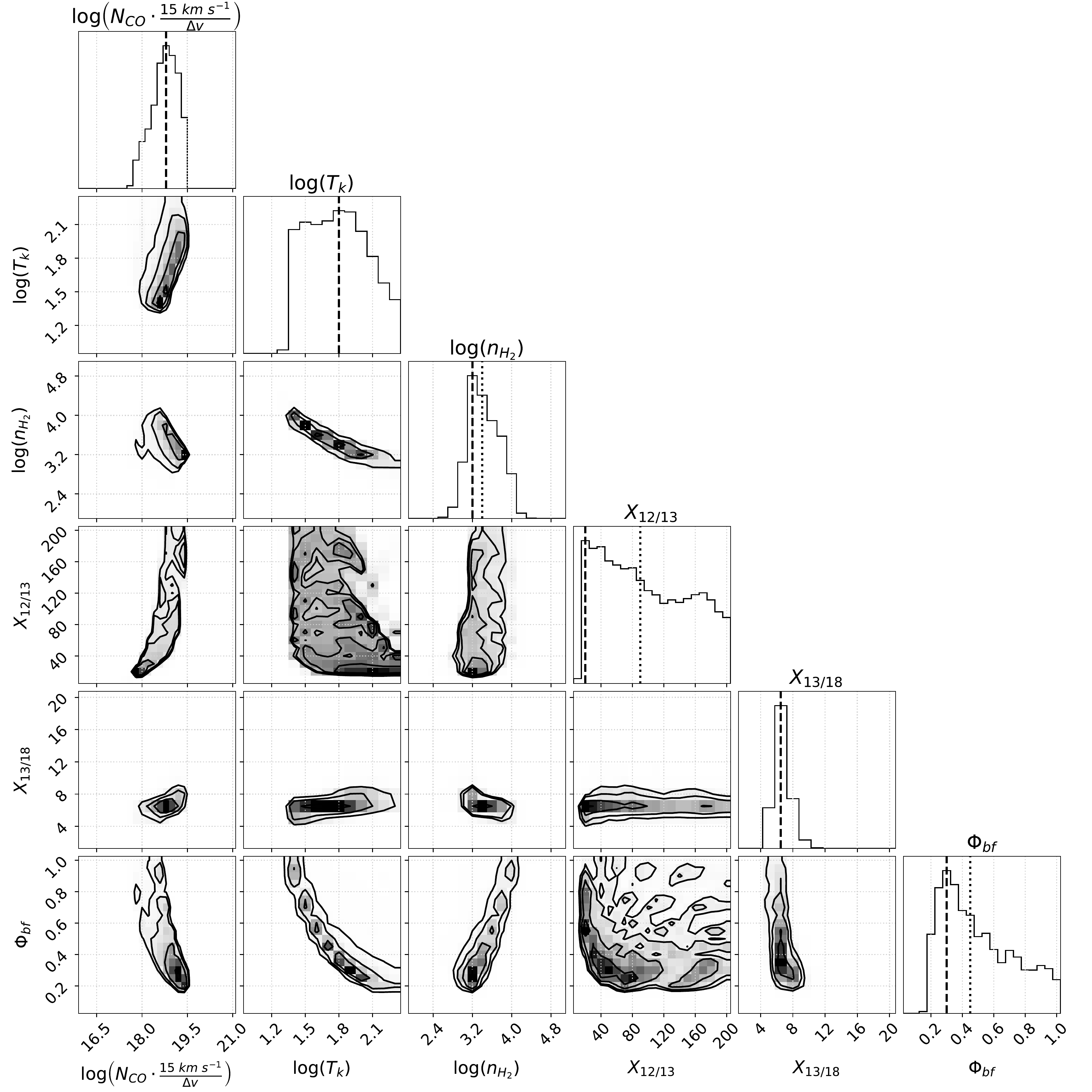}
\caption{Marginalized 1D and 2D likelihood distributions of a bright pixel at $(\mathrm{R.A.}, \mathrm{Decl.}) = (10^\mathrm{h}43^\mathrm{m}57\fs9, 11^\circ42\arcmin19\farcs5)$ in the northern contact point. The dashed lines on the 1D likelihood plots (in diagonal) represent the ``1DMax'' values that correspond to the peaks of each 1D likelihood. The dotted lines show the medians of the cumulative probability density function. For a less-constrained parameter like $X_{12/13}$, the median (${\sim}90$) can substantially deviate from the 1DMax value (${\sim}20$). However, the 1DMax solutions of $X_{12/13}$ are found to be consistent over the central ${\sim}1$~kpc region, as shown in Figure~\ref{fig:1dmax_rmcor_los50}.}
\label{fig:corner_bright}
\end{figure*}

\begin{figure*}
\centering
\includegraphics[width=\linewidth]{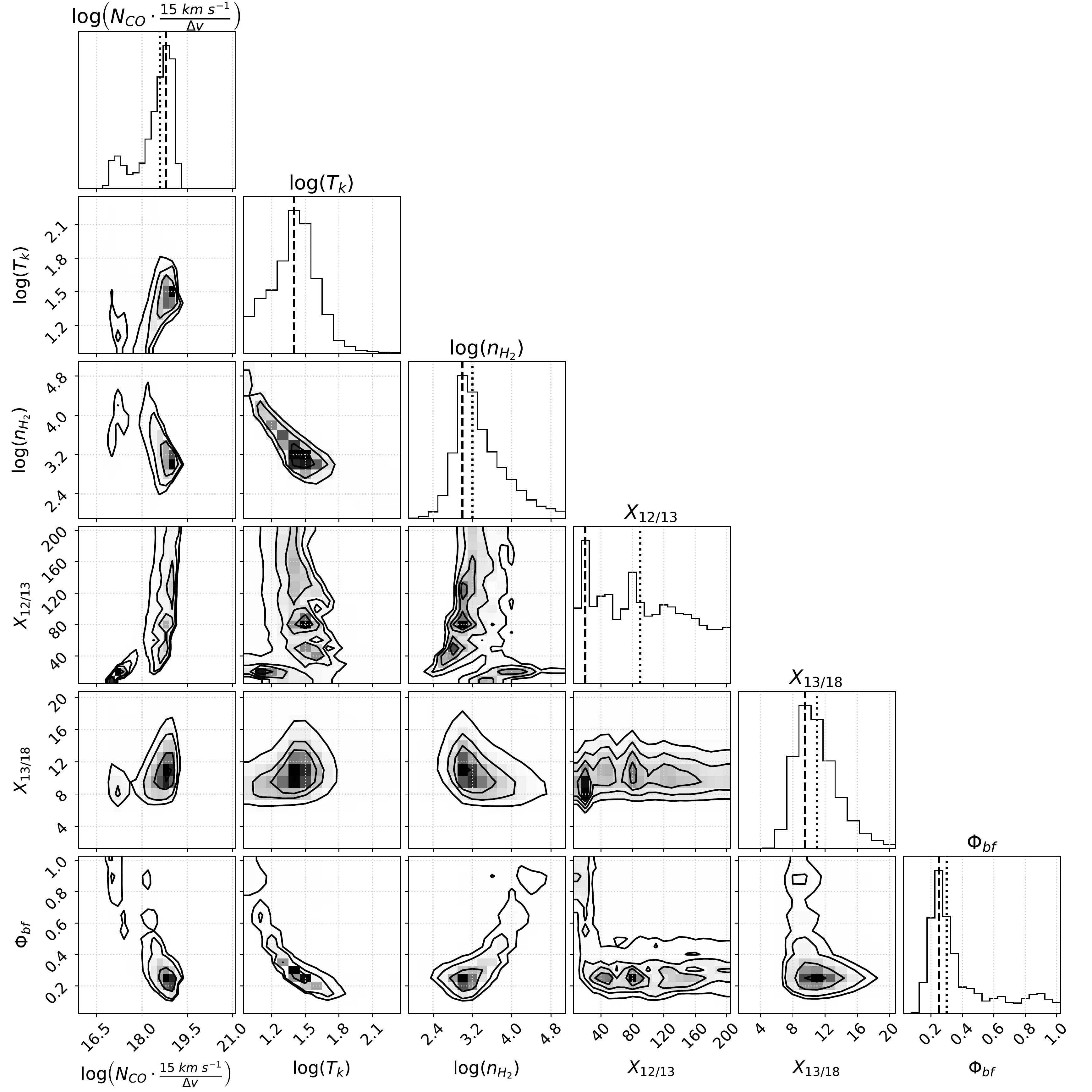}
\caption{Marginalized 1D and 2D likelihood distributions of a faint pixel at $(\mathrm{R.A.}, \mathrm{Decl.}) = (10^\mathrm{h}43^\mathrm{h}57\fs8, 11^\circ42\arcmin15\farcs5)$ in the gap region between the ring and the nucleus.
See the caption of Figure~\ref{fig:corner_bright} for more information.}
\label{fig:corner_faint}
\end{figure*}

\begin{figure*}
\begin{minipage}{.49\linewidth}
\centering
\includegraphics[width=\linewidth]{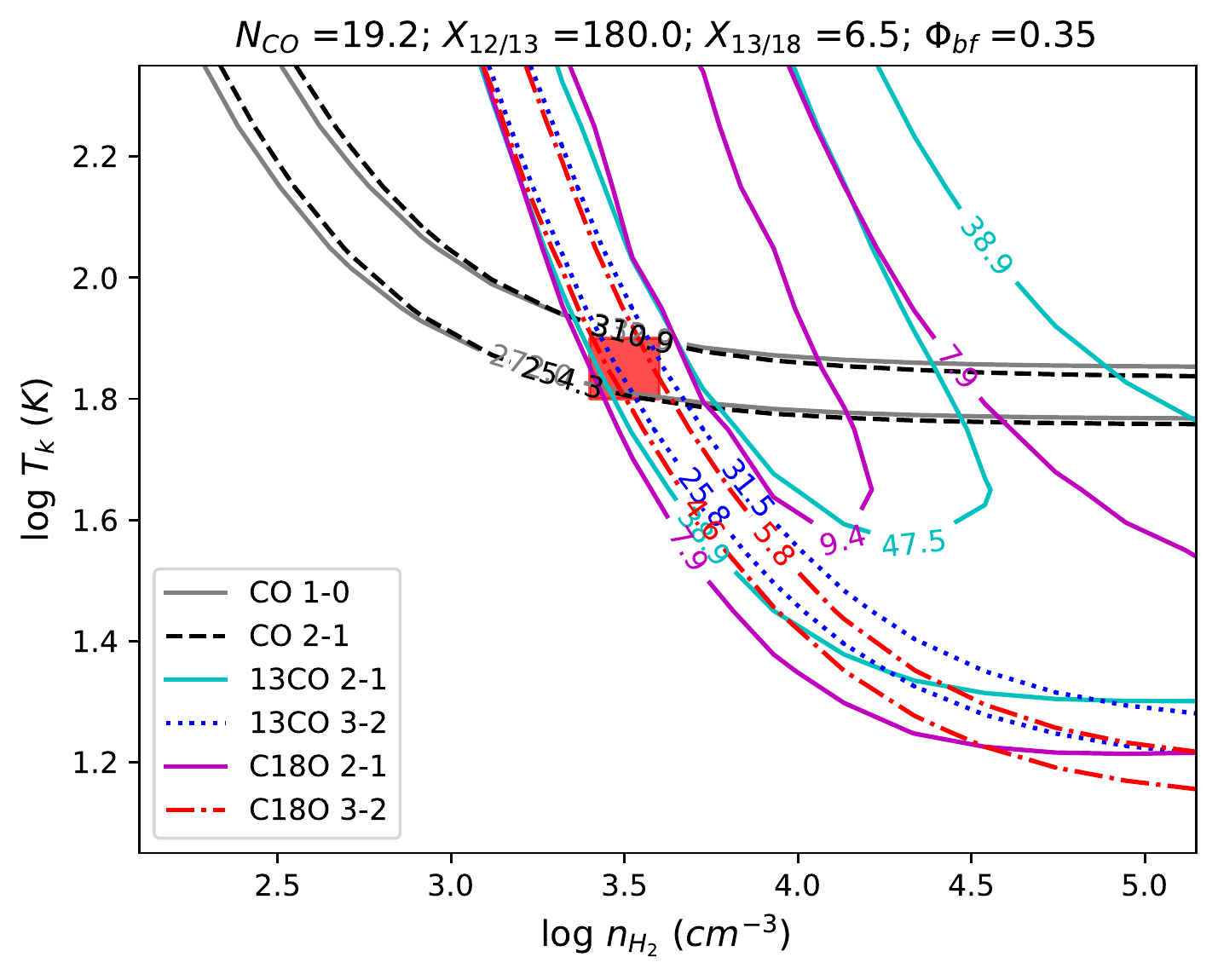}\\
(a)
\end{minipage}
\hfill
\begin{minipage}{.49\linewidth}
\centering
\includegraphics[width=\linewidth]{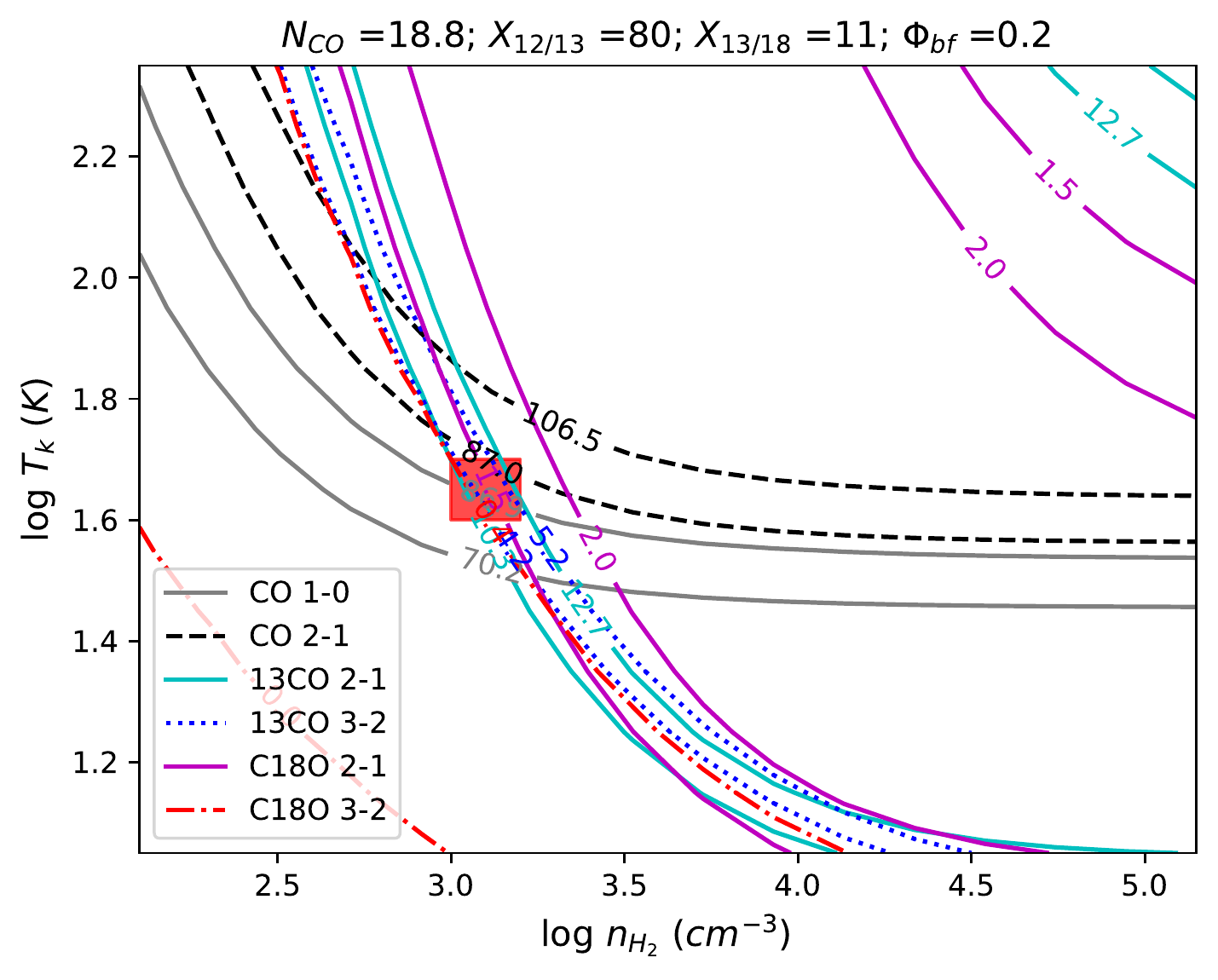}\\
(b)
\end{minipage}
\caption{Best-fit (i.e. highest likelihood) constraints from the six observed line fluxes at the same pixel as in (a)~Figure~\ref{fig:corner_bright} and (b)~Figure~\ref{fig:corner_faint}. Contours show the ranges of observed line intensities $\pm1\sigma$ uncertainties, including the measurement uncertainty from the data and a 10\% flux calibration uncertainty. Red boxes represent the best-fit solutions. Note that these are the solutions with the lowest $\chi^2$ value in the full grid, not the 1DMax solutions based on marginalized probability distributions, and thus the $X_{12/13}$ values here deviate from the lower $X_{12/13} \sim 20$ suggested by 1DMax solutions. Except for $X_{12/13}$, other parameters are similar to the 1DMax solutions as their 1D likelihoods are single-peaked and well-constrained.}
\label{fig:flux_contour}
\end{figure*}

\begin{figure*} \centering
\includegraphics[width=\linewidth]{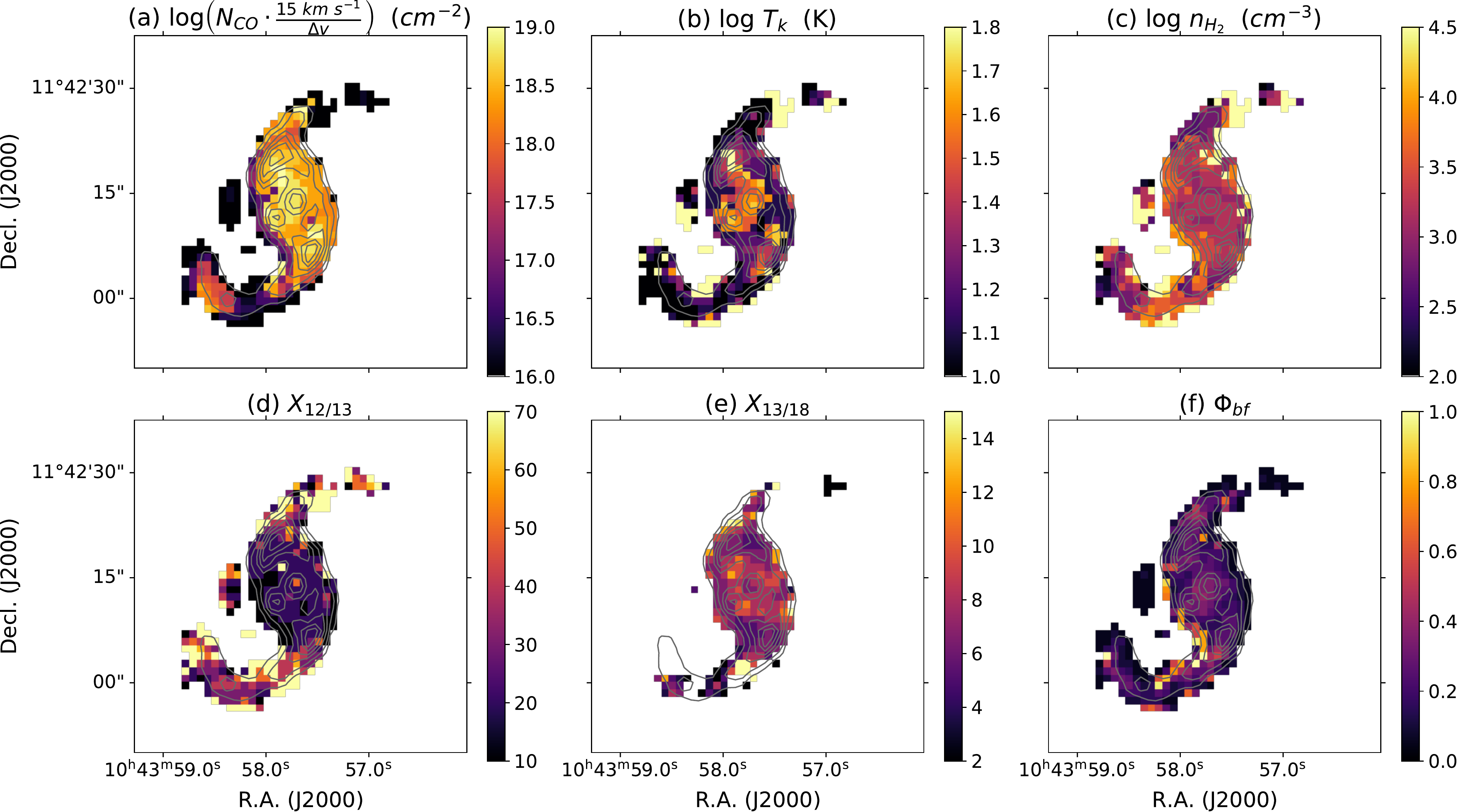} 
\caption{Maps of the 1DMax physical conditions derived from the one-component model. Panel~(a) shows $\log(N_\mathrm{CO})$ assuming a constant line width of $15$~km~s$^{-1}$ over the whole region. Contours represent the CO 2--1 emission shown in Figure~\ref{fig:mom0}b. A $1\sigma$ mask of the $\rm C^{18}O$ 2--1 image is applied to (e). These results show that $N_\mathrm{CO}$ is higher in the center, $T_\mathrm{k}$ is higher near the nucleus and the contact points, and $n_\mathrm{H_2}$ is overall ${\sim}2\times10^3$~$\mathrm{cm^{-3}}$. The $X_{12/13}$ and $X_{13/18}$ abundances are found to be consistent with those in the center of Milky Way.}
\label{fig:1dmax_rmcor_los50}
\end{figure*}

Assuming that the gas along each line of sight is uniform and that the emission can be described by a single physical condition ($n_\mathrm{H_2}$, $T_\mathrm{k}$, $N_\mathrm{CO}/\Delta v$, $X_{12/13}$, $X_{13/18}$) with an identical filling factor $\Phi_\mathrm{bf}$ across all lines, we create intensity model grids for each observed transition. As listed in Table~\ref{tab:radex}, the model can be represented by a six-dimensional grid with $\log(n_\mathrm{H_2}\,[\mathrm{cm}^{-3}])$ varied from $2$ to $5$ in steps of $0.2$~dex, $T_\mathrm{k}$ from $10$ to $200$~K in steps of $0.1$~dex, $N_\mathrm{CO}/\Delta v$ from $10^{16}/15$ to $10^{21}/15$ $\mathrm{cm^{-2}}\ \mathrm{(km~s^{-1})^{-1}}$ in steps of $0.2$~dex, $X_{12/13}$ from $10$ to $200$ in steps of~$10$, $X_{13/18}$ from $2$ to $20$ in steps of~$1.5$, and $\Phi_\mathrm{bf}$ from $0.05$ to $1$ in steps of~$0.05$. The range of $N_\mathrm{CO}/\Delta v$ covers reasonable $N_\mathrm{H_2}$ ranges of ${\sim}10^{20}{-}10^{25}$ $\mathrm{cm}^{-2}$ with typical $\mathrm{CO}/\mathrm{H_2}$ abundance ratios of ${\sim}10^{-4}$, and all the parameter ranges were optimized from representative pixels such that the shapes of the 1D likelihood distributions are well-covered for typical bright and faint regions in the galaxy. 

We apply a prior on the path length of the molecular gas along the line of sight. The idea of applying priors on line-of-sight lengths to avoid unrealistic solutions was also adopted in \citet{2014ApJ...795..174K} and is supported by \citet{2016ApJ...819..161T}. 
The line-of-sight path length is calculated by $\ell_\mathrm{los} = N_\mathrm{CO} (\sqrt{\Phi_\mathrm{bf}}\ n_\mathrm{CO})^{-1}$, where $n_\mathrm{CO} = n_\mathrm{H_2}\,x_\mathrm{CO}$ and $x_\mathrm{CO}$ is the $\mathrm{CO}/\mathrm{H_2}$ abundance ratio which we assume a typical value of $3\times10^{-4}$ for starburst regions \citep{1994ApJ...428L..69L,2003ApJ...587..171W,2014ApJ...796L..15S}. The $\sqrt{\Phi_\mathrm{bf}}$ factor can be interpreted as the one-dimensional filling factor along the line of sight to match the area filling factor $\Phi_\mathrm{bf}$, and the same equation was also adopted by \citet{2003ApJ...587..171W} and \citet{2012ApJ...753...70K,2014ApJ...795..174K}. With an angular resolution of ${\sim}100$~pc, we require all parameter sets in our model grids to have line-of-sight path length $\ell_\mathrm{los} \leq 100$~pc. The $100$~pc path length is also consistent with the vertical height (thickness) of CO observed in the central few kpc of Milky Way and other nearby edge-on galaxies \citep{2014AJ....148..127Y,2015ARA&A..53..583H}. We have checked that relaxing the constraint to $200$~pc results in similar best-fit and 1DMax solutions, and thus the result is not sensitive to factor of ${\sim}2$ changes in the line-of-sight constraint. The change in the gas thickness along the line of sight due to galaxy inclination would not alter the results. We note that all the parameters for calculating $\ell_\mathrm{los}$ are the varied inputs of our model except for the constant $x_\mathrm{CO}$ and the observed line width to obtain $N_\mathrm{CO}$ from $N_\mathrm{CO}/\Delta v$. In other words, we adopt $\ell_\mathrm{los} < 100$~pc as a prior by setting a zero probability to all the parameter sets that violate this constraint, and then investigate the resultant likelihood distributions. This prior generally rules out solutions with $N_\mathrm{CO} \gtrsim 10^{19}$~cm$^{-2}$ or $n_\mathrm{H_2} \lesssim 3\times10^2$~cm$^{-3}$, as either a high column density or a low volume density would lead to large~$\ell_\mathrm{los}$. We also find that the primary effect of allowing larger path lengths would be to shift the solutions to higher $N_\mathrm{CO}$ while $n_\mathrm{H_2}$ and $T_\mathrm{k}$ stay roughly constant, and we do not think that such large $N_\mathrm{CO}$ values and path lengths are reasonable solutions.   

Figures~\ref{fig:corner_bright} and~\ref{fig:corner_faint} are two corner plots \citep{Foreman-Mackey2016} showing the marginalized likelihood distributions of a bright pixel at $(\mathrm{R.A.}, \mathrm{Decl.}) = (10^\mathrm{h}43^\mathrm{m}57\fs9, 11^\circ42\arcmin19\farcs5)$ in the northern contact point and a faint pixel at ($10^\mathrm{h}43^\mathrm{m}57\fs8, 11^\circ42\arcmin15\farcs5$) in the gap region between the ring and the nucleus. The 1D likelihoods of $N_\mathrm{CO}$, $T_\mathrm{k}$, $n_\mathrm{H_2}$, and $X_{13/18}$ are well-constrained in both pixels.
On the other hand, $X_{12/13}$ is loosely constrained compared to other parameters, and the peaks in both 1D likelihoods (i.e. 1DMax, indicated by the dashed lines) are consistently at a lower $X_{12/13}$ of ${\sim}20{-}30$. This leads to a deviation between the 1DMax solutions and the medians (median $X_{12/13} \sim 90{-}100$, indicated by the dotted lines). Even though the 1D likelihoods of $X_{12/13}$ along each sight line may be less-constrained than other parameters, we will show that the 1DMax solutions of $X_{12/13}$ are consistent over the central ${\sim}1$~kpc region. 

We note that some parameters show correlations in their 2D likelihoods, including $N_\mathrm{CO}{-}n_\mathrm{H_2}$, $T_\mathrm{k}{-}n_\mathrm{H_2}$, and $N_\mathrm{CO}{-}\Phi_\mathrm{bf}$. Below the critical densities of the optically thin lines (see Table~\ref{tab:obs}), the emissivity depends on the density. Therefore, the anti-correlation between $N_\mathrm{CO}$ and $n_\mathrm{H_2}$ is expected, as a larger column density is needed to produce the same emission if we decrease the volume density with all other parameters fixed. Similarly, as $n_\mathrm{H_2}$ decreases, $T_\mathrm{k}$ has to increase to produce the same intensity if all other parameters remain unchanged. The inversely correlated $N_\mathrm{CO}{-}\Phi_\mathrm{bf}$ may imply well-constrained total mass of CO, as $M_\mathrm{mol} \propto N_\mathrm{CO}\,\Phi_\mathrm{bf}$ (see Equation~\ref{def_Mtot}).
In Figure~\ref{fig:flux_contour}, we show how well the best-fit (i.e. highest likelihood in the grid) solutions for the two corresponding pixels in Figures~\ref{fig:corner_bright} and~\ref{fig:corner_faint} match the constraints given by the observed line fluxes. The contour values are set as the ranges of observed line intensities $\pm1\sigma$ uncertainties, which include the measurement uncertainty from the data and a $10$\% flux calibration uncertainty. Note that the best-fit solutions of $X_{12/13}$ also deviate from the 1DMax solutions at ${\sim}20{-}30$. 

We present the pixel-by-pixel maps and regional statistics of all six 1DMax parameters in Figure~\ref{fig:1dmax_rmcor_los50} and Table~\ref{tab:1comp}, respectively. It is clear that the CO column densities are highest in the center and the ring, while the average $\mathrm{H_2}$ densities are similar in all regions. Although the average temperatures from Table~\ref{tab:1comp} are also similar in each region, Figure~\ref{fig:1dmax_rmcor_los50}b reveals a higher kinetic temperature of ${\sim}30{-}60$~K near the nucleus and both contact points. High-resolution studies at pc scales toward the inner ${\sim}500$~pc (Central Molecular Zone, CMZ) of the Milky Way suggested a dominant gas component with a temperature of ${\sim}25{-}50$~K \citep{2013MNRAS.429..987L,2016A&A...586A..50G,2017ApJ...850...77K}, which is consistent with the temperatures we find near the nucleus and contact points. In addition, we find that these peaks in the $T_\mathrm{k}$ map cover several circumnuclear star-forming regions identified by previous UV, H$\alpha$, near-infrared or radio observations \citep{1997ApJ...484L..41C,1997A&A...325...81P,2007MNRAS.378..163H,2020ApJS..248...25L,2021ApJ...913...37C}, and thus higher temperatures could be related to the heating from young stars. We also find consistent 1DMax values of $X_{12/13} \sim 20{-}30$ and $X_{13/18} \sim 6{-}10$ inside the central ${\sim}1$~kpc region. Unlike $X_{12/13}$, the 1DMax solutions for $X_{13/18}$ are consistent with the medians and are very well constrained over the central region. Both the 1DMax $X_{12/13}$ and $X_{13/18}$ abundance ratios are consistent with those found in the central kpc of the Milky Way \citep{1990ApJ...357..477L,1994ARA&A..32..191W,2005ApJ...634.1126M,2008A&A...487..237W,2018A&A...612A.117A}. We note that only the ``center'' pixels defined in Figure~\ref{fig:velocity}c have $\mathrm{S/N} > 3$ in the $\rm{C}^{18}O$ images, and thus pixel-by-pixel estimation beyond this region could be subject to larger uncertainties. Therefore, to examine our pixel-based results in the arm regions, we conduct similar analysis using stacked spectra in Section~\ref{subsec:arms}.

\subsection{Two-Component Models} \label{subsec:model_4d_2comp}

\begin{table*}
\centering
\caption{1DMax Solutions from Two-Component Modeling \label{tab:2comp}}
\begin{tabular}{lccccccccc}
  \hline\hline
  Region & & $\log \left(N_1\,\frac{\mathrm{15~km~s^{-1}}}{\Delta v}\right)$ & $\log T_1$ & $\log n_1$ & $\log \Phi_1$ & $\log \left(N_2\,\frac{\mathrm{15~km~s^{-1}}}{\Delta v}\right)$ & $\log T_2$ & $\log n_2$ &$ \log \Phi_2$ \\
  & &($\mathrm{cm}^{-2}$) &(K) &($\mathrm{cm}^{-3}$) & &($\mathrm{cm}^{-2}$) &(K) &($\mathrm{cm}^{-3}$) & \\
  \hline
  Whole &Median  &17.00 &1.1 &3.75 &-0.6 &17.00 &2.0 &3.25 &-0.8 \\ 
  &Mean &17.01 &1.27 &3.55 &-0.68 &17.17 &1.74 &3.00 &-0.79 \\
  &Std.~Dev. &0.85 &0.31 &0.74 &0.42 &1.00 &0.36 &0.64 &0.40 \\ 
  \hline
  Center &Median  &17.75 &1.3 &3.50 &-0.5 &18.25 &2.0 &3.25 &-0.7 \\ 
  &Mean &17.80 &1.38 &3.61 &-0.55 &18.11 &1.90 &3.15 &-0.71 \\
  &Std.~Dev. &0.53 &0.25 &0.48 &0.30 &0.45 &0.24 &0.39 &0.26 \\ 
  \hline
  Ring &Median  &17.75 &1.4 &3.75 &-0.5 &18.00 &2.0 &3.25 &-0.7 \\ 
  &Mean &17.83 &1.42 &3.65 &-0.57 &18.11 &1.89 &3.16 &-0.74 \\
  &Std.~Dev. &0.58 &0.26 &0.94 &0.34 &0.47 &0.27 &0.39 &0.29 \\ 
  \hline
  Arms &Median  &16.25 &1.0 &3.50 &-0.6 &16.25 &1.9 &3.25 &-0.8 \\ 
  &Mean &16.38 &1.13 &3.36 &-0.71 &16.39 &1.67 &2.88 &-0.77 \\
  &Std.~Dev. &0.45 &0.25 &0.83 &0.48 &0.55 &0.37 &0.70 &0.47 \\ 
  \hline
\end{tabular} 
\tablecomments{The denser component is represented as the first component, i.e., ($N_1$, $T_1$, $n_1$, $\Phi_1$).}
\end{table*} 

\begin{figure*} \centering
\includegraphics[width=\linewidth]{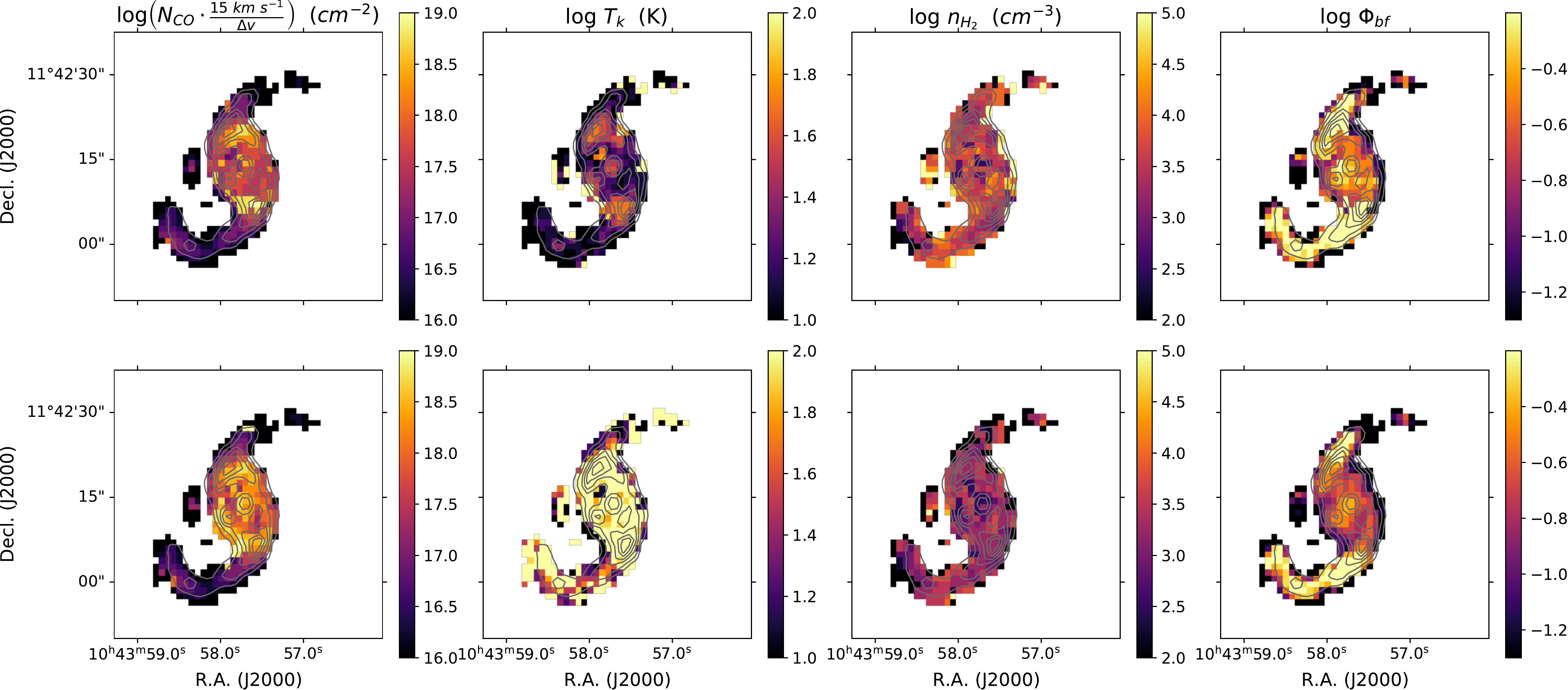} 
\caption{Maps of the 1DMax physical conditions derived from the two-component model. The top row shows the denser component having higher $n_\mathrm{H_2}$ and the bottom row shows the more diffuse gas phase. The two components have similar physical properties, with a $\leq$0.5~dex difference in each parameter. This is likely the reason why the two-component model results in similar solutions and spatial variations compared to the one-component model.}
\label{fig:1dmax_2comp}
\end{figure*}

While single-component modeling has been widely used to determine physical properties of molecular gas in various galaxies \citep[e.g.,][]{2014MNRAS.437.1434T,2017ApJ...840....8S,2018ApJ...856..143I,2020ApJ...893...63T}, it is likely that the molecular line emission comes from multiple gas components with different physical conditions \citep[e.g.,][]{2016A&A...586A..50G,2017MNRAS.469.2263B,2017ApJ...850...77K}. Some studies have modeled the gas with smoothly varying distributions of temperatures and densities \citep[e.g.,][]{2017ApJ...835..217L,2019MNRAS.485.3097B,2021ApJ...909...56L}, while others modeled the emission with a warm and cold or dense and diffuse component \citep[e.g.,][]{2014ApJ...795..174K,2014ApJ...781..101S,2015ApJ...810L..14L}. In addition, theoretical expectation for isolated and virialized molecular clouds and suggestively some observational studies of galaxy centers imply a two-phase molecular gas with a high-$\alpha_\mathrm{CO}$ dense component and a low-$\alpha_\mathrm{CO}$ diffuse component \citep[e.g.][]{2012ApJ...751...10P,2015ApJ...801...25L}. \citet{2020A&A...635A.131I} also conducted a two-phase analysis toward the centers of ${\sim}$100 galaxy samples with single dish measurements, although NGC~3351 was not included.
To test how multiple gas components could change our results, we construct a two-component model of $N_\mathrm{CO}/\Delta v$, $T_\mathrm{k}$, $n_\mathrm{H_2}$, and $\Phi_\mathrm{bf}$. This is primarily to see if we get dramatically different results if we change our assumptions about the gas components.

Based on the consistent 1DMax values of $X_{12/13}$ and $X_{13/18}$ from the one-component modeling and their consistency with the values of the Milky Way center, we assume fixed isotopologue abundances of $X_{12/13} = 25$ and $X_{13/18} = 8$ for both gas phases. The two-component model can thus be described by eight parameters, namely, four parameters for each gas phase. We remind the readers that even if the 1DMax values of $X_{12/13}$ are consistently at ${\sim}25$ in the one-component modeling, the 1D likelihoods of $X_{12/13}$ are loosely constrained; the $X_{13/18}$ likelihoods, on the other hand, are well-constrained at 6--10 (see Figure~\ref{fig:corner_bright} and~\ref{fig:corner_faint}). 
\citet{2017ApJ...836L..29J} have observed the central kpc of NGC~3351 at a $8\arcsec$ resolution with $^{13}$CO and C$^{18}$O 1--0 and suggested a $X_{13/18}$ value that is also consistent with our one-component modeling as well as the Galactic Center value. 
Nevertheless, we note that applying fixed, Galactic Center-like abundance ratios to the two-component modeling may be a rather strict assumption. We emphasize that the primary goal of our two-component model is to test if the solutions will differ substantially after altering the assumption in the numbers of gas components, and this goal can be achieved by comparing the solutions of $N_\mathrm{CO}/\Delta v$, $T_\mathrm{k}$, or $n_\mathrm{H_2}$ while fixing the isotopologue abundances to remain consistent with the one-component modeling.

Due to large grid size of the 8D model, we slightly adjust the parameter ranges and step sizes (see Table~\ref{tab:radex}): $N_\mathrm{CO}/\Delta v$ ranges from $10^{16}/15$ to $10^{19}/15$ $\mathrm{cm^{-2}~(km~s^{-1})^{-1}}$ in steps of $0.25$~dex, $T_\mathrm{k}$ ranges from $10$ to $100$~K in steps of $0.1$~dex, $\log(n_\mathrm{H_2}\,[\mathrm{cm}^{-3}])$ ranges from~$2$ to~$5$ in steps of $0.25$~dex, and $\log(\Phi_\mathrm{bf})$ ranges from $-1.3$ to $-0.1$ in steps of~$0.1$~dex. 
We lower the upper limit of $T_\mathrm{k}$ to 100~K in the two-component model grids because the sensitivity to distinguish different temperatures above 100~K is very weak with $J$=3--2 as the highest transition. In addition, the one-component modeling results show that $T_\mathrm{k}$ is well below 100~K across the whole region.
We also lower the upper limit of $N_\mathrm{CO}/\Delta v$ to $10^{19}/15$ because a higher value is normally ruled out by the line-of-sight constraint of $\ell_\mathrm{los} < 100$~pc as mentioned in Section~\ref{subsec:model_6d}. 

The 8D model grid consists of all combinations of either two components from the 4D parameter space. By summing up the predicted intensities of both components, each grid point has a prediction of the total integrated intensity. Since the two components are inter\-change\-able under summation, nearly half of the grid points are redundant, and we thus require the first component to have higher $T_\mathrm{k}$ than the second component. This not only allows us to distinguish between warm and cold or dense and diffuse components, but also avoids getting the exact same marginalized likelihoods for both components due to symmetric model grids. Similar to the 6D one-component modeling, we include a $10$\% flux calibration uncertainty and the measurement uncertainties for fitting.  

When deriving the values of $N_\mathrm{CO}$, we assume both components to have the same line widths as the observed CO 1--0 line width. This assumption has negligible effect on our $\alpha_\mathrm{CO}$ results in Section~\ref{sec:alpha_co}, as $N_\mathrm{CO}$ and $\Phi_\mathrm{bf}$ are being fit simultaneously to determine $\alpha_\mathrm{CO}$ values, which means that the two parameters can adjust themselves to fit best with the line intensities.  
In addition, we mostly observed a single-peaked spectrum which can be well represented by a single line width (see Figure~\ref{fig:specfit_center}), so assuming the same line widths as the measured line width is likely the most feasible thing we could do in the two-component analysis. If both components have Gaussian spectra, this assumption would imply that both components are at the same velocity. 

Figure~\ref{fig:1dmax_2comp} shows the 1DMax solutions from the two-component modeling, where the component with higher $n_\mathrm{H_2}$ is shown in the top row. Regional statistics are also summarized in Table~\ref{tab:2comp} with the denser component being presented as the first component. We note that the separation by density is made for visualization purposes and has no role in the subsequent calculation of $\alpha_\mathrm{CO}$. In general, the dense components correspond to lower $T_\mathrm{k}$ and $N_\mathrm{CO}$ values than the diffuse components, which is similar to the correlations observed in the one-component result within the probability distribution of a single pixel. From the regional averages of both components (see Table~\ref{tab:2comp}), we find that the differences in $N_\mathrm{CO}$, $n_\mathrm{H_2}$ and $\Phi_\mathrm{bf}$ between the two components are all less than ${\sim}0.5$~dex. Moreover, the difference between their masses, which are proportional to their $N_\mathrm{CO}$ and $\Phi_\mathrm{bf}$ (see Equation~\ref{def_Mtot}), is even found to be less than ${\sim}0.3$~dex. This means that the two components have similar physical properties. We also find these properties to be similar to those suggested by the one-component model.
For example, the 1DMax $n_\mathrm{H_2}$ distribution from the two-component model suggests a density of ${\sim}2{-}3 \times 10^3$~cm$^{-3}$ in the center, which is consistent with ${\sim}2 \times 10^3$~cm$^{-3}$ suggested by the one-component model. These are also similar to the average gas density of ${\sim}5\times10^3$~$\mathrm{cm^{-3}}$ found in the Milky Way's CMZ \citep[e.g.,][]{2013MNRAS.429..987L}. The only substantial difference between our two-component solutions is the kinetic temperature. 
While the denser component has a similar range of temperature and region-by-region variations as observed in the one-component model (see Figure~\ref{fig:1dmax_rmcor_los50}b), the other component shows evidently higher kinetic temperature of ${\sim}100$~K over the whole region. These temperatures are quantitatively similar to previous two-component studies toward the GMCs in our Galactic Center, which suggest a dominant component by mass with $T_\mathrm{k} \sim$25--50~K and a less-dominant warm component with $T_\mathrm{k} \gtrsim$100~K \citep[e.g.,][]{1993A&A...280..255H,2017ApJ...850...77K}.
We thus conclude that the dominant component in our two-component model is well-represented by the one-component model alone, but there might be a secondary, warmer component unaccounted for in the one-component model, as previously seen in the Galactic CMZ.

\subsection{Marginalized 1D Likelihoods} \label{subsec:marginalize}

In this subsection, we describe the unified procedure of how we generate and deal with the marginalized likelihoods and then derive plausible solutions for parameters like $\alpha_\mathrm{CO}$. This is fundamental to the results presented in Section~\ref{sec:alpha_co}. 

In Sections~\ref{subsec:model_6d} and~\ref{subsec:model_4d_2comp}, we derived the marginalized 1D likelihoods for each input parameter. In both cases, we have a full 6D or 8D grid, with each grid point having an associated probability. Since our input parameters are sampled uniformly, their marginalized 1D likelihood distribution is simply a probability-weighted histogram, where the probability value of each grid point is reflected proportionally in the number of counts toward each bin. 
On the other hand, to derive the marginalized likelihoods of parameters that are functions of the intrinsic grid parameters (e.g., the modeled line intensities or $\alpha_\mathrm{CO}$), an additional normalization on the histograms would be needed due to the possibility of irregular sampling on the derived parameter grid space. For instance, given the grids of modeled line integrated intensities, we could determine 1D likelihoods for each line intensity with some chosen bin spacing and range. However, since multiple combinations of parameters in the 6D or 8D grid space could produce the same integrated intensity, and some line intensities may only be produced by a small subset of the parameter combinations, the resulting 1D likelihoods will have an additional non-uniform weighting due to the starting grid. 
To properly deal with the grid irregularity, we can generate a uniformly weighted histogram for line intensities under the same parameter range and bin size, treating that as a normalization. To obtain the marginalized 1D likelihood of any derived parameter from the original grid, we divide the probability-weighted histogram by the uniformly weighted histogram to normalize out any grid irregularity. The marginalized likelihoods of $\alpha_\mathrm{CO}$, which is a function of $N_\mathrm{CO}$, $\Phi_\mathrm{bf}$, and the CO 1--0 line intensity, can also be derived in this way (see Section~\ref{sec:alpha_co}).

Once the 1D likelihood distributions of a parameter are obtained, we can determine either the 1DMax solutions by finding the values that correspond to the peaks, or the medians, and $\pm1\sigma$ values by finding the 16th and 84th percentiles of the cumulative 1D likelihood distribution. Then, maps of the 1DMax solutions or medians $\pm1\sigma$ can be generated by iterating over each pixel. While we could look individually into the likelihood distributions at each pixel, comparing between these maps is a more simple and feasible way to get a sense of the probability distribution in different regions of the galaxy.
With this procedure, it is important to note that our solutions for $\alpha_\mathrm{CO}$ in Section~\ref{sec:alpha_co} do not rely on the individually derived $N_\mathrm{CO}$ and $\Phi_\mathrm{bf}$ distributions presented in Sections~\ref{subsec:model_6d} and~\ref{subsec:model_4d_2comp}, since those parameters are being fit simultaneously within the full grid before we obtain the marginalized likelihoods of $\alpha_\mathrm{CO}$.

\section{CO-to-\texorpdfstring{$\mathrm{H_2}$}{H2} Conversion Factor} \label{sec:alpha_co}

The CO-to-$\mathrm{H_2}$ conversion factor ($\alpha_\mathrm{CO}$) is defined as the ratio of molecular gas mass to CO 1--0 luminosity \citep{1986ApJ...309..326D,co-to-h2}, which is also equivalent to the ratio of molecular gas mass surface density to CO 1--0 intensity:
\begin{equation}
\alpha_\mathrm{CO} = \frac{M_\mathrm{mol}}{L_\mathrm{CO(1-0)}} = \frac{\sum_\mathrm{mol}}{I_\mathrm{CO(1-0)}}\,\left(\frac{\mathrm{M_\odot}} {\mathrm{K~km~s^{-1}~pc^2}}\right)~.
\label{def_alpha}
\end{equation}
The CO 1--0 luminosity is the CO 1--0 intensity multiplied by the area in square parsecs, and the molecular gas mass is proportional to the product of CO column density and the filling factor times the area \citep{2014ApJ...795..174K}:
\begin{equation}
M_\mathrm{mol} = 1.36\,m_\mathrm{H_2}\,N_\mathrm{CO}\,A\,\Phi_\mathrm{bf}\,x_\mathrm{CO}^{-1}~,
\label{def_Mtot}
\end{equation}
where $A$ is the same area as that in CO 1--0 luminosity. Therefore, Equation~\ref{def_alpha} can be rewritten as
\begin{equation}\label{eqn_xco2alpha}
\begin{split}
\alpha_\mathrm{CO} &= \frac{1.36\,m_\mathrm{H_2}\,(\mathrm{M}_\odot)\,N_\mathrm{CO}\,(\mathrm{cm}^{-2})\,A\,(\mathrm{cm}^2)\,\Phi_\mathrm{bf}} {x_\mathrm{CO}\,I_\mathrm{CO(1-0)}\,(\mathrm{K~km~s}^{-1})\,A\,(\mathrm{pc}^2)} \\
&= \frac{1}{4.5\times10^{19}} \cdot \frac{N_\mathrm{CO}\,(\mathrm{cm}^{-2})\,\Phi_\mathrm{bf}} {x_\mathrm{CO}\,I_\mathrm{CO(1-0)}\,(\mathrm{K~km~s^{-1})}}~,
\end{split}
\end{equation}
where $m_\mathrm{H_2}$ is the mass of molecular hydrogen, $1.36$ is the factor after including the mass contribution from Helium, and $x_\mathrm{CO}$ is the CO-to-$\mathrm{H_2}$ abundance ratio. The $x_\mathrm{CO}$ value in different galaxies or metallicity conditions can vary from $0.5{-}5\times10^{-4}$ \citep[e.g.,][]{1982ApJ...262..590F,1990ApJ...358..459B,1998ApJ...507..615D,2012ApJ...753...46S} and is commonly assumed as ${\sim}10^{-4}$. In our analysis, we assume a higher $x_\mathrm{CO}$ value of $3\times10^{-4}$, which is typically found and/or adopted in starburst regions \citep{1994ApJ...428L..69L,2003ApJ...587..171W,2012ApJ...753...70K,2014ApJ...795..174K,2014ApJ...796L..15S,2017ApJ...840....8S}.
This means that the absolute value of our derived $\alpha_\mathrm{CO}$ is accurate only when $x_\mathrm{CO} = 3\times10^{-4}$, and the true $\alpha_\mathrm{CO}$ can be represented as $\alpha_\mathrm{CO}^\mathrm{true} = \alpha_\mathrm{CO} \times (3\times10^{-4} / x_\mathrm{CO})$. Therefore, we note that our results on the $\alpha_\mathrm{CO}$ values depend inversely on any variation of $x_\mathrm{CO}$, and that the spatial variation of our derived $\alpha_\mathrm{CO}$ could partially result from spatial variations in $x_\mathrm{CO}$.   

With Equation~\ref{eqn_xco2alpha}, we derive the spatial distribution of $\alpha_\mathrm{CO}$ using the approach described in Section~\ref{subsec:marginalize}: for each pixel, we first calculate an $\alpha_\mathrm{CO}$ value for every grid point in the model using the corresponding $N_\mathrm{CO}$, $\Phi_\mathrm{bf}$, and predicted CO intensity $S^\mathrm{mod}$, and then marginalize over the whole grid to obtain the 1DMax value of $\alpha_\mathrm{CO}$. We apply this method to both one- and two-component modeling results and compare the derived $\alpha_\mathrm{CO}$ distributions. 

\subsection{\texorpdfstring{$\alpha_\mathrm{CO}$}{alphaCO} from the Modeling Results} \label{subsec:alpha_mod}

\begin{figure*}
\begin{minipage}{.45\linewidth}
\centering
\includegraphics[width=\linewidth]{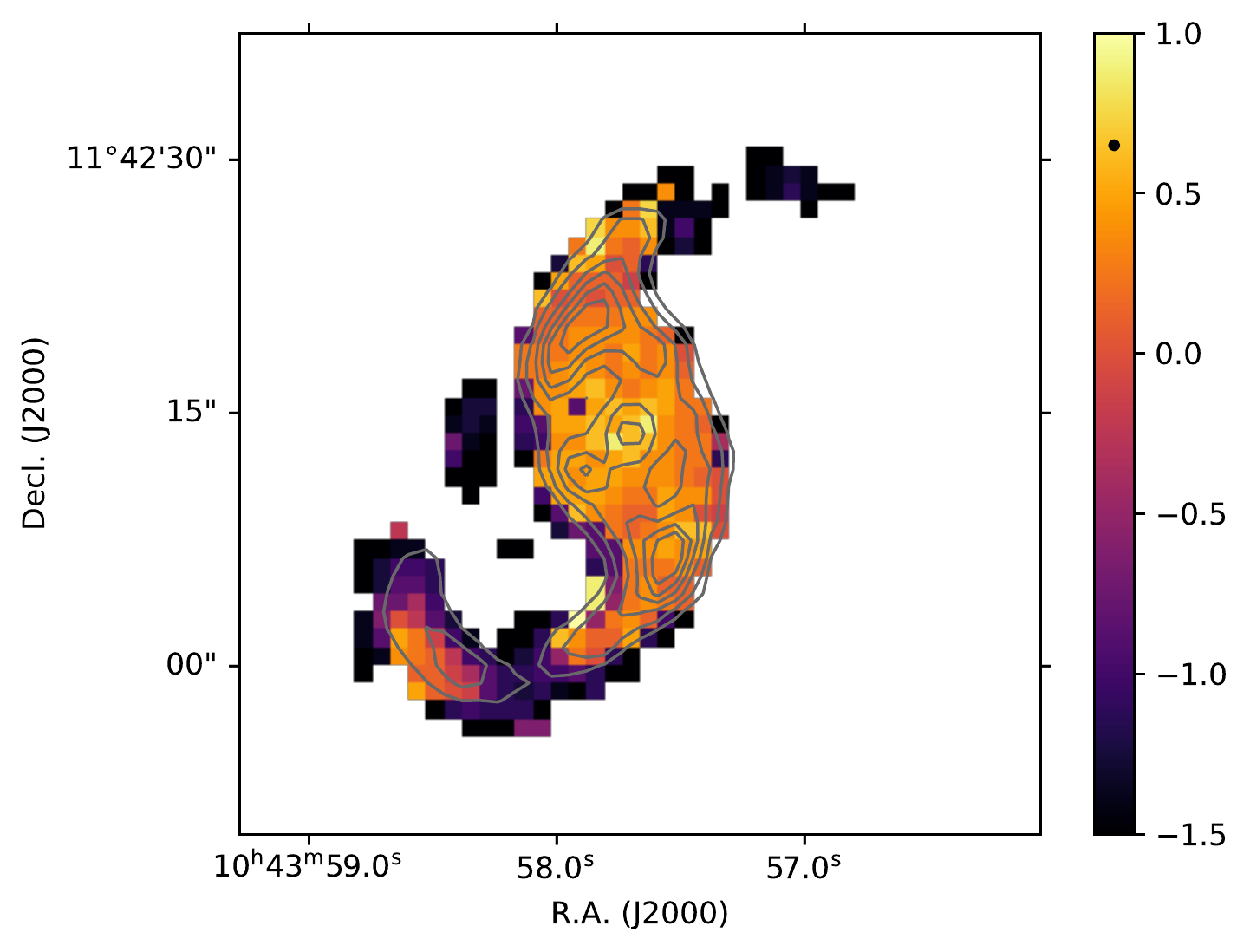}\\
(a)
\end{minipage}
\hfill
\begin{minipage}{.5\linewidth}
\centering
\includegraphics[width=\linewidth]{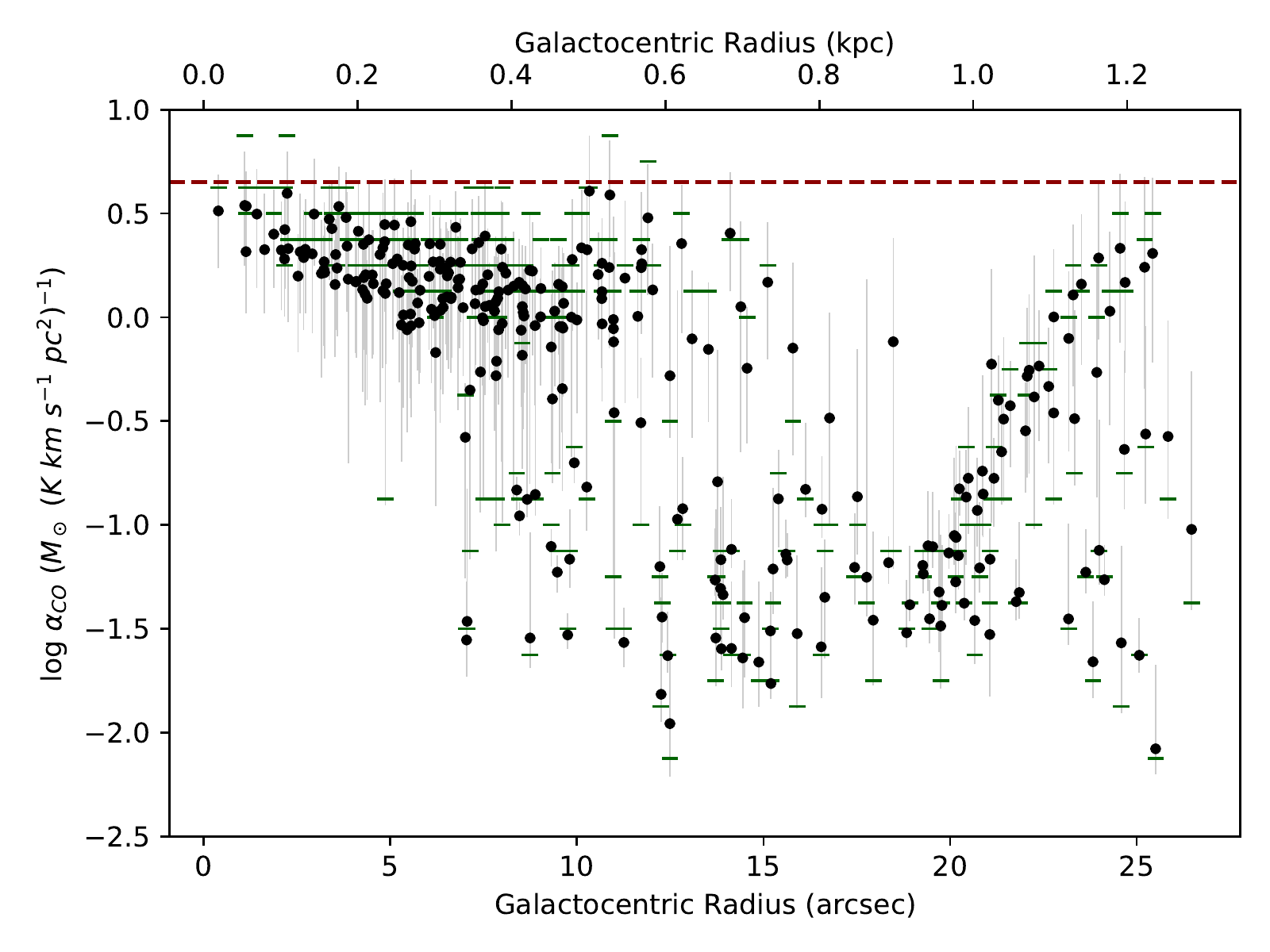}\\
(b)
\end{minipage}
\caption{$\alpha_\mathrm{CO}$ variations determined by the one-component modeling results. (a) 1DMax $\log (\alpha_\mathrm{CO})$ map in units of $\mathrm{M_\odot\ (K~km~s^{-1}~pc^2)^{-1}}$; the contours show zeroth moment of CO 1--0 and the black dot on the color bar indicates the Galactic average value of $\alpha_\mathrm{CO} \sim 4.4$ $\mathrm{M_\odot\ (K~km~s^{-1}~pc^2)^{-1}}$. (b) Relation between the median $\alpha_\mathrm{CO}$ (black dots) and galactocentric radius; the green horizontal lines represent the 1DMax $\alpha_\mathrm{CO}$ and the vertical lines show the $1\sigma$ ranges (16th--84th percentiles); the red dashed line shows the Galactic disk average value of $\alpha_\mathrm{CO} \sim 4.4$ $\mathrm{M_\odot\ (K~km~s^{-1}~pc^2)^{-1}}$. In the center, the median and 1DMax $\alpha_\mathrm{CO}$ values are only slightly below the Galactic disk average, and decrease slowly from the nucleus to the ring. Beyond a galactocentric radius of ${\sim}10\arcsec$, $\alpha_\mathrm{CO}$ drops significantly to a value that is approximately an order of magnitude lower than the Galactic disk average.}
\label{fig:factor_1comp_marg}
\end{figure*}

\begin{figure*}
\begin{minipage}{.45\linewidth}
\centering
\includegraphics[width=\linewidth]{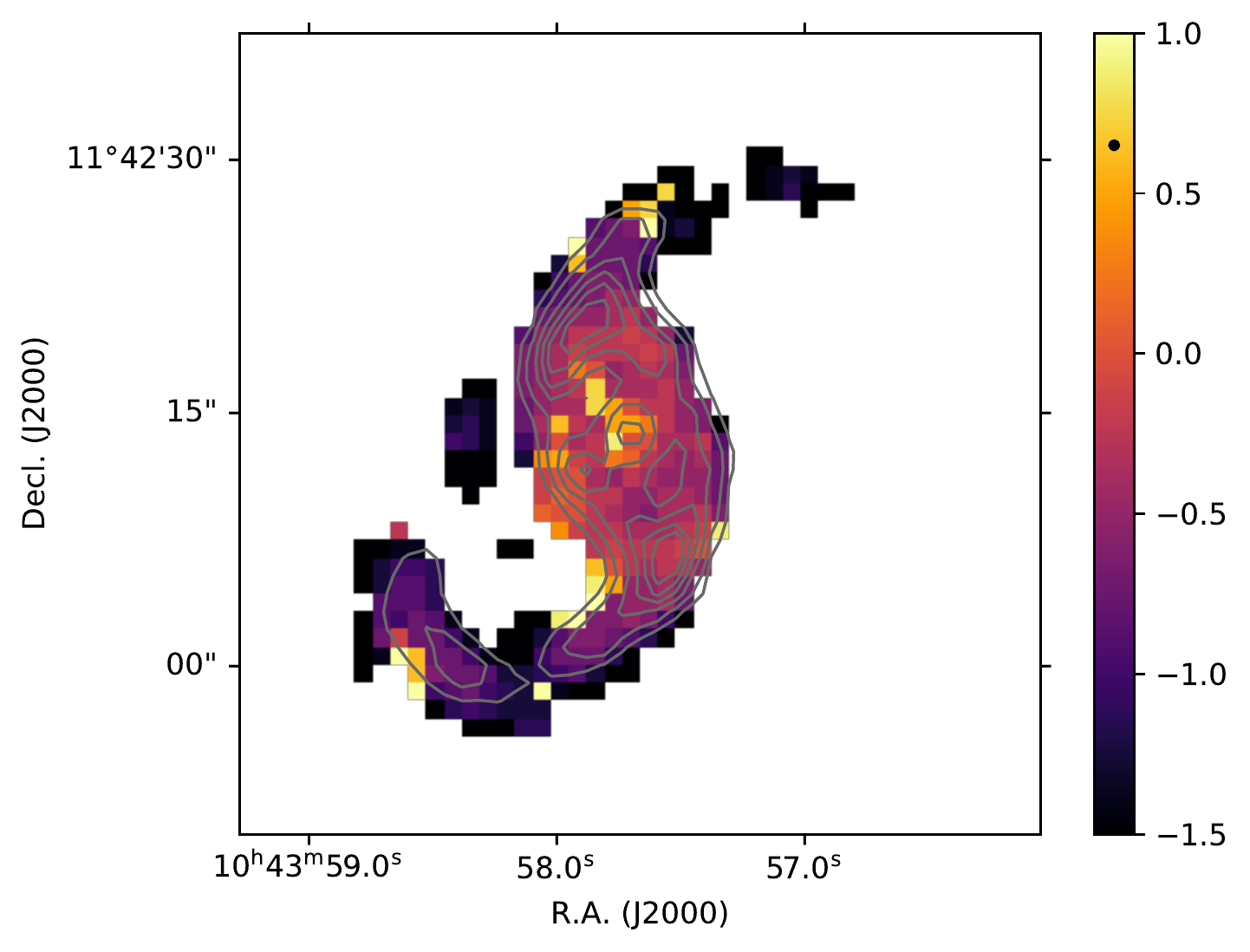}\\
(a)
\end{minipage}
\hfill
\begin{minipage}{.5\linewidth}
\centering
\includegraphics[width=\linewidth]{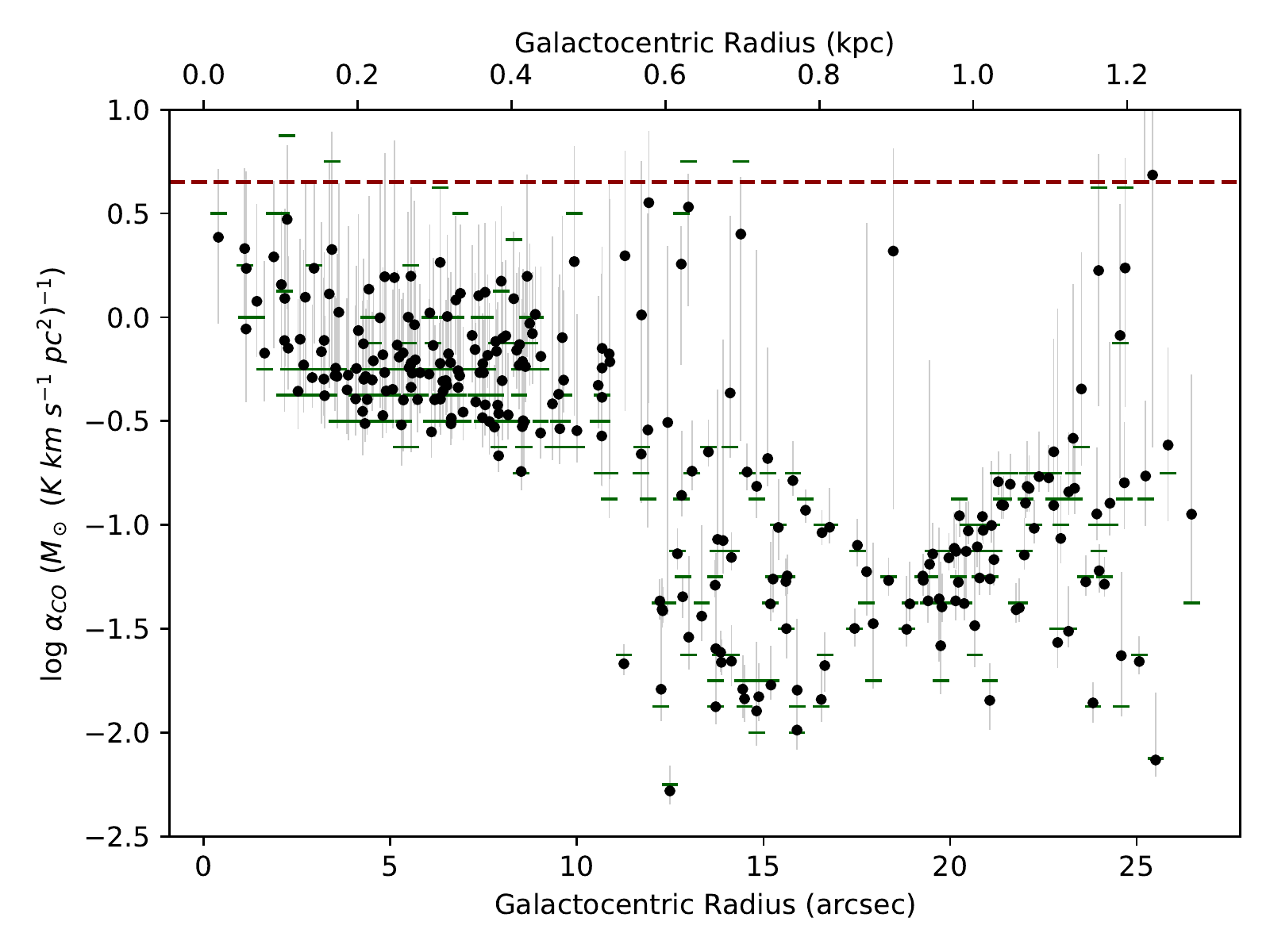}\\
(b)
\end{minipage}
\caption{$\alpha_\mathrm{CO}$ variations determined by the two-component modeling results. (a) 1DMax $\log (\alpha_\mathrm{CO})$ map in units of $\mathrm{M_\odot\ (K~km~s^{-1}~pc^2)^{-1}}$. (b) Relation between the median/1DMax $\alpha_\mathrm{CO}$ and galactocentric radius, determined by the marginalized $\alpha_\mathrm{CO}$ likelihoods from the two-component models. See the caption of Figure~\ref{fig:factor_1comp_marg} for more information. The spatial variation of $\alpha_\mathrm{CO}$ and the values in the arms are similar to that derived from the one-component model, while the values in the center are a factor of 2--3 lower.}
\label{fig:factor_2comp_marg}
\end{figure*}

Using the method described in Section~\ref{subsec:marginalize}, we derive the marginalized 1D likelihoods of $\log(\alpha_\mathrm{CO})$ for each pixel with $\log(\alpha_\mathrm{CO})$ ranging from $-2.5$ to $2.5$ in steps of $0.125$. Figures~\ref{fig:factor_1comp_marg} and~\ref{fig:factor_2comp_marg} present the $\alpha_\mathrm{CO}$ variations derived from the one-component and two-component modeling, respectively. The marginalized 1DMax $\alpha_\mathrm{CO}$ shown in Figure~\ref{fig:factor_1comp_marg}a reveals an average of ${\sim}2$ $\mathrm{M_\odot\ (K~km~s^{-1}~pc^2)^{-1}}$ in the inner $1$~kpc region, while the $\alpha_\mathrm{CO}$ along the arms is a factor of ${\sim}10$ lower than the center.
Figure~\ref{fig:factor_1comp_marg}b shows how the median and 1DMax $\alpha_\mathrm{CO}$ vary with the galactocentric radius of NGC~3351. Both the median and 1DMax $\alpha_\mathrm{CO}$ values exhibit a slow decrease from the nucleus to the ring and a significant drop at a radius of ${\sim}10\arcsec$ (${\sim}0.5$~kpc), which is where the ring and the arms intersect. Interestingly, while $\alpha_\mathrm{CO}$ remains roughly constant at a radius of ${\sim}10{-}20\arcsec$, it starts to increase beyond a radius of ${\sim}20\arcsec$. We find that the $\gtrsim 20\arcsec$ region corresponds to the curved-up feature in the southern arm (see Figure~\ref{fig:velocity}c). This region is called the ``transverse dust lane'' in \citet{2019MNRAS.488.3904L}, where they suggested the feature as evidence for an outflow pushing gas/dust away from the nuclear ring via stellar feedback processes.  

Overall, all the pixels in our observed region show at least factor of $\sim$2--3 (0.3--0.5~dex) lower $\alpha_\mathrm{CO}$ values than the Milky Way average at the assumed $x_\mathrm{CO}$ of $3\times10^{-4}$. If $x_\mathrm{CO}$ were set to be $10^{-4}$, $\alpha_\mathrm{CO}$ would increase by a factor of three. In such a case, the central $1$~kpc region would have a nearly Galactic $\alpha_\mathrm{CO}$, whereas the $\alpha_\mathrm{CO}$ is still substantially lower in the inflow arms since a global change in $x_\mathrm{CO}$ would not change the relative drop in $\alpha_\mathrm{CO}$. Note that if there were spatial variations of $x_\mathrm{CO}$ across this region, it could alter our $\alpha_\mathrm{CO}$ variation results. 
\citet{2019MNRAS.488.3904L} estimated the circular rotational velocity to be 150--200~$\mathrm{km~s^{-1}}$ on the circumnuclear ring of NGC~3351, implying a gas rotation period of $15{-}20$~Myr. Since this orbital time scale is much shorter than the stellar evolution time scale that can cause a difference in chemical abundances, we expect the abundances of carbon and oxygen to be well-mixed in the central kpc of NGC~3351, and thus sub-kpc scale variations of $x_\mathrm{CO}$ should be small if $x_\mathrm{CO}$ is essentially determined by the availability of carbon and oxygen. Despite such expectation, we emphasize that disentangling $x_\mathrm{CO}$ variations and $\alpha_\mathrm{CO}$ is infeasible under current modeling and measurements. Therefore, our estimated $\alpha_\mathrm{CO}$ distribution can be affected if there is any $x_\mathrm{CO}$ variation on sub-kpc scales.     

The 1DMax $\alpha_\mathrm{CO}$ distributions derived from the two-component modeling results, as shown in Figure~\ref{fig:factor_2comp_marg}a, suggest a similar spatial variation to that determined by the one-component results, while the $\alpha_\mathrm{CO}$ values in the center are generally lower than those in Figure~\ref{fig:factor_1comp_marg}a.
Figure~\ref{fig:factor_2comp_marg}b shows the median and 1DMax solutions determined from the marginalized $\alpha_\mathrm{CO}$ likelihoods and their variation with galactocentric radius. The $\alpha_\mathrm{CO}$ solutions inside the central $1$~kpc region are shifted ${\sim}0.3{-}0.5$~dex downward compared with Figure~\ref{fig:factor_1comp_marg}b. On the other hand, the $\alpha_\mathrm{CO}$ solution in the inflow arms is consistent with Figure~\ref{fig:factor_1comp_marg}b, and we can see a clear distinction between $\alpha_\mathrm{CO}$ values in the center and inflow regions, which are separated at the $10\arcsec$ or $0.5$~kpc radius. 
We note that the possible cause for a ${\sim}0.3{-}0.5$~dex shift of $\alpha_\mathrm{CO}$ solutions in the center may originate from the assumption of a fixed $X_{12/13}$ at 25. The 2D correlations in Figure~\ref{fig:corner_bright} and \ref{fig:corner_faint} show that $X_{12/13} = 25$ corresponds to a lower $N_\mathrm{CO}/\Delta v$ and a higher $\Phi_\mathrm{bf}$ compared with the 1DMax solutions, which may altogether result in a $0.3{-}0.5$~dex offset in $\alpha_\mathrm{CO}$.  
In general, both our one- and two-component models result in lower-than-Galactic $\alpha_\mathrm{CO}$ values for the entire observed region and significantly ($\gtrsim 10\times$) lower $\alpha_\mathrm{CO}$ values in the inflow arms. Furthermore, by comparing with the derived $T_\mathrm{k}$ distributions shown in Figure~\ref{fig:1dmax_rmcor_los50}b, there is no sign that lower $\alpha_\mathrm{CO}$ are associated with higher temperatures (see discussion in Section~\ref{sec:discussion}). In Section~\ref{subsec:arms}, we will discuss possible scenarios that cause the $\alpha_\mathrm{CO}$ variation and such low $\alpha_\mathrm{CO}$ values in the arms. 

To compare the derived $\alpha_\mathrm{CO}$ with those observed previously at kpc scales, we compute the intensity-weighted mean $\alpha_\mathrm{CO}$ over our entire observed region of ${\sim}2$~kpc. The intensity-weighted mean $\alpha_\mathrm{CO}$ is calculated as the ratio of total molecular mass to the total luminosity over the whole region. Based on Equation~\ref{eqn_xco2alpha}, this can be done with the full grids of the modeled $N_\mathrm{CO}$, $\Phi_\mathrm{bf}$, and intensity of CO 1--0 ($S^\mathrm{mod}$) at every pixel. For each pixel, we first perform a likelihood-weighted random draw from the model grids to obtain $N_\mathrm{CO}$, $\Phi_\mathrm{bf}$, and the corresponding modeled $I_\mathrm{CO(1-0)}$ value. Next, we sum up the values over the pixels to derive the total molecular mass and CO intensity, and then calculate the resulting intensity-weighted mean $\alpha_\mathrm{CO}$. We then perform another random draw and repeat $2000$ times. Finally, the mean and standard deviation over the $2000$ measurements of $\alpha_\mathrm{CO}$ are obtained. This approach is a combination of Monte Carlo sampling with likelihood weighting of the model grids, and is similar to the ``realize'' method used by \citet{gordon_2014} for dust spectral energy distribution (SED) fitting.       

The intensity-weighted mean $\alpha_\mathrm{CO}$ of the entire region is $1.79\pm0.10$ and $1.11\pm0.09$ $\mathrm{M_\odot\ (K~km~s^{-1}~pc^2)^{-1}}$ based on the one- and two-component models, respectively. Since the sum of our $I_\mathrm{CO(1-0)}$ and $I_\mathrm{CO(2-1)}$ over the entire observed region are similar, weighting either by the CO 1--0 or 2--1 intensities give approximately the same intensity-weighted mean $\alpha_\mathrm{CO}$. With an angular resolution of ${\sim}40\arcsec$, \citet{2013ApJ...777....5S} found an $\alpha_\mathrm{CO(2-1)}$ of $1.0^{+0.4}_{-0.3}$ $\mathrm{M_\odot\ (K~km~s^{-1}~pc^2)^{-1}}$ in the central $1.7$~kpc of NGC~3351 using dust modeling with CO 2--1 intensities. This is similar to the intensity-weighted mean $\alpha_\mathrm{CO}$ values derived from both our one- and two-component models.
However, since they assumed a constant CO 2--1/1--0 ratio ($R_{21}$) of $0.7$, they concluded a lower mean $\alpha_\mathrm{CO(1-0)}$ value of $0.7^{+0.27}_{-0.20}$ $\mathrm{M_\odot\ (K~km~s^{-1}~pc^2)^{-1}}$ appropriate for CO 1--0 intensities. While $R_{21} = 0.7$ may be a good approximation across galaxy disks \citep[e.g.,][]{2021MNRAS.504.3221D}, studies on nearby galaxy centers have suggested higher $R_{21}$ (${\gtrsim}0.9$) in the central kpc region of disk galaxies \citep{2020A&A...635A.131I,2021PASJ...73..257Y}. As presented in Table~\ref{tab:ratio}, the mean $R_{21}$ over our entire observed region is $1.0$, and both our one- and two-component models predict a mean $R_{21}$ higher than $0.9$. Thus, if a higher $R_{21}$ of $> 0.9$ had been adopted by \citet{2013ApJ...777....5S}, their $\alpha_\mathrm{CO}$ estimates with CO 1--0 would be in good agreement with our modeling results.

\subsection{\texorpdfstring{$\alpha_\mathrm{CO}$}{alphaCO} in the Inflow Arms} \label{subsec:arms}

\begin{figure*} \centering
\includegraphics[width=\linewidth]{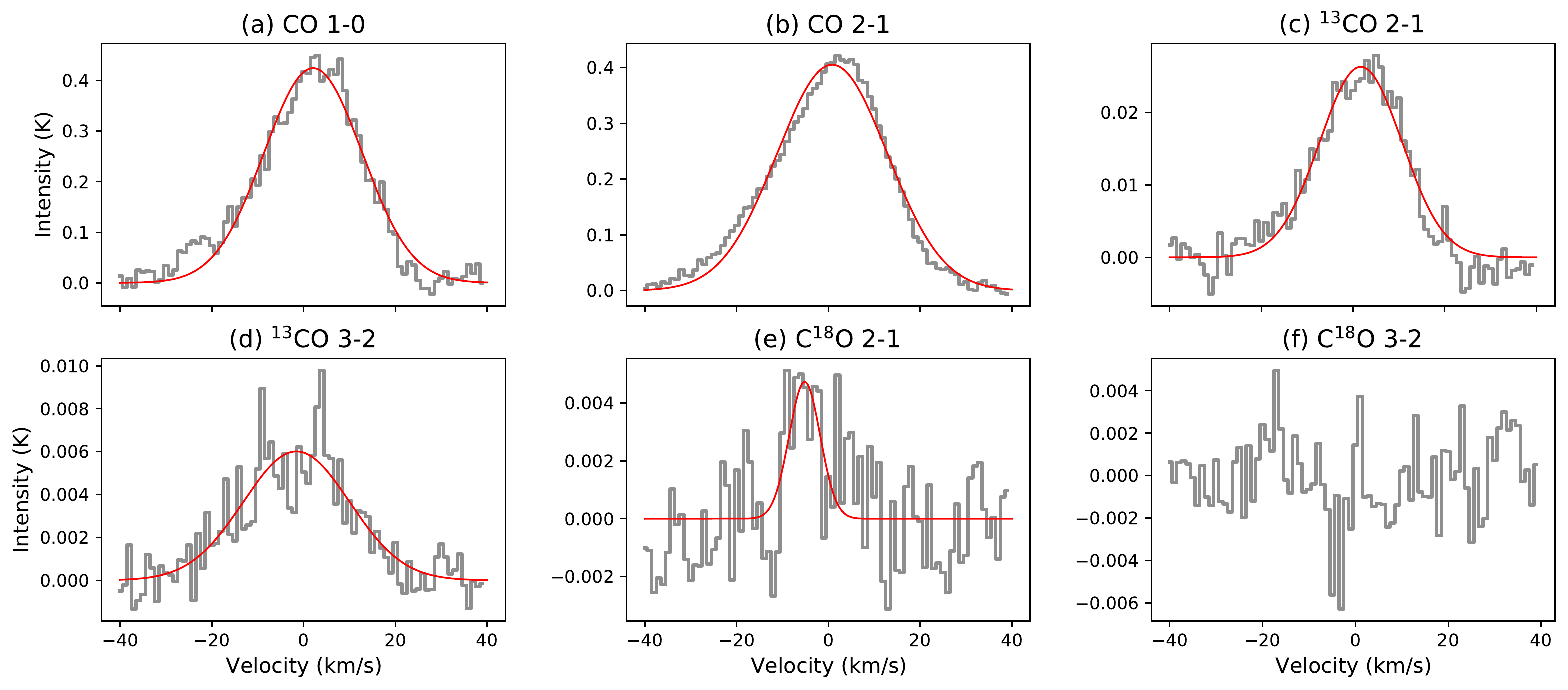} 
\caption{Shifted and averaged spectra over the inflow arms, overlaid with the best-fit Gaussian profiles. This stacking is used for the purpose of recovering faint emission from the arms and comparing with the pixel-based analysis. The integrated intensities of each line is estimated from the averaged spectra and then input to the multi-line modeling to determine the best-fit/1DMax solutions for environmental parameters and $\alpha_\mathrm{CO}$ in the arm regions. The results of the modeling are shown in Figure~\ref{fig:corner_arms} and \ref{fig:bestfit_arms}.}
\label{fig:specfit_arms}
\end{figure*}

\begin{figure*}
\centering
\includegraphics[width=\linewidth]{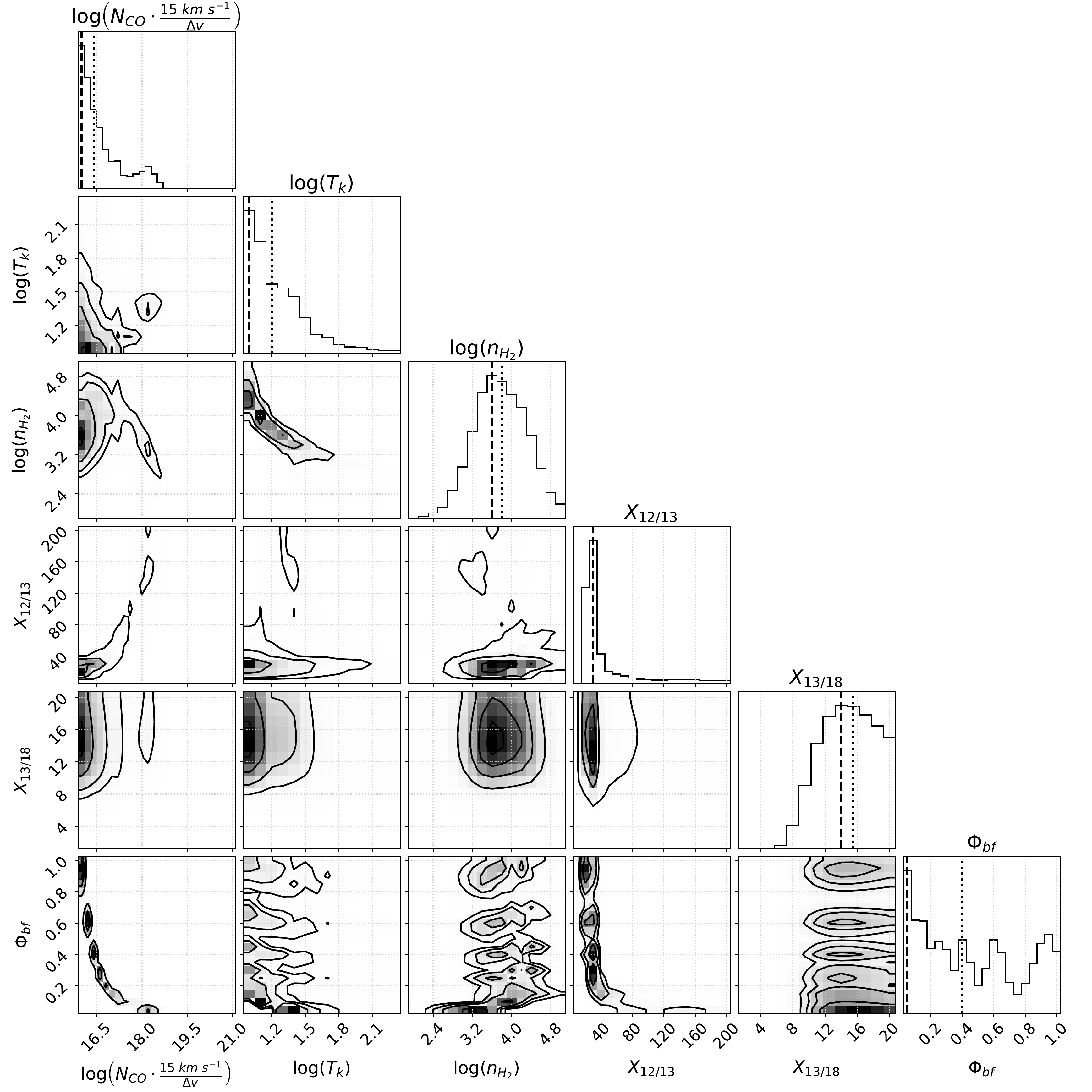}
\caption{Marginalized 1D and 2D likelihood distributions of the stacked spectra from inflow arms. See the caption of Figure~\ref{fig:corner_bright} for more information. Note that some of the parameters are less-constrained in these faint regions and thus pushing the solutions to the boundaries of the grid.}
\label{fig:corner_arms}
\end{figure*}

\begin{figure}
\centering
\includegraphics[width=\linewidth]{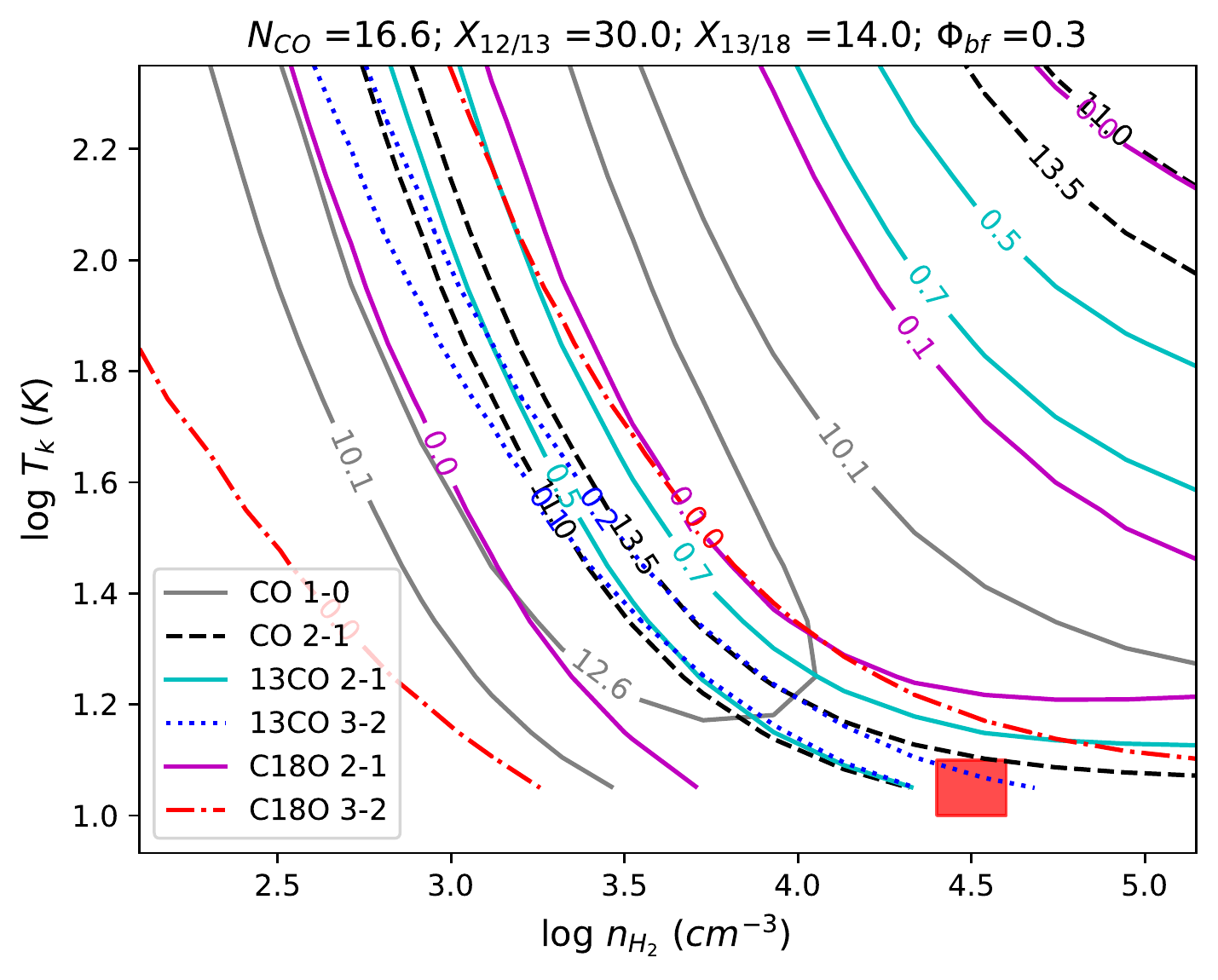}
\caption{Constraints on the $T_\mathrm{k}{-}n_\mathrm{H_2}$ slice where the best-fit parameter (red box) lies assuming a $10$\% flux uncertainty. See the caption of Figure~\ref{fig:flux_contour} for more information.}
\label{fig:bestfit_arms}
\end{figure}

\begin{figure}
\centering
\includegraphics[width=\linewidth]{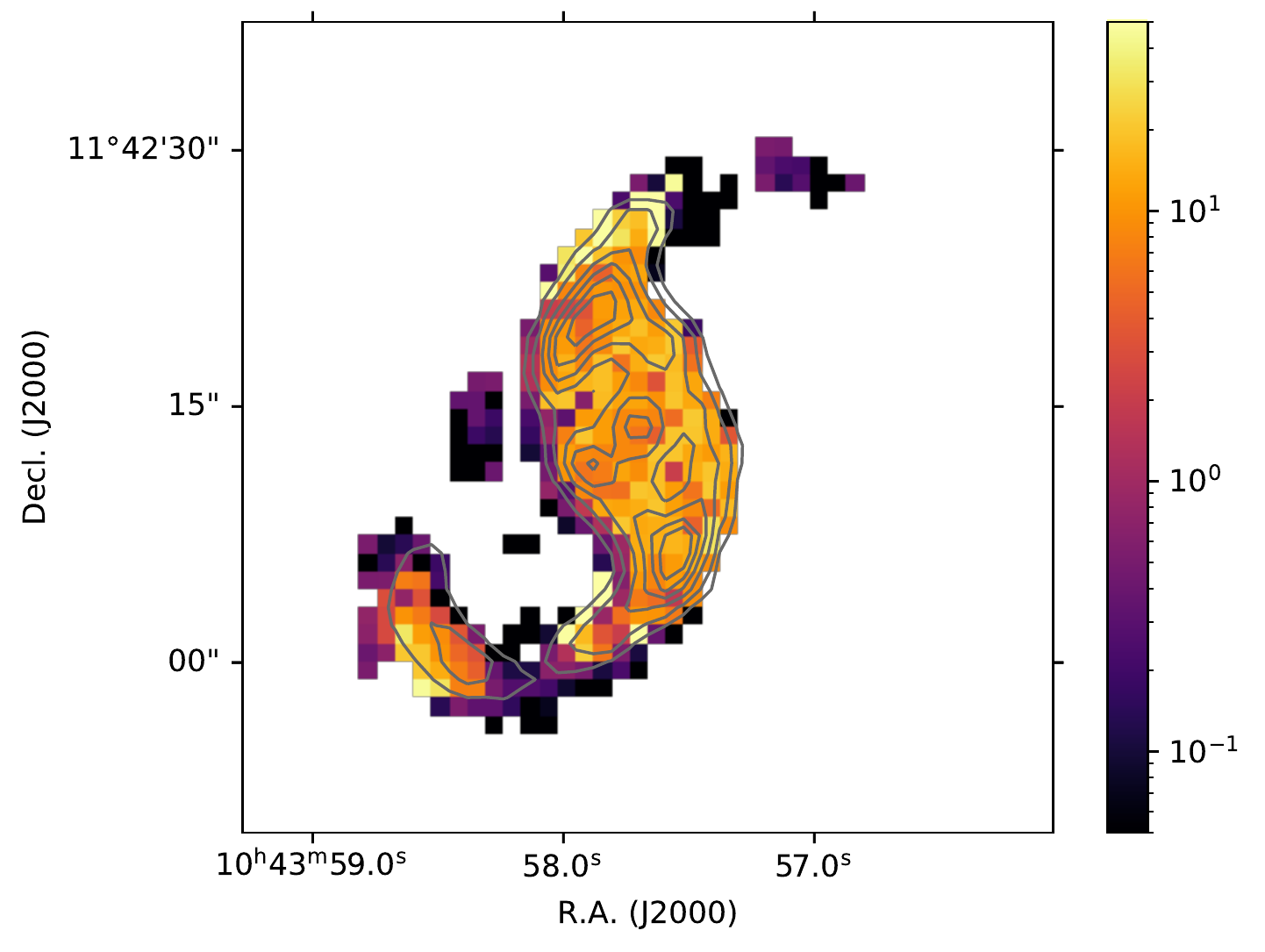}
\caption{Optical depth ($\tau$) map of CO 1--0 determined from the one-component 1DMax physical conditions. The arm regions have lower optical depths than the center, which may originate from higher velocity dispersions. The optically thin regions with $\tau < 1$ are found to have similar $\alpha_\mathrm{CO}$ values to those predicted under LTE assumption \citep{co-to-h2}.}
\label{fig:tau}
\end{figure}

As pixel-based analyses in the arm regions may have higher uncertainties due to low $\mathrm{S/N}$, we stack the spectra from inflow regions to recover low brightness emission applying the approach by \citet{stacking}. First, we set the moment~1 velocities of CO 2--1 as the reference velocity (i.e., $v = 0$) for each pixel and shift the spectra from all other lines to $v = 0$. Next, we extract the spectra in the $v = \pm100$ $\mathrm{km~s}^{-1}$ range and sum them up to obtain a stacked spectrum for each line. Figure~\ref{fig:specfit_arms} presents the shifted and stacked spectra over the arms (defined in Figure~\ref{fig:velocity}c), overlaid with the best-fit Gaussian function. The spectra of the $^{13}$CO lines show much improvement in $\mathrm{S/N}$ after stacking. However, the spectrum of $\rm C^{18}O$ 3--2 is still noisy, as most pixels in the arms are below $1\sigma$ in C$^{18}$O 3--2 (see Figure~\ref{fig:mom0}) and therefore do not have reliable measurements even after stacking. 
Thus, we should be aware that C$^{18}$O 3--2 is not a significant detection in the stacked spectrum of inflow arms, and it is included in the fitting as an upper limit. 

We measure the observed integrated flux of each line and use it as input to the multi-line modeling technique described in Section~\ref{subsec:model_6d}. Figures~\ref{fig:corner_arms} and \ref{fig:bestfit_arms} show the likelihood distributions and the $T_\mathrm{k}{-}n_\mathrm{H_2}$ slice where the best-fit parameter set lies. The 1DMax and best-fit physical conditions to the stacked spectra of inflow arms are found at ($N_\mathrm{CO}/\Delta v$, $T_\mathrm{k}$, $n_\mathrm{H_2}$, $X_{12/13}$, $X_{13/18}$, $\Phi_\mathrm{bf}$) = ($\frac{10^{16}}{15}$ $\mathrm{cm^{-2}\ (km~s^{-1})^{-1}}$, $10$~K, $10^{3.6}$ $\mathrm{cm}^{-3}$, $30$, $14$, $0.05$) and ($\frac{10^{16.6}}{15}$ $\mathrm{cm^{-2}\ (km~s^{-1})^{-1}}$, $10$~K, $10^{4.4}$ $\mathrm{cm}^{-3}$, $30$, $14$, $0.3$), respectively. Both solutions imply lower $N_\mathrm{CO}$, lower $T_\mathrm{k}$, and higher $n_\mathrm{H_2}$ in the arms compared with those in the central ${\sim}1$~kpc. As shown in Figure~\ref{fig:bestfit_arms}, the constraints from five of the six lines have similar shapes, while CO 1--0 is the main constraint that pushes the solution to the lower-right corner. Therefore, even if we exclude the constraints from the C$^{18}$O lines that have low $\mathrm{S/N}$ in the arms, the best-fit solution would remain, which means that C$^{18}$O lines are not the key data in the arm regions. The line width determined by the best-fit Gaussian to the stacked CO 
1--0 spectrum is $\Delta v_\mathrm{FWHM} = 25.18 \pm 0.58$ $\mathrm{km~s}^{-1}$. By substituting the best-fit or 1DMax solutions of $N_\mathrm{CO}/\Delta v$ and $\Phi_\mathrm{bf}$, and the fitted CO 1--0 line width into Equation~\ref{eqn_xco2alpha}, the $\alpha_\mathrm{CO}$ of inflow arms is estimated to be $0.01{-}0.1$ $\mathrm{M_\odot\ (K~km~s^{-1}~pc^2)^{-1}}$. This matches our pixel-based estimation in the arms despite with lower $\mathrm{S/N}$, and thus supports the idea that the inflow regions have substantially lower $\alpha_\mathrm{CO}$ values than the center/ring. 

From Figure~\ref{fig:ratio}d and~e, we find that the $\rm CO/{}^{13}CO$ and $\rm CO/C^{18}O$ line ratios in the inflow arms are ${\sim}2\times$ higher than in the center. This means that we observe more CO emission in the arms compared to the total molecular gas mass, which by definition indicates a lower $\alpha_\mathrm{CO}$ value. One possible scenario that could cause such high line ratios are high CO/$^{13}$CO and CO/C$^{18}$O abundance ratios. It is likely that the circumnuclear star formation activities have enriched $^{13}$C or $^{18}$O in the center but not in the inflows. This could lower the $X_{12/13}$ or $X_{13/18}$ abundances in the center and cause lower CO/$^{13}$CO and CO/C$^{18}$O line ratios than those in the arms. However, both the best-fit and 1DMax solutions based on the stacked spectra of inflow arms suggest $X_{12/13} \sim 30$ and $X_{13/18} \sim 10$, which are similar to those found in the center region as shown in Figure~\ref{fig:1dmax_rmcor_los50}d and~e. Note that the 1D likelihood of $X_{12/13}$ is very well-constrained for the arm regions with stacking (see Figure~\ref{fig:corner_arms}), which is different from the loosely constrained $X_{12/13}$ in the individual pixels of the center region. Even if the center pixels have higher $X_{12/13}$ (e.g., median $X_{12/13} \sim 90$ in Figure~\ref{fig:corner_bright} and~\ref{fig:corner_faint}), this would contradict the expectation of star formation enrichment and would instead enhance CO/$^{13}$CO and CO/C$^{18}$O line ratios in the center. Therefore, changes in CO isotopologue abundances may not be the main reason that causes higher line ratios in the inflow regions.

The high line ratios observed in $\rm CO/{}^{13}CO$ and $\rm CO/C^{18}O$ 2--1 could also be explained if the optical depths of CO changes significantly. Figure~\ref{fig:velocity}b reveals a broader line width in the arms, which may indicate a more turbulent environment or the existence of shear in these bar-driven inflows. This could lead to lower optical depths in the gas inflows. As shown in Figure~\ref{fig:tau}, the optical depths derived from the 1DMax physical conditions are found to be lower in the arms compared to the center, which is likely a combined effect of low $N_\mathrm{CO}$ and high $\Delta v$ (see Equation~\ref{eqn_tau}). Since higher velocity dispersions and consequently lower optical depths allow a larger fraction of CO emission to escape, these may explain why $\alpha_\mathrm{CO}$ is low in the inflow arms. 

In optically thin regions, the average $\alpha_\mathrm{CO}$ is $0.04^{+0.13}_{-0.03}$ $\mathrm{M_\odot\ (K~km~s^{-1}~pc^2)^{-1}}$ based on the marginalized 1DMax $\alpha_\mathrm{CO}$ map shown in Figure~\ref{fig:factor_1comp_marg}a. The optically thin regions were determined by selecting pixels having $\tau < 1$ in Figure~\ref{fig:tau}. \citet{co-to-h2} derived an LTE equation to determine $\alpha_\mathrm{CO}$ as a function of the $\mathrm{CO}/\mathrm{H_2}$ abundance ($x_\mathrm{CO}$) and excitation temperature ($T_\mathrm{ex}$):
\begin{equation}
\alpha_\mathrm{CO} \approx \frac{1.6{\times}10^{19}}{4.5{\times}10^{19}}\,\frac{10^{-4}}{x_\mathrm{CO}}\,\frac{T_\mathrm{ex}} {30\,\mathrm{K}}\,\exp\left(\frac{5.53\,\mathrm{K}}{T_\mathrm{ex}} - 0.184\right)
\label{eqn_lte}
\end{equation}
where $4.5\times10^{19}$ is the factor to convert $X_\mathrm{CO} = N_\mathrm{H_2} / I_\mathrm{CO(1-0)}$ $\mathrm{cm^{-2}\ (K~km~s^{-1})^{-1}}$ to $\alpha_\mathrm{CO}$ (see Equation~\ref{eqn_xco2alpha}).
By plugging our modeled $T_\mathrm{k}$ (Figure~\ref{fig:1dmax_rmcor_los50}b) and the assumed $x_\mathrm{CO}$ of $3\times10^{-4}$ into Equation~\ref{eqn_lte}, we get an average $\alpha_\mathrm{CO}$ of $0.12^{+0.16}_{-0.07}$ $\mathrm{M_\odot\ (K~km~s^{-1}~pc^2)^{-1}}$ in the optically thin regions. This overlaps well with the average derived from our modeled $\alpha_\mathrm{CO}$ distribution, showing that the solutions from multi-line modeling are physically reasonable.
We have also compared the excitation temperatures between CO and $^{13}$CO predicted by the best-fit physical conditions of the averaged spectra in each region. In the arm regions, the predicted excitation temperatures of CO and $^{13}$CO 2--1 lines are at 9.0 and 8.8~K, implying that the conditions in the arms are very close to LTE. On the other hand, the excitation temperatures of $^{13}$CO in the center/ring are lower than those of CO by a factor of ${\sim}1.5{-}2$, indicating a clear departure from LTE that may originate from radiative trapping affecting the CO excitation. Radiative trapping lowers the effective critical density for the CO lines, leading to differences in the excitation for CO and $^{13}$CO. We also note that temperature inhomogeneities in the gas will  tend to bias CO lines to higher temperatures since warmer gas has a higher luminosity and, unlike $^{13}$CO or optically thin tracers, not all CO emission from the full line-of-sight contributes to the measured integrated intensity due to optical depth effects.
It is also important to note that our derived values of $\alpha_\mathrm{CO}$ vary inversely with $x_\mathrm{CO}$. Thus, any increase/decrease to the assumed $x_\mathrm{CO}$ of $3\times10^{-4}$ would lower/raise the numerical values of $\alpha_\mathrm{CO}$.

\subsection{Comparison with Other CO Isotopologue Transitions}

\begin{figure*}
\begin{minipage}{.5\linewidth}
\centering
\includegraphics[width=\linewidth]{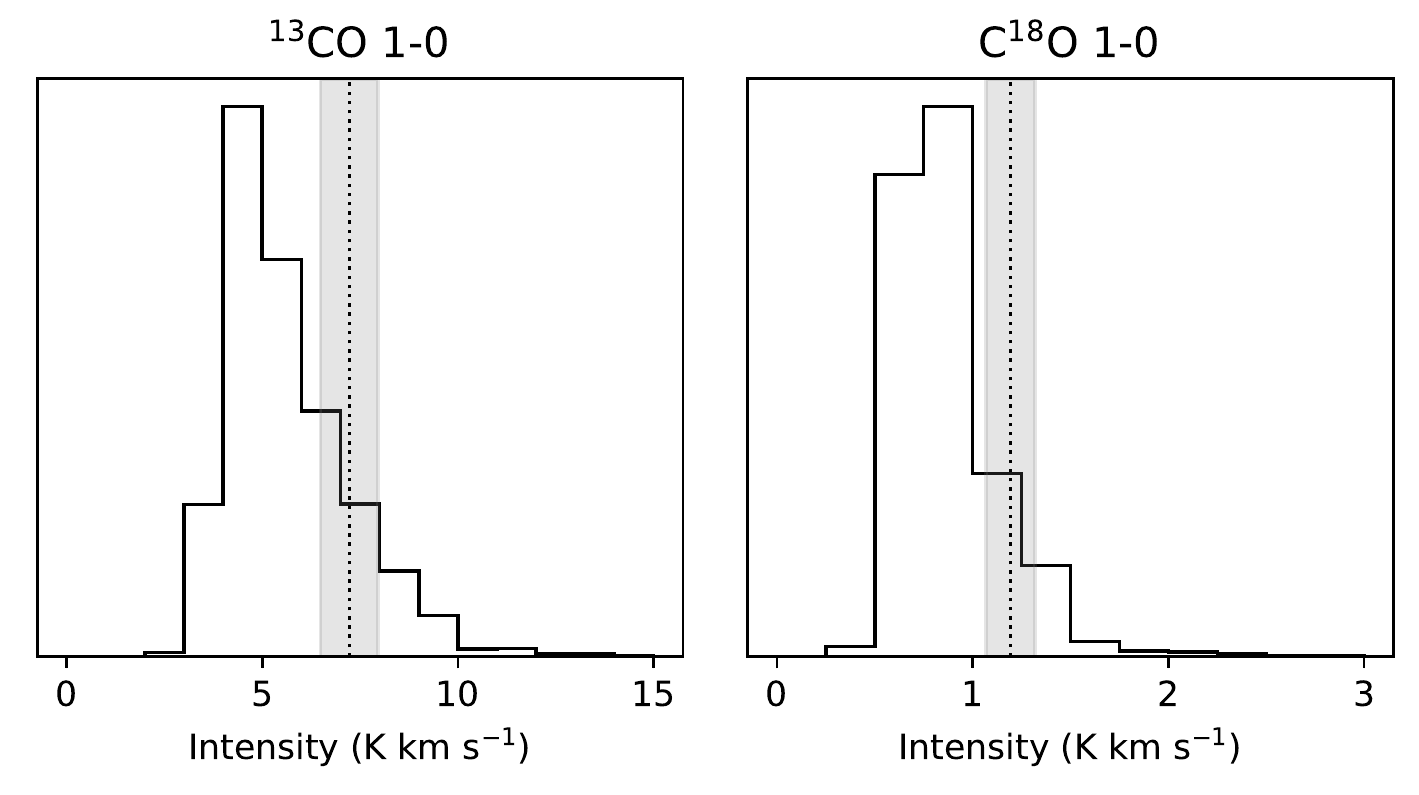}\\
(a)
\end{minipage}
\hfill
\begin{minipage}{.5\linewidth}
\centering
\includegraphics[width=\linewidth]{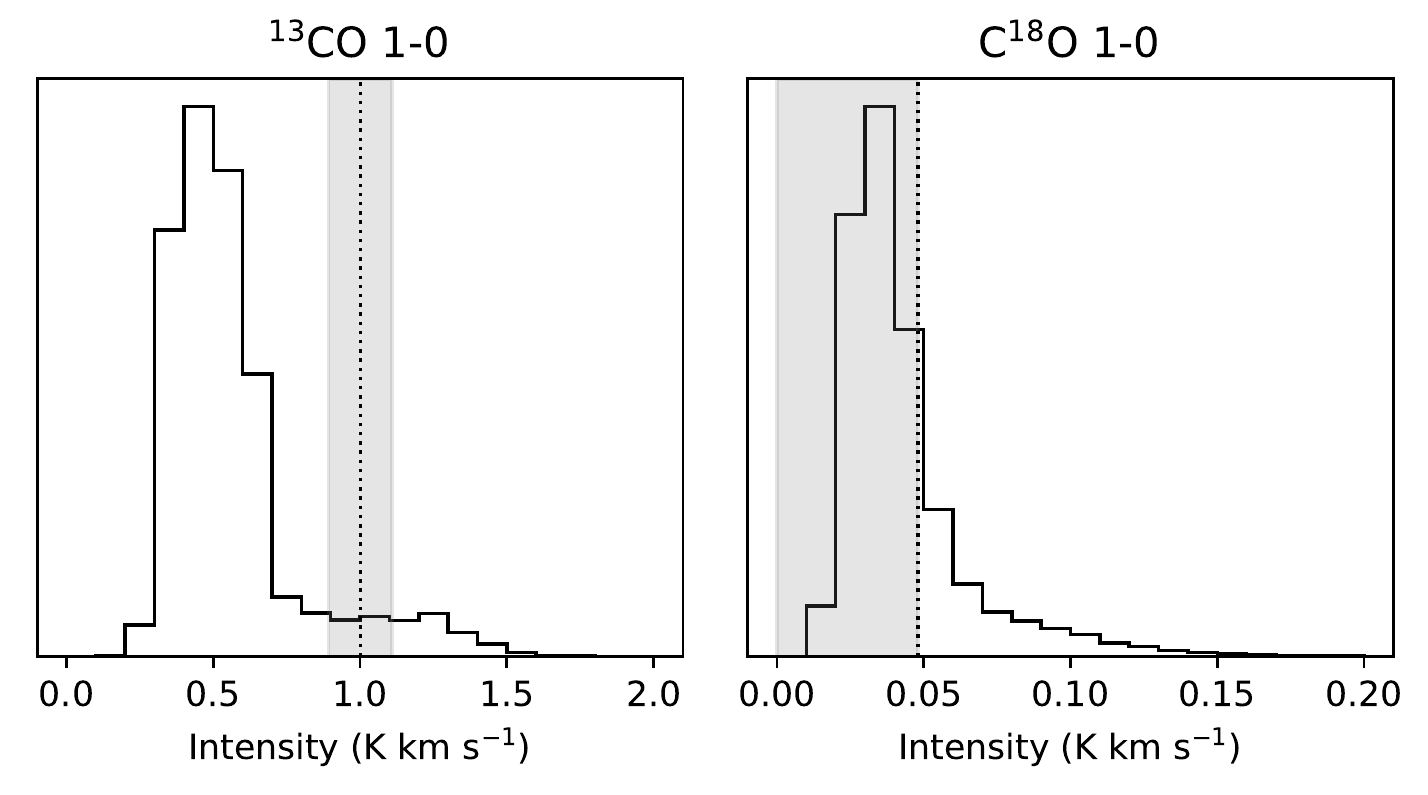}\\
(b)
\end{minipage}
\caption{Marginalized 1D likelihoods of the $^{13}$CO and C$^{18}$O 1--0 intensities in the (a) center and (b) arms regions. The dotted lines mark the data measurements and the shaded regions show the $\pm 1 \sigma$ uncertainties including the measurement uncertainty and a 10\% calibration uncertainty. Since there is no detection of C$^{18}$O 1--0 in the arms, the dotted line in the far right panel marks the $1 \sigma$ upper limit. Except for $^{13}$CO 1--0 showing a factor of ${\sim}2$ higher intensity than the predicted 1DMax value, the observed intensities of all the lines in other regions are within ${\sim}50\%$ of the predicted 1DMax intensities.}
\label{fig:flux10_marg}
\end{figure*}

To compare our modeling results with other observations, we also acquired ALMA data of $^{13}$CO and C$^{18}$O 1--0 with an angular resolution of $8\arcsec$ towards the center of NGC~3351. These data were analyzed in \citet{2017ApJ...836L..29J} and \citet{2018ApJ...858...90G}. Since the resolution is lower than the matched $2.1\arcsec$ resolution in our analysis, we only extract regional averages in the center, ring, and arms for comparison. First, we input the integrated intensities measured from the stacked spectra of each region (e.g., Figure~\ref{fig:specfit_center}) into the six-line modeling described in Section~\ref{subsec:model_6d}. This gives us a full probability grid for each region. Then, we generate the modeled intensity grids of $^{13}$CO and C$^{18}$O 1--0 using RADEX, such that the modeled intensity at each grid point has an associated probability. Finally, we obtain the predicted 1D likelihoods for both lines using the method described in Section~\ref{subsec:marginalize}.

Figure~\ref{fig:flux10_marg}a shows the predicted 1D likelihoods for the center region, covering an intensity range of 0 to 15~K~km~$\mathrm{s^{-1}}$ in steps of 1~K~km~$\mathrm{s^{-1}}$ for $^{13}$CO 1--0 and 0 to 3~K~km~$\mathrm{s^{-1}}$ in steps of 0.25~K~km~$\mathrm{s^{-1}}$ for C$^{18}$O 1--0. The dotted lines mark measured intensities and the shaded regions show the $\pm 1 \sigma$ uncertainties which include the measurement uncertainty and a 10\% calibration uncertainty. Following \citet{stacking}, we estimate the measurement uncertainties using the signal-free part of the spectra and the fitted line widths. In the center region, both the observed intensities are within ${\sim}50\%$ of the predicted 1DMax intensities. The 1D likelihoods and observed intensities for the ring region are similar to those in the center. On the other hand, Figure~\ref{fig:flux10_marg}b shows that both the predicted and observed emission in the arms are fainter than the center/ring. For the arms, the 1D likelihoods range from 0 to 2~K~km~$\mathrm{s^{-1}}$ in steps of 0.1~K~km~$\mathrm{s^{-1}}$ for $^{13}$CO and 0 to 0.2~K~km~$\mathrm{s^{-1}}$ in steps of 0.01~K~km~$\mathrm{s^{-1}}$ for C$^{18}$O. As there is no detection of C$^{18}$O 1--0 in the arms, the dotted line in the right panel of Figure~\ref{fig:flux10_marg}b marks the $1\sigma$ upper limit, which is also within $50\%$ of the 1DMax intensity. Notably, the observed $^{13}$CO 1--0 intensity is a factor of ${\sim}2$ higher than the prediction. This may indicate a more optically-thick CO or a lower $X_{12/13}$ abundance ratio in the arms compared with our modeling results. It could also imply that the excitation temperatures of $^{13}$CO is underestimated in our models.

\section{Discussion} \label{sec:discussion}

\begin{figure*}
\begin{minipage}{.5\linewidth}
\centering
\includegraphics[width=\linewidth]{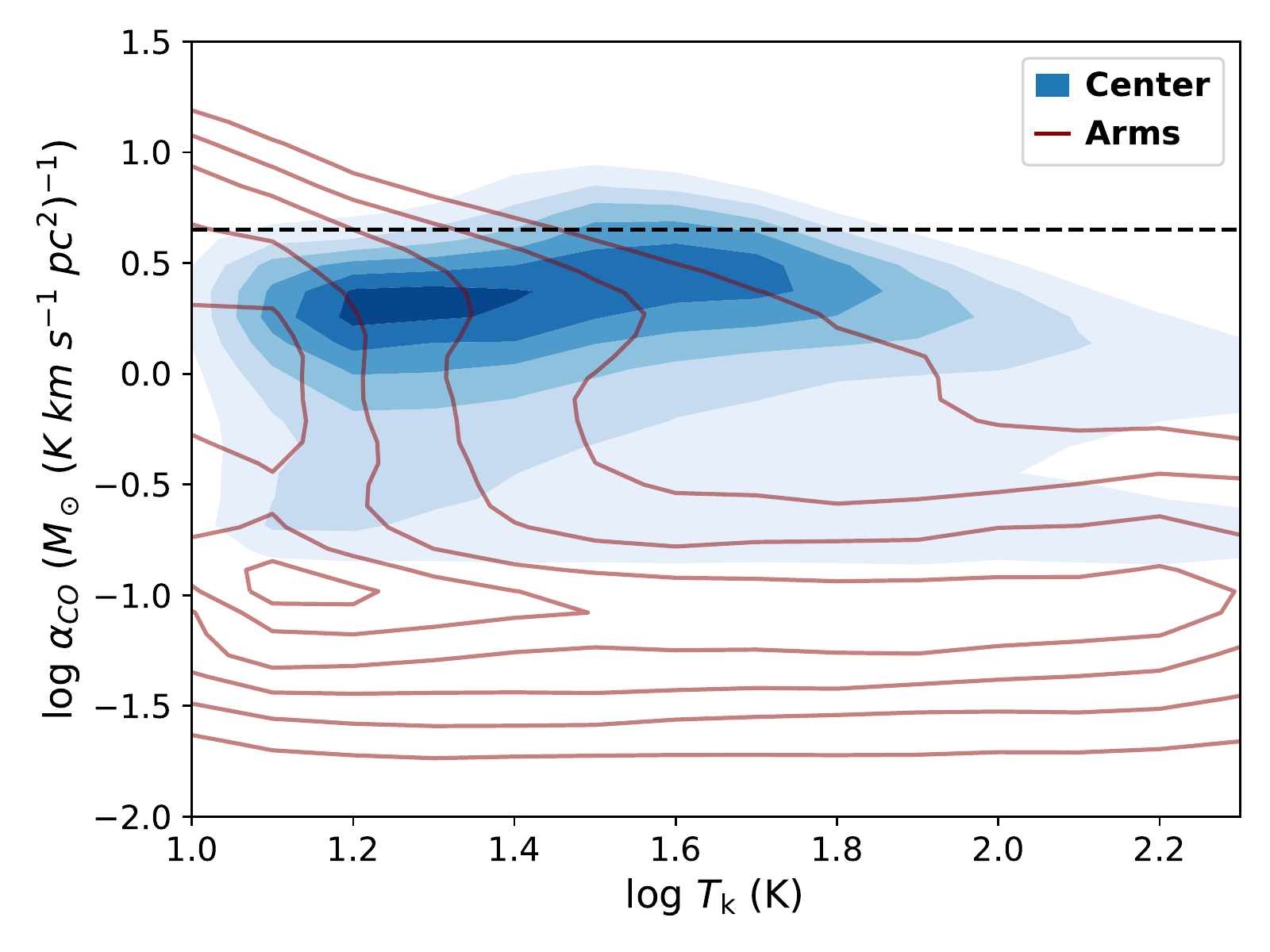}\\
(a)
\end{minipage}
\hfill
\begin{minipage}{.5\linewidth}
\centering
\includegraphics[width=\linewidth]{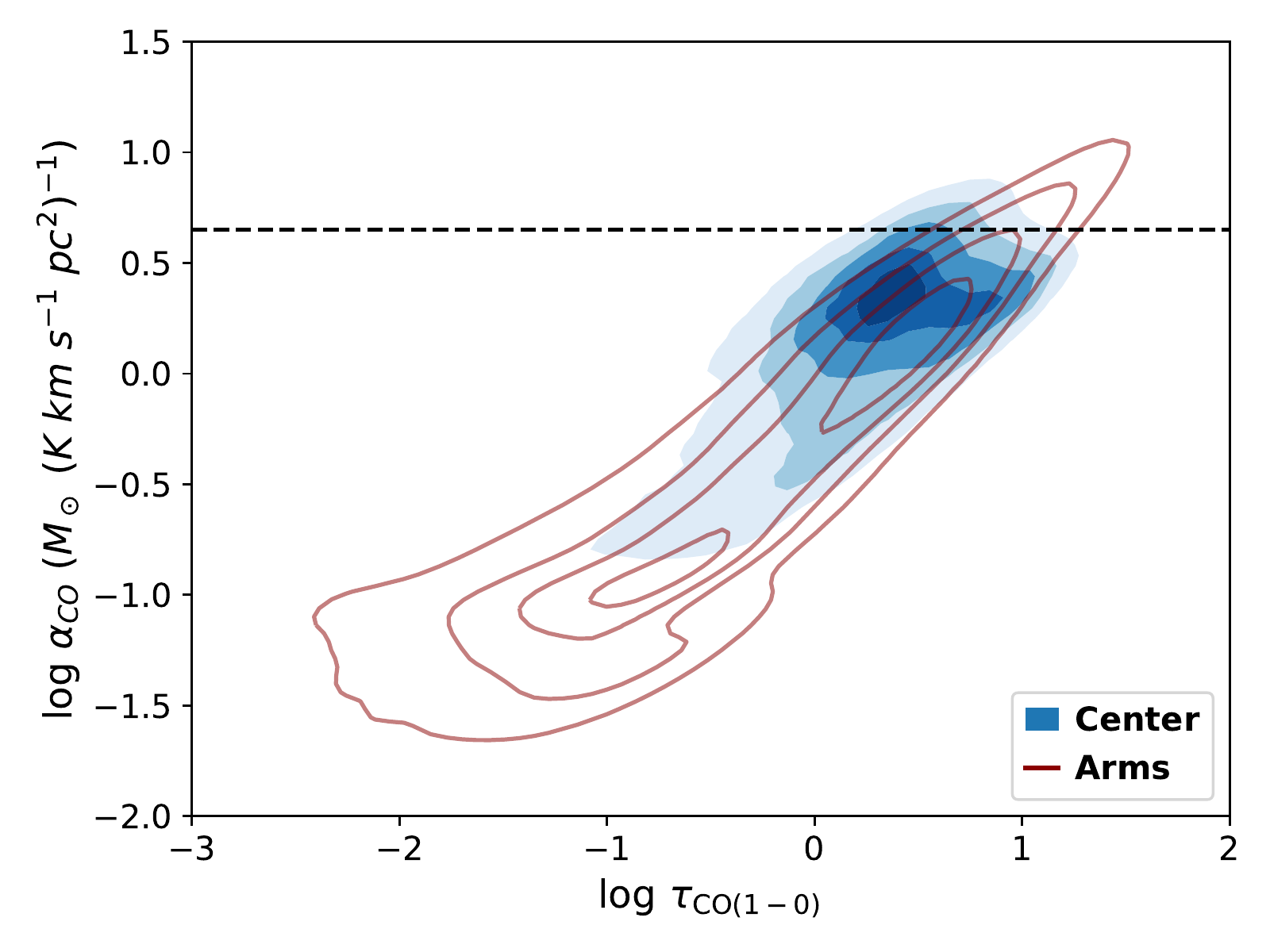}\\
(b)
\end{minipage}
\caption{Relation between $\alpha_\mathrm{CO}$ and (a) $T_\mathrm{k}$ or (b) $\tau_\mathrm{CO(1-0)}$. The contours show density of points from 1000 likelihood-weighted random draws from the one-component model grids for all pixels in each region. The blue/red contours represent the pixels in the center/arms. The spread in the contours reflects both uncertainties and the relationship between the parameters. The dashed lines mark the Galactic $\alpha_\mathrm{CO}$ value. There is no clear signs of correlation between $\alpha_\mathrm{CO}$ and $T_\mathrm{k}$, while $\alpha_\mathrm{CO}$ and $\tau_\mathrm{CO(1-0)}$ potentially shows a positive correlation.}
\label{fig:alpha_tau}
\end{figure*}

Our multi-line modeling, either under the assumption of one- or two-component gas, results in similar distribution of environmental parameters. First, it reveals different physical conditions between the center (i.e., central ${\sim}1$~kpc) of NGC~3351 and its adjacent inflow arms - the $N_\mathrm{CO}/\Delta v \sim 10^{18}/15$~$\mathrm{cm^{-2}}~(\mathrm{km~s^{-1}})^{-1}$ found in the center is much higher than $N_\mathrm{CO}/\Delta v \sim 2 \times 10^{16}/15$~$\mathrm{cm^{-2}}~(\mathrm{km~s^{-1}})^{-1}$ in the arms. Second, the temperature is ${\sim}30{-}60$~K near the nucleus and both contact points, while other regions show lower temperatures of ${\sim}10{-}20$~K (see Figure~\ref{fig:1dmax_rmcor_los50}b). Third, the $\mathrm{H_2}$ volume densities are ${\sim}2{-}3 \times 10^{3}$~$\mathrm{cm^{-3}}$ across the whole region. The derived temperature and density ranges are consistent with those found in the Milky Way's CMZ \citep{2013MNRAS.429..987L,2016A&A...586A..50G,2017ApJ...850...77K}. Finally, the arm regions have lower optical depths than the center, which could originate from the broader line widths, shear, and bulk gas flows in the arms. The lower optical depth in the arms allows more CO emission to escape and is possibly the reason why higher CO/$^{13}$CO and CO/C$^{18}$O ratios are observed in the arms (see Figure~\ref{fig:ratio}d and e). While a high $X_{12/13}$ value can also cause such high line ratios, this may not be the main reason as our modeling of the arm regions with stacking suggests a normal and well-constrained $X_{12/13}$ similar to the center. 
Observational studies of other barred galaxies have also found significantly higher CO/$^{13}$CO line ratio in their centers and shown that it is more likely to result from the dramatic change in optical depths or line widths than a high $X_{12/13}$ abundance \citep[e.g., NGC~3627:][NGC~7465: \citealt{2021ApJ...909...98Y}]{2015PASJ...67....2M}.
In addition, we think that turbulence and/or shear in the inflows is likely the cause of higher velocity dispersion in the arms, which is similar to the shear-driven inflows and turbulence in the CMZ of the Milky Way \citep[e.g.][]{2016A&A...586A..50G,2019MNRAS.484.5734K,2021arXiv210608461H}. 

Based on the one- and two-component models, the derived $\alpha_\mathrm{CO}$ also differs significantly between the center and the arms. Assuming a $x_\mathrm{CO}$ of $3 \times 10^{-4}$, we find $\alpha_\mathrm{CO} \sim 0.5{-}2$~$\mathrm{M_\odot\ (K~km~s^{-1}~pc^2)^{-1}}$ in the center, which is lower than the Galactic value at $4.4$~$\mathrm{M_\odot\ (K~km~s^{-1}~pc^2)^{-1}}$. In the arms, the value of $\alpha_\mathrm{CO}$ is even found to be approximately an order of magnitude lower than the center. The center and the arms altogether gives an intensity-weighted mean $\alpha_\mathrm{CO}$ of ${\sim}1.5$~$\mathrm{M_\odot\ (K~km~s^{-1}~pc^2)^{-1}}$ over the entire observed region of ${\sim}2$~kpc. Using the dust mass surface density derived from infrared SED modeling and lower resolution CO and H\,{\small I} observations, \citet{2013ApJ...777....5S} found a similar $\alpha_\mathrm{CO}$ value of $1.0^{+0.4}_{-0.3}$~$\mathrm{M_\odot\ (K~km~s^{-1}~pc^2)^{-1}}$ in the central 1.7~kpc of NGC~3351 at $40\arcsec$ scales. Their method assumes that the dust-to-gas ratio is constant across the center and finds the $\alpha_\mathrm{CO}$ which best reproduces that result by minimizing scatter in dust-to-gas. Since the method used by \citet{2013ApJ...777....5S} is completely independent of the multi-line modeling we use, the fact that both methods reach a similar $\alpha_\mathrm{CO}$ value gives us more confidence in both results. 

With the results of environmental conditions and $\alpha_\mathrm{CO}$ summarized above, we conclude that the low $\alpha_\mathrm{CO}$ in the central few kpc of NGC~3351 results from a combination of two factors. First, the central ring/nucleus has a slightly lower-than-Galactic $\alpha_\mathrm{CO}$. Secondly, the inflow arms have a substantially lower $\alpha_\mathrm{CO}$, because they consist mainly of optically-thin gas. 
\citet{2018ApJ...860..172S,2020ApJ...901L...8S} have analyzed cloud-scale molecular gas properties in nearby galaxy samples and clearly showed the increase of velocity dispersion and turbulent pressure toward barred galaxy centers. For the centers of barred galaxies, they reported a mass-weighted velocity dispersion that is ${\sim}5$ times higher than the disk regions. This could be the cause for the ${\sim}2{-}9$ times lower-than-Galactic $\alpha_\mathrm{CO}$ in the central ${\sim}1$~kpc of NGC~3351, as the Galactic $\alpha_\mathrm{CO}$ value was based on molecular cloud measurements in the disk of the Milky Way. 

Nevertheless, it is also possible given a different $x_\mathrm{CO}$, that the central ${\sim}1$~kpc of NGC~3351 has a Galactic $\alpha_\mathrm{CO}$, so that the significantly lower $\alpha_\mathrm{CO}$ from the arms would become the main reason for a lower-than-Galactic $\alpha_\mathrm{CO}$ observed over the central few kpc region. As mentioned in Section~\ref{sec:alpha_co}, the assumed $x_\mathrm{CO}$ value of $3\times10^{-4}$ is appropriate for starburst regions, and thus it is higher than the common assumption of ${\sim}10^{-4}$. If $x_\mathrm{CO}$ is in fact a factor of two or three lower than our assumption in the center, the derived value of $\alpha_\mathrm{CO}$ would increase linearly and lead to a nearly-Galactic $\alpha_\mathrm{CO}$ in this region. This resulting situation of a nearly-Galactic $\alpha_\mathrm{CO}$ in the center and a significantly lower $\alpha_\mathrm{CO}$ in the inflow arms would then be similar to that found in \citet{2012ApJ...751...10P} and \citet{2015ApJ...801...25L}, where they reported a nearly-Galactic $\alpha_\mathrm{CO}$ for GMCs and a much lower $\alpha_\mathrm{CO}$ value for non-GMC associated gas. 

Many explanations for low $\alpha_\mathrm{CO}$ values have been proposed in the literature. From a theoretical perspective, \citet{co-to-h2} have shown that enhanced temperatures may lead to lower $\alpha_\mathrm{CO}$ for isolated and virialized GMCs. The magnetohydrodynamics simulation conducted by \citet{2020ApJ...903..142G} also showed that $\alpha_\mathrm{CO}$ decreases with increasing heating from cosmic ray ionization. On the other hand, \citet{2012ApJ...751...10P} showed the dependence of $\alpha_\mathrm{CO}$ on optical depths for thermalized, optically thick emission. In such a case, $\alpha_\mathrm{CO}$ is expected to be approximately proportional to $\tau / [1-\exp(-\tau)]$ if the excitation temperature is kept constant. 
To understand if the $\alpha_\mathrm{CO}$ variations in the galaxy center of NGC~3351 are more related to increased temperature or to decreased optical depth, we investigate the correlation between the inferred $\alpha_\mathrm{CO}$ and the kinetic temperatures or CO optical depths derived from the one-component modeling.
For each pixel in the center and arms, we conduct 1000 likelihood-weighted random draws from the 6D full grids of the predicted $T_\mathrm{k}$, $\tau_\mathrm{CO(1-0)}$ and $\alpha_\mathrm{CO}$. Figure~\ref{fig:alpha_tau} shows the density of the points from the random draw. The spread of these points reflects both the parameter uncertainties and the overall relationship between $\alpha_\mathrm{CO}$ and $T_\mathrm{k}$ or $\tau_\mathrm{CO(1-0)}$. 
In Figure~\ref{fig:alpha_tau}a, the center region spans a wide range in $T_\mathrm{k}$ and shows a roughly constant $\alpha_\mathrm{CO}$ value. For the arms, there are two distinct regions with a higher or lower $\alpha_\mathrm{CO}$ values, while the low-$\alpha_\mathrm{CO}$ region extending to the highest temperatures indicates large $T_\mathrm{k}$ uncertainties in these pixels. Thus, we do not see clear correlation between $\alpha_\mathrm{CO}$ and $T_\mathrm{k}$ in either the center or the arm regions.
In contrast, Figure~\ref{fig:alpha_tau}b clearly shows that $\alpha_\mathrm{CO}$ and $\tau_\mathrm{CO(1-0)}$ are more correlated, with both the center and arm regions spanning a range in optical depths and $\alpha_\mathrm{CO}$. The distribution in Figure~\ref{fig:alpha_tau}b suggests a positive correlation that is consistent with the thermalized $\alpha_\mathrm{CO}$ expression in \citet{2012ApJ...751...10P}, where $\alpha_\mathrm{CO}$ is expected to increase linearly with $\tau$ at $\tau \gg 1$ but barely rise at $\tau \ll 1$ if $T_\mathrm{ex}$ is constant. Furthermore, the data points from the center and arms are aligned in the $\alpha_\mathrm{CO}$-$\tau_\mathrm{CO(1-0)}$ parameter space, showing a continuous transition between the optically-thin and optically-thick regimes.

\section{Conclusion} \label{sec:conclusion}

We present ALMA observations of multiple CO, $\rm ^{13}CO$, and $\rm C^{18}O$ rotational lines at ${\sim}100$~pc resolution toward the central ${\sim}2$~kpc region of NGC~3351. We constrain the distributions of multiple environmental parameters using multi-line radiative transfer modeling and a Bayesian likelihood analysis along each sight line. With the probability distribution of physical parameters at each pixel, we derive the spatial variation of $\alpha_\mathrm{CO}$. We construct models with a one-component or two-component assumption on gas phases and compare the results. To recover faint emission from the inflow arms, we also conduct spectral stacking and compare with the pixel-based analysis. Our main results are summarized as follows:
  
\begin{enumerate}

\item All of the CO, $\rm ^{13}CO$, and $\rm C^{18}O$ images resolve a compact nucleus at the galaxy center, a circumnuclear ring of star formation in the central ${\sim}1$~kpc, and an emission-faint gap region inbetween. Except for $\rm C^{18}O$ 3--2, the images reveal two bar-driven inflow arms connected to the northern and southern part of the ring (i.e., contact points). The emission is brightest in all the lines at both contact points. 

\item The 1DMax solutions from multi-line modeling show no clear variation in CO isotopologue abundance ratios. In the central ${\sim}1$~kpc, we find $X_{12/13} \sim 20{-}30$ and $X_{13/18} \sim 6{-}10$, which are consistent with the values of the Galactic Center. However, we note that $X_{12/13}$ is the least well-constrained property with a broad 1D likelihood distribution in each pixel.

\item Both the one- and two-component modeling results suggest a dominant gas phase with $n_\mathrm{H_2} \sim 2{-}3\times10^3~\mathrm{cm}^{-3}$ in the center and a higher $T_\mathrm{k}$ of ${\sim}30{-}60$~K near the nucleus and contact points than in the rest of the regions. This density and temperature are consistent with those found in the Milky Way's Central Molecular Zone at pc resolutions. 

\item The derived $\alpha_\mathrm{CO}$ distributions based on the one- and two-component models reveal similar spatial variations: in the central $20\arcsec$ (${\sim}1$~kpc), $\alpha_\mathrm{CO}$ ${\sim}0.5{-}2$~($3\times10^{-4}/x_\mathrm{CO}$)~$\mathrm{M_\odot\ (K~km~s^{-1}~pc^2)^{-1}}$, which is a factor of 2--9 lower than the Galactic value, and it slowly decreases with increasing galactocentric radius; on the other hand, a substantially lower $\alpha_\mathrm{CO}$ value of ${\lesssim}0.1$~($3\times10^{-4}/x_\mathrm{CO}$)~$\mathrm{M_\odot\ (K~km~s^{-1}~pc^2)^{-1}}$ is found in the inflow arms. We also derive similarly low $\alpha_\mathrm{CO}$ values from a stacking analysis in the arms and using the LTE expression for $\alpha_\mathrm{CO}$ in optically thin regions.

\item The substantially lower $\alpha_\mathrm{CO}$ in the arms can be explained by lower optical depths which implies a higher escape probability for CO emission. The significantly higher CO/$\rm ^{13}CO$ and CO/$\rm C^{18}O$ 2--1 ratios in the arms support this scenario, and the large velocity dispersions observed in the arms, likely due to turbulence or shear in the inflows, may be the reason behind such low optical depths. 

\item The $\alpha_\mathrm{CO}$ in the center/ring does not show correlation with temperature. The lower-than-Galactic $\alpha_\mathrm{CO}$ in this region may be due to higher velocity dispersions in barred galaxy centers than in the disk of the Milky Way. Nevertheless, if our assumption of $x_\mathrm{CO} \sim 3\times10^{-4}$ is a factor of few higher than reality, the $\alpha_\mathrm{CO}$ in the center/ring could have nearly Galactic $\alpha_\mathrm{CO}$.

\item We derive an intensity-weighted mean $\alpha_\mathrm{CO}$ of $1.79\pm0.10$ and $1.11\pm0.09$~($3\times10^{-4}/x_\mathrm{CO}$)~$\mathrm{M_\odot\ (K~km~s^{-1}~pc^2)^{-1}}$ over the observed ${\sim}2$~kpc region, based on one- and two-component models, respectively. This result is a combination of the central ${\sim}1$~kpc region with a slightly lower-than-Galactic $\alpha_\mathrm{CO}$ and the inflow arm regions with a significantly lower $\alpha_\mathrm{CO}$. The derived values of the overall $\alpha_\mathrm{CO}$ are similar to that determined by \citet{2013ApJ...777....5S} using dust modeling at kpc scales. 
\end{enumerate}

Overall, our results suggest that dynamical effects and non-GMC associated gas can be important factors that cause $\alpha_\mathrm{CO}$ variations. In the inflow arms of NGC~3351, $\alpha_\mathrm{CO}$ is approximately an order of magnitude lower than in the center, possibly due to molecular gas with low optical depths resulting from broader line widths. Therefore, when using CO to trace molecular gas in galaxies with dynamical features that drive inflows (e.g., bars, ovals, interactions, mergers), it is important to account for $\alpha_\mathrm{CO}$ variations originating from changes in the velocity dispersion and therefore optical depth in specific regions.

\acknowledgments
We thank the referee for helpful comments that improved the manuscript. Y.-H.T. thanks I-D. Chiang and J. Chastenet for helpful discussion at group meetings. Y.-H.T. and K.S. acknowledge funding support from NRAO Student Observing Support Grant SOSPADA-012 and from the National Science Foundation (NSF) under grant No. 2108081.
The work of J.S. and A.K.L. is partially supported by the National Science Foundation (NSF) under Grants No.\ 1615105, 1615109, and 1653300.
A.D.B. acknowledges partial support from grant NSF-AST210814.
J.M.D.K.\ gratefully acknowledges funding from the Deutsche Forschungsgemeinschaft (DFG, German Research Foundation) through an Emmy Noether Research Group (grant number KR4801/1-1) and the DFG Sachbeihilfe (grant number KR4801/2-1), as well as from the European Research Council (ERC) under the European Union's Horizon 2020 research and innovation programme via the ERC Starting Grant MUSTANG (grant agreement number 714907).
A.U. acknowledges support from the Spanish funding grants PGC2018-094671-B-I00 (MCIU/AEI/FEDER) and PID2019-108765GB-I00 (MICINN).
A.T.B. and F.B. acknowledge funding from the European Research Council (ERC) under the European Union’s Horizon 2020 research and innovation programme (grant agreement No. 726384/Empire).
E.R. acknowledges the support of the Natural Sciences and Engineering Research Council of Canada (NSERC), funding reference number RGPIN-2017-03987.

This paper makes use of the following ALMA data:
ADS/JAO.ALMA\#2013.1.00634.S,\\  
ADS/JAO.ALMA\#2013.1.00885.S,\\  
ADS/JAO.ALMA\#2015.1.00956.S,\\  
ADS/JAO.ALMA\#2016.1.00972.S. \\     
ALMA is a partnership of ESO (representing its member states), NSF (USA), and NINS (Japan), together with NRC (Canada), NSC and ASIAA (Taiwan), and KASI (Republic of Korea), in cooperation with the Republic of Chile. The Joint ALMA Observatory is operated by ESO, AUI/NRAO, and NAOJ. The National Radio Astronomy Observatory is a facility of the National Science Foundation operated under cooperative agreement by Associated Universities, Inc.

We acknowledge the usage of NASA's Astrophysics Data System (\url{http://www.adsabs.harvard.edu}) and \texttt{ds9}, a tool for data visualization supported by the Chandra X-ray Science Center (CXC) and the High Energy Astrophysics Science Archive Center (HEASARC) with support from the JWST Mission office at the Space Telescope Science Institute for 3D visualization.

\vspace{5mm}
\facilities{ALMA}

\software{\texttt{CASA} \citep{2007ASPC..376..127M}, \texttt{ds9} \citep{2000ascl.soft03002S,2003ASPC..295..489J}, \texttt{matplotlib} \citep{Hunter:2007}, \texttt{numpy} \citep{harris2020array}, \texttt{scipy} \citep{Virtanen_2020}, \texttt{astropy} \citep{2013A&A...558A..33A,2018AJ....156..123A}, \texttt{RADEX} \citep{radex}, \texttt{corner} \citep{Foreman-Mackey2016} \texttt{spectral-cube} \citep{2019zndo...3558614G}}

\bibliography{sample63}{}
\bibliographystyle{aasjournal}

\end{document}